# Semileptonic Decays of Heavy Flavors in QCD

by

Lev A. Koyrakh, M.S.

**Dissertation**

Presented to the Faculty of the Graduate School of

The University of Minnesota

in Partial Fulfillment

of the Requirements

for the Degree of

**Doctor of Philosophy**

**The University of Minnesota**

May 1996

This work is dedicated to:
my parents Abram and Martha,
my wife Natalia
and children Roman and Alice.

# Acknowledgments

I am greatly indebted to a number of individuals who have helped me during my graduate studies. Many members of the Univerity of Minnesota physics department faculty have been extremely helpful. I am particularly grateful for the insights and assistance shared with me by my advisor Prof. A. Vainshtein, and Prof. M. Voloshin. Different parts of this work were done together with B. Blok, N. Uraltsev, M. Shifman and A. Vainshtein. Chapter 2 is based on the work [13], and Chapters 4 and 6 – on the work [34].

I would like to thank my parents, Martha Slobodkina and Abram Koyrakh for their support during my graduate studies.

I am especially thankful to my wife Natalia and children Roman and Alice, whose continued support and sacrifice have proved vital throughout my graduate career.

I also would like to akcnowledge the great help from *MATHEMATICA* by Wolfram Research, Inc. in making a lots of lengthy calculations fun to play with.

And I am grateful to the people of the United States who made it possible for me to come to this beautiful country and study the science I love so much.

My dissertation research was supported, in part, by the US Department of Education Fellowships.

<div style="text-align:right">Lev A. Koyrakh</div>

*The University of Minnesota*
*May 1996*

# Semileptonic Decays of Heavy Flavors in QCD

Lev A. Koyrakh, Ph.D.

The University of Minnesota, 1996

Supervisor: A. Vainshtein

A model independent approach based on a generalization of the operator product expansion is used to describe semileptonic decays of heavy flavors. In the first part of the dissertation we calculate differential distributions in the inclusive semileptonic weak decays of heavy flavors in QCD. In particular, the double distribution in electron energy and invariant mass of the lepton pair is calculated. The distributions are calculated as series in $m_Q^{-1}$ where $m_Q$ is the heavy quark mass. All effects up to $m_Q^{-2}$ are included.

Also calculated are the energy distribution and semileptonic decay width for the case of a heavy lepton in the final state. In the case of $B$-meson decays, for $b \to u\tau\bar{\nu}$ transitions the nonperturbative corrections decrease the decay rate by 6% of its perturbative value, while for $b \to c\tau\bar{\nu}$ they decrease it by 10%.

Based on the results of the first part of the work, the full up to order $1/m_Q^2$ set of the OPE sum rules for the heavy flavor transitions and radiative corrections to them up to order $\alpha_s \Lambda_{\rm QCD}/m_Q$ are calculated.

A new model is proposed for the inclusive semileptonic decays of the $B$ mesons $B \to l\bar{\nu}X_c$, which is *defined* by the requirement that it should satisfy the QCD consistency conditions. These conditions are imposed in the form of the OPE sum rules. It is shown that under some natural assumptions the OPE sum rules provide sufficient number of constraints to fully determine exclusive contributions of a number of resonances in the final state of the decay. The proposed model can be used for experimental measurements of the heavy meson matrix element $\mu_\pi^2$ which describes the kinetic energy of the heavy quark inside the hadron, as well as for extraction of value of $|V_{cb}|$ from experimental data.

# Table of Contents

# List of Figures

# Chapter 1

# Introduction

The problem of theoretical understanding of semileptonic decays of heavy particles have long attracted theorists's attention. One of the most challenging aspects of the problem has to do with the strong interaction. Namely it is difficult to theoretically calculate different formfactors for exclusive decay channels. The problem is that the exclusive channels are formed by the strong interaction at large distances when the perturbation theory is not applicable. Phenomenological models have been employed in the situations when theoretical calculation from the first principles was not possible. Different models of the semileptonic decays of heavy quarks have been suggested [2],[28]. Those models are based on different sides of theoretical understanding of the problem. The ACCMM (Altarelli, Cabibbo, Corbò, Maiani and Martinelli [2]) model takes into account the so called primordial motion of the heavy quark inside the hadron and assigns Gaussian distribution of the heavy quark spatial momentum with the so called Fermi momentum $\mathbf{p}_F$ being the width (dispersion) of that distribution. Of course, one of immediate consequences of this picture is that the lifetime of the meson becomes longer than in a free quark decay due to the Lorentz time dilation. The other, ISGW (Isgur, Scora, Grinstein, Wise [28],[47],[46]) model uses quark-antiquark potential in order to calculate the formfactors, which are obtained as overlap integrals between final and initial states wave functions of the so called light degrees of freedom of the corresponding mesons.

In the recent years model independent predictions for the lepton energy distributions as well as double distributions in the lepton energy and invariant mass of the lepton pair in the decays have been obtained [5],[6],[13],[38]. These predictions are based on the QCD operator product expansion (OPE) in which two new phenomenological parameters describing heavy particles were introduced: matrix element of the chromomagnetic operator $\mu_G^2$ which is responsible for the hyperfine mass splitting within heavy quark doublets, and $\mu_\pi^2$ - average kinetic energy of the heavy quark inside the hadron. The results of the operator product expansion are related to measurable quantities via sum rules [49],[48],[9],[10]. One can also view the sum rules as model independent constraints which should be satisfied by any reasonable phenomenological model of the decay. In the present work an attempt is made to introduce a phenomenological model of semileptonic decays of $B$ mesons which is consistent with the QCD in the sense that it satisfies the OPE sum rules. In the proposed model the final state of the decay is represented as sum of different resonance contributions along with continuum states. Then the resulting hadronic tensor is restrained by the OPE sum rules. It is shown that with limited number of resonances in the final state the constraints could be solved exactly yielding predictions for different formfactors for the decay thus satisfying the QCD conditions. Once the constraints are solved, one can use them to calculate the hadronic tensor and differential distributions in the semileptonic decays. In this way the differential distributions in the decay could be described without $\delta$-function singularities which appear in the pure OPE approach. One can also relate the parameter $\mu_\pi^2$ to the quantities which could be (at least in principal) measured by experimentalists (such as hadronic structure functions, or at least the lepton spectrum) and in turn measure the $\mu_\pi^2$. This approach could be also useful in a more accurate determination of $|V_{cb}|$ and $|V_{ub}|$ from experimental data.

In order to perform a lot of calculations done in this work a special program package was written in *MATHEMATICA* (Wolfram Research, Inc.). The package includes a small but effective "Lorentz calculator" intended for performing calculations and simplifications of expressions involving various tensor and gamma matrix structures, and a special program which effectively does the operator product expansion described in this work.

The dissertation is organized as follows. In Chapter 2, based on the work [13], the operator product expansion is used to derive the differential distributions in semileptonic decays of heavy flavors with a massless final state lepton (electron), then in Chapter 3, based on the paper [33], the lepton energy distributions are derived for the case of a massive lepton ($\tau$–lepton) in the final state. In mostly review Chapter 4 the results of Chapter 2 are used to derive the OPE sum rules (in that we follow works [48],[9],[10]), and the full list of the sum rules is calculated. In Chapter 6 a new model of semileptonic decays of heavy flavors is proposed. The model is based on the QCD consistency conditions following from the OPE sum rules. In Chapter 6 the model is formulated in the heavy quark limit. Perturbative corrections to the sum rules are calculated in Chapter 5. In Chapter 7 the model is generalized to take into account perturbative and nonperturbative corrections to the OPE sum rules. Numerical analysis of the proposed model is made in Section 7.4.

# Chapter 2

# Differential distributions in semileptonic decays of heavy flavors in QCD

## 2.1 Introduction

Differential distributions in semileptonic decays of heavy flavors are used for measurements of the CKM matrix elements, key phenomenological parameters of the standard model. To extract the CKM matrix elements from data one needs to disentangle the effects of strong interactions at large distances from the quark-lepton lagrangian known at short distances.

Up to now essentially two approaches are applied to describe nonperturbative strong interaction effects in the inclusive weak decays: the naive parton model amended to include the motion of the heavy quark inside the decaying meson [2]; and the 'exclusive variant' based on summation of different channels, one by one [28]. Both approaches are admittedly model-dependent, neither their accuracy nor the connection to the fundamental parameters of QCD are clear *a priori*. Each of them needs an input from constituent quark model to parametrize nonperturbative effects. The latter play an especially important role in the form of the spectra near the endpoints.

The need for the model-independent QCD-based predictions is apparent. Consid-erable progress achieved recently in the theory of preasymptotic effects (proportional to powers of $1/m_Q$ where $m_Q$ is the heavy quark mass) allows one to make these predictions.

The theoretical construction presented in this chapter is, in a sense, a generalization and combination of the formalisms which are used in deep inelastic scattering and total cross section of $e^+e^-$ annihilation. The expansion parameter in deep inelastic scattering is $Q^{-1}$ where $Q$ is the momentum transfer. In the problem at hand the expansion parameter is $m_Q^{-1}$ or, more exactly, the inverse energy released in the final hadronic state (in the rest frame of the decaying quark).

In the classical problems of this type, like $e^+e^-$ – annihilation, there are two alternative ways to get predictions. The first approach having a solid theoretical justification in terms of OPE [57] is based on calculations in the euclidean domain where one can apply OPE. The contact with the observable quantities is made through the dispersion relations and in this way predictions for certain integrals are obtained. In the second approach we perform the calculations directly in Minkowski domain. Although formally this calculation refers to large distances, from the first approach we know that in specific integrals large distance contributions drop out. Therefore the results obtained in this way, although not valid literally, should be understood in the sense of duality: being smeared over some duality interval the theoretical prediction should coincide with the smeared experimental curve. The inclusive weak decays will be treated within the second approach. The averaging mainly refers to the invariant mass of the inclusive hadronic state produced in the decay considered.

If the invariant mass of the final hadronic state is large this is not a constraint at all since the theory 'itself' takes care of the averaging required by duality. In the opposite limit, near a spectral endpoint, the smearing is not provided for free. The boundary of the distribution corresponds to low momentum of the quark produced (low momentum of the hadronic final state). At this point the OPE blows up, therefore we do not have any specific prediction for the distributions near the boundary. Nevertheless the integrals taken over the domain from the kinematical boundary up to a new boundary, defined by the requirement that the OPE is convergent, are predicted. In particular this integration domain should include the resonances range (when $m_Q$ is large a parametrically stronger limitation is imposed by the fact that the heavy quark and meson masses are different). An example of the safe integration is the total decay width where the integration domain is maximal.

Although we explicitly work in the Minkowski kinematics we always keep in mind the relationship to the euclidean domain and the corresponding operator product expansion. The first analysis of this type has been outlined in [54] for inclusive heavy flavor decay rates. A general analysis of the semileptonic inclusive spectra along this line is presented in Ref. [18]. In that work it was observed, in particular, that the leading operator and those appearing at the next-to-leading order have a gap in dimensions of two units, and, consequently, the $\mathcal{O}(m_Q^{-1})$ term should be absent in certain quantities. The analysis presented in [18] was not backed up, however, by concrete calculations of the preasymptotic effects. Recently this formalism has been systematically developed and applied to the non-leptonic decays of heavy flavors [6, 15] and the charged-lepton energy spectrum in the semileptonic decays [7] (see also [5]).

We generalize the results of Ref. [7] to find the complete inclusive distributions in the semileptonic decays. The leptonic variables — $E_e, q^2$ and $q_0$, where $E_e$ is the charged lepton energy and $q$ is the momentum of the lepton pair [1] — are kept fixed which automatically fixes the invariant mass of the inclusive hadronic state. Integrating over $q_0$ we obtain the double spectral distribution in $E_e$ and $q^2$.

At the first stage we construct the transition *operator* $T(Q \to X \to Q)$ describing the forward scattering amplitude of the heavy quark $Q$ on a weak current. Our focus is the influence of the 'soft' modes (background fields) on the transition operator $T_{\mu\nu}$ which is expressed as an infinite series in the local operators built from gluon and quark fields and bilinear in $Q, \bar{Q}$.

The local operators are ordered according to their dimensions; the coefficient functions contain the corresponding powers of $1/m_Q$ (or $1/E_h$, where $E_h$ is the energy released into the hadronic system). At sufficiently large $m_Q$ or $E_h$ the operators with the lowest dimensions dominate, and the infinite series can be truncated. Generically, we will refer to the power expansion as $1/m_Q$ expansion, although strictly speaking it is an expansion in $1/E_h$. At the next stage the matrix elements of the relevant operators over the initial heavy hadron $H_Q$ must be evaluated. Unfortunately, in the present-day QCD the matrix elements over the hadronic states are not theoretically calculable. In some instances they can be related, through heavy quark symmetries, to measurable quantities [24],[28]; in other cases they have to be parametrized. These parameters play the role analogous to the gluon condensate [49]. As a matter of fact, at the level of the leading preasymptotic corrections only two operators are relevant. The matrix element of the first one can be related to the mass splittings of the vector and pseudoscalar heavy mesons. The matrix element of the second one has the meaning of the average square of the spatial momentum of the heavy quark $Q$ in $H_Q$ and the state must be treated as a parameter.

---

[1] The charged massless lepton produced will be generically called 'electron' hereafter.

Finally, the observed decay rates and spectra are obtained by taking the discontinuity of the hadronic tensor $\langle H_Q|T_{\mu\nu}|H_Q\rangle$ and convoluting the result with the lepton currents and appropriate kinematic factors.

In this work we consider the differential distributions in the semileptonic decays at the level of $O(m_Q^{-2})$. The differential distributions are measured experimentally in the $B$ meson decays and will be used for more precise determination of $V_{ub}$, for example. This was a primary motivation for our investigation. We would like to make it as close to the fundamental QCD as possible.

The organization of this chapter is as follows. In Section 2.2 we describe the kinematics and in Section 2.3 we present the operator product expansion. In Section 2.4 we derive the differential distributions for the massless lepton case. Section 2.5 is devoted to analysis of our distributions and limitations on the their use. Then we apply our results to derivation of the heavy lepton energy distribution and semileptonic width in decays $H_b \to \tau\bar{\nu}X$. Our results are summarized in Section 3.3. Appendix B contains expressions for hadronic invariant functions.

## 2.2 Kinematical analysis

We will consider the inclusive weak decays of the mesons (or baryons) with the open heavy flavor into the lepton pair plus (inclusive) hadronic state

$$H_Q(p_H) \to l(p_l) + \bar{\nu}(p_\nu) + hadrons.$$

Our goal in this chapter is to calculate the differential decay rate

$$\frac{d^3\Gamma}{dE_e dq^2 dq_0}, \tag{2.1}$$

where $E_e$ is the energy of the emitted electron and $q^\mu = p_l^\mu + p_\nu^\mu$ is the 4-momentum of the lepton pair. In order to find the differential distributions we need to know the amplitude of the process, which is given by the expression

$$\mathcal{M} = V_{qQ} \frac{G_F}{\sqrt{2}} \, \bar{e} \, \Gamma_\nu \, \nu \, \langle X | j_\nu | H_Q \rangle. \tag{2.2}$$

Here $V_{qQ}$ is the corresponding Cabibbo-Kobayashi-Maskawa matrix element, $j_\mu = \bar{q} \Gamma_\mu Q$ is the electroweak currents, $\Gamma_\mu = \gamma_\mu(1 + \gamma_5)$. (Although our theory is general we will keep in mind the $b \to c$ and $b \to u$ decays, so that $Q = b$ and $q = c$ or $u$). The differential distributions we are interested in are given by the modulus squared of the amplitude (2.2) summed over the final hadronic states.

The modulus squared of the amplitude summed over the final hadronic states can be written as

$$|\mathcal{M}|^2 = |V_{qQ}|^2 G_F^2 M_{H_Q} \, l^{\mu\nu} \, W_{\mu\nu}, \tag{2.3}$$

where $M_{H_Q}$ is the mass of hadron $H_Q$, $W_{\mu\nu}$ is the hadronic tensor

$$W_{\mu\nu} = \sum_X (2\pi)^4 \delta^4(p_{H_Q} - q - p_X) \frac{1}{2M_{H_Q}} \langle H_Q | j_\mu^\dagger(0) | X \rangle \langle X | j_\nu(0) | H_Q \rangle, \tag{2.4}$$

and $l^{\mu\nu}$ is the lepton tensor

$$l^{\mu\nu} = 8 \, [ \, (p_e)^\mu (p_\nu)^\nu + (p_e)^\nu (p_\nu)^\mu - g^{\mu\nu}(p_e \cdot p_\nu) + i\epsilon^{\mu\nu\alpha\beta}(p_e)^\alpha (p_\nu)^\beta ]. \tag{2.5}$$

Let us introduce the hadronic structure functions $w_i$ and parametrize the hadronic tensor in the following way:

$$W_{\mu\nu} = -w_1 \, g_{\mu\nu} + w_2 \, v_\mu \, v_\nu - i \, w_3 \, \epsilon_{\mu\nu\alpha\beta} \, v_\alpha q_\beta + w_4 \, q_\mu \, q_\nu + w_5 \, (q_\nu \, v_\mu + q_\mu \, v_\nu). \tag{2.6}$$

Here $q_\mu = (p_e + p_\nu)_\mu$ is the 4 - momentum of the lepton pair, $v_\mu = (p_{H_Q})_\mu/M_{H_Q}$ is the 4-velocity of the initial *hadron* (not that of the $Q$–quark). Note that we have omitted the structure $q_\mu v_\nu - q_\nu v_\mu$ which can not appear because of the $T$–invariance. The structure functions $w_i$ depend on two invariant variables, $q \cdot v$ and $q^2$. In the rest frame of $H_Q$ which will be used throughout the paper $q \cdot v = q_0$, so $w_i = w_i(q_0, q^2)$. The convolution of $W_{\mu\nu}$ with the lepton tensor (2.5) is given by the expression:

$$W_{\mu\nu}\, l^{\mu\nu} = 4\, \{2\, q^2\, w_1 + [\, 4\, E_e\, (\, q_0 - E_e\,) - q^2\,]\, w_2 + 2\, q^2\, (\, 2\, E_e - q_0)\, w_3\, \}. \qquad (2.7)$$

We see that only three structure functions are relevant for the processes we are considering in this paper. At this step we encounter the third variable, the electron energy $E_e = p_e \cdot p_{H_Q}/M_{H_Q}$, entering through the leptonic tensor.

Finally the formulae for the differential width takes the form

$$\frac{d^3\Gamma}{dE_e\, dq^2\, dq_0} = |V_{qQ}|^2 \frac{G_F^2}{32\, \pi^4} [2q^2 w_1 + [4\, E_e(q_0 - E_e) - q^2]\, w_2 + 2\, q^2 (2\, E_e - q_0)\, w_3]. \qquad (2.8)$$

This expression concludes the kinematical analysis. Our task is, of course, the calculation of the structure functions $w_i(q_0, q^2)$. We will proceed to this calculation in the next section.

The phase space of the decay is described in details in Appendix A. Here let us write down expressions for different Lorentz invariant phase space integrations which we will need in the future.

$$
\begin{aligned}
LIPS &= \int\int\int dE_e\, dq^2\, dq_0 \\
&= \int_0^{\frac{M_B^2 - M_D^2}{2M_B}} dE_e \int_0^{\frac{2E_e(M_B^2 - M_D^2 - 2M_B E_e)}{M_B - 2E_e}} dq^2 \int_{E_e + \frac{q^2}{4E_e}}^{\frac{M_B^2 - M_D^2 + q^2}{2M_B}} dq_0 \\
&= \int_0^{(M_B - M_D)^2} dq^2 \int_{\sqrt{q^2}}^{\frac{M_B^2 - M_D^2 + q^2}{2M_B}} dq_0 \int_{\frac{q_0 - \sqrt{q_0^2 - q^2}}{2}}^{\frac{q_0 - \sqrt{q_0^2 - q^2}}{2}} dE_e \\
&= \int_0^{\frac{(M_B^2 - M_D^2)^2}{4M_B^2}} d\mathbf{q}^2 \int_{\sqrt{\mathbf{q}^2}}^{M_B - \sqrt{M_D^2 + \mathbf{q}^2}} dq_0 \int_{\frac{q_0 - |\mathbf{q}|}{2}}^{\frac{q_0 - |\mathbf{q}|}{2}} dE_e. \qquad (2.9)
\end{aligned}
$$

The limits of integration are easily established by considering kinematical boundaries for the particles taking part in the decay.

We now can perform the integration over the electron energy and obtain distribution in lepton pair energy and momentum (the distribution is in fact the integrand in the following expression for the total width):

$$\Gamma = |V_{qQ}|^2 \frac{G_F^2}{16\,\pi^4} \int_0^{\frac{(M_B^2-M_D^2)^2}{4M_B^2}} d\mathbf{q}^2 \int_{\sqrt{\mathbf{q}^2}}^{M_B-\sqrt{M_D^2+\mathbf{q}^2}} dq_0 \{|\mathbf{q}|(q_0^2-\mathbf{q}^2)w_1 + \frac{\mathbf{q}^3}{3}w_2\}. \quad (2.10)$$

Integrating equation (2.8) over $q_0$ we obtain the double distribution in the electron energy and the invariant mass of the lepton pair:

$$\frac{d^2\Gamma}{dE_e\,dq^2} = |V_{qQ}|^2 \frac{G_F^2}{32\,\pi^4} \int_{q_0^{\min}}^{q_0^{\max}} dq_0 \{2q^2 w_1 + [4\,E_e(q_0-E_e) - q^2]\,w_2 + 2\,q^2(2\,E_e - q_0)\,w_3\},$$
$$(2.11)$$

where

$$q_0^{\min} = E_e + q^2/4E_e,$$
$$q_0^{\max} = (M_B^2 - M_D^2 + q^2)/2M_B. \quad (2.12)$$

Let us notice that the last distribution (2.11) contains weighted integrals of $w_i$'s over $q_0$:

$$\int_{q_0^{\min}}^{q_0^{\max}} q_0^k w_i(q_0, q^2) dq_0,$$

in other words, moments of the structure functions, where $q_0^{\min}$ and $q_0^{\max}$ are kinematical boundaries of $q_0$. This is important, because it is the moments of the structure functions which appear in the inclusive distributions are the quantities which could be reliably predicted in QCD. We will show this in Section 4.1 devoted to derivation of the OPE sum rules.

As it is well known, not all values of $q_0$ correspond to physical processes. The ones which do lie on the physical cuts of the forward scattering amplitude $T_{\mu\nu}$ (which will be introduced in the next section) in the complex plain $q_0$. The picture of the cuts is shown on Fig. 2.1. This picture could be readily understood. First, all the cuts lie on the real axis for only real $q_0$ are physical. Different cuts correspond to different physical processes. $B$ meson decay processes correspond to $0 < q_0 < M_B - \sqrt{M_D^2 + \mathbf{q}^2}$, this cut will be called $cut_1$. There is also a cut corresponding to collision of $B$ and $\bar{D}$ mesons, which lies in $M_B + \sqrt{M_D^2 + \mathbf{q}^2} < q_0 < +\infty$, this cut will be important in considering operator product expansion sum rules, we call it $cut_2$. All negative $q_0$ are physical, they correspond to a $W$ boson hitting $B$ meson with formation of some final hadronic state $X$, we call it $cut_3$. Let us note that separation between the decay $cut_1$ and collision $cut_2$ is $2\sqrt{M_D^2 + \mathbf{q}^2} \gg \Lambda_{\rm QCD}$.

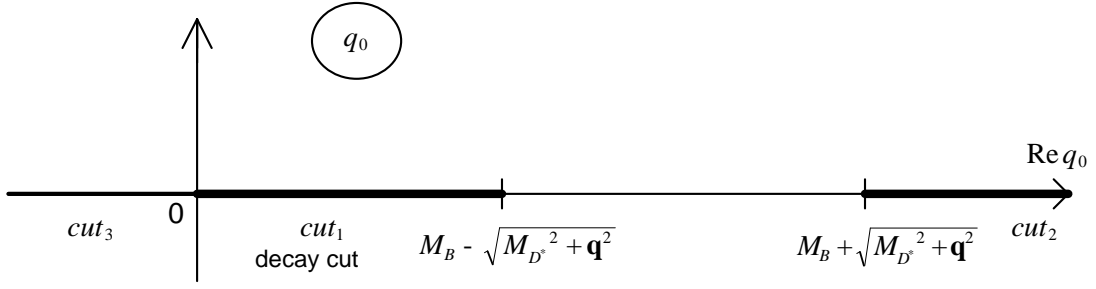

Figure 2.1: Cuts of the forward scattering amplitude in the complex plane $q_0$.

## 2.3 Operator product expansion

In this section we will discuss the derivation of the tensor $W_{\mu\nu}$. The Operator Product Expansion (OPE) is similar to that in the deep inelastic scattering. It is convenient to introduce the hadronic tensor $h_{\mu\nu}$ (forward scattering amplitude) as follows:

$$h_{\mu\nu} = i \int d^4 x e^{-iqx} \frac{1}{2\, M_{H_Q}} \langle H_Q | T\{j_\mu^+(x) j_\nu(0)\} | H_Q \rangle. \tag{2.13}$$

The absorptive part of this tensor reduces to $W_{\mu\nu}$ discussed above

$$W_{\mu\nu} = (1/i)\, {\rm disc}\, h_{\mu\nu}. \tag{2.14}$$

Here $\mathrm{disc}(h_{\mu\nu})$ is the discontinuity of the forward scattering amplitude $h_{\mu\nu}$ on the physical cut in the complex plane of the variable $q_0$. Of course, $h_{\mu\nu}$ can be expanded into the same set of structures as $W_{\mu\nu}$ (see. Eq. (2.6))

$$\begin{aligned} h_{\mu\nu} &= -h_1\, g_{\mu\nu} + h_2\, v_\mu\, v_\nu - i\, h_3\, \epsilon_{\mu\nu\alpha\beta}\, v_\alpha q_\beta + \\ &\quad h_4\, q_\mu\, q_\nu + h_5\, (q_\nu\, v_\mu + q_\mu\, v_\nu), \end{aligned} \tag{2.15}$$

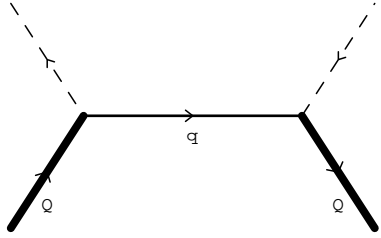

Figure 2.2: The tree diagram determining the transition operator $T_{\mu\nu}$ in the leading approximation. The dashed lines correspond to the weak currents, the solid internal line describes the propagation of the quark $q$ and the bold external lines represent the heavy quark $Q$.

and the relation (2.14) implies that

$$w_i = 2\,\mathrm{Im}\,h_i. \tag{2.16}$$

Let us remind that $h_{\mu\nu}$ is the matrix element of the transition operator $T_{\mu\nu}$

$$h_{\mu\nu} = \frac{1}{2\,M_{H_Q}} \langle H_Q | T_{\mu\nu} | H_Q \rangle, \tag{2.17}$$

$$T_{\mu\nu} = i \int d^4x\, e^{-iqx}\, T\{j_\mu^+(x) j_\nu(0)\}, \tag{2.18}$$

so below we will construct OPE for the product of currents in equation (2.18). Having in mind the relationship to euclidean analysis discussed above we will treat our expansion in the same way as a normal euclidean OPE. In the asymptotic limit $m_Q \to \infty$ the hadronic tensor $h_{\mu\nu}$ is given by the tree graph of Fig. 2.2. This graph defines the matrix element of the transition operator $T_{\mu\nu}$ over the heavy quark state,

$$\langle Q | T_{\mu\nu} | Q \rangle = -\bar{u}_Q \Gamma_\mu \frac{1}{\slashed{P} - \slashed{q} - m_q} \Gamma_\nu\, u_Q. \tag{2.19}$$

The latter expression represents nothing else but the free quark decay. In the asymptotic regime $m_Q \to \infty$ the interaction of the heavy quark with the gluon/light quark medium, as well as its intrinsic motion inside the hadron can be neglected. Then

$$P_\mu = P_{0\,\mu} \equiv m_Q\, v_\mu, \tag{2.20}$$

where $v_\mu$ is the 4-velocity of hadron $H_Q$.

Equation (2.19) allows one to immediately write down the operator form in the approximation at hand (only operators bilinear in $Q$, $\bar{Q}$ are considered, see the discussion of other operators at the end of this section):

$$T_{\mu\nu} = -\bar{Q}\Gamma_\mu \frac{1}{\slashed{k} - m_q} \Gamma_\nu Q = -\frac{2}{(k^2 - m_q^2)}[g_{\alpha\mu}k_\nu + g_{\alpha\nu}k_\mu - g_{\mu\nu}k_\alpha - i\epsilon_{\mu\nu\alpha\beta}k_\beta]\bar{Q}\gamma^\alpha(1+\gamma_5)Q, \tag{2.21}$$

where $k = P_0 - q$.

As we see, the two operators $\bar{Q}\gamma_\alpha Q$ and $\bar{Q}\gamma_\alpha\gamma_5 \bar{Q}$ showed up in the operator expansion at the level considered. Note, that the $\bar{Q}\gamma_\alpha\gamma_5\bar{Q}$ term vanishes after averaging over the unpolarized hadronic states.

In this paper the perturbative corrections in $\alpha_s$ are not touched upon at all. As for nonperturbative corrections they appear due to interactions with the soft medium of the light cloud in $H_Q$. By taking these interactions into account we isolate two types of effects. First, the fast quark $q$ produced does not propagate as a free one, but interacts with the background fields; these corrections will be included explicitly into the OPE coefficients. Second, the heavy quark $Q$ also does not live in the empty space; it is surrounded by the light cloud. In particular, due to this fact the heavy quark momentum does not coincide with $m_Q v_\mu$. This large distance effect will not be calculated explicitly, but implicitly it will be reflected in the $H_Q$ matrix elements of the operators in $T_{\mu\nu}$. This is in full analogy with what people usually do in deep inelastic scattering. The influence of the background field on the transition operator is summarized by the following expression

$$T_{\mu\nu} = -\int dx\, e^{-iqx}\bar{Q}(x)\Gamma_\mu S_q(x,0)\Gamma_\nu Q(0), \tag{2.22}$$

where $S_q(x,0)$ is the propagator of the quark $q$ in an external gluon field $A_\mu^a$. It is convenient to use the Schwinger technique of treating the motion in an external field (for a review of QCD adaptation see, e.g. Ref. [43]). Within that formalism the propagator $S_q$ is presented by the following expression

$$S_q(x,0) = (x|\frac{1}{\slashed{P} - m_q}|0). \tag{2.23}$$

Here $\slashed{P} = \gamma^\mu(p_\mu + A_\mu(X))$, $A_\mu = gA_\mu^a T^a$ is the gluon field in the matrix representation. Furthermore, the operator of coordinate $X_\mu$ and momentum $p_\mu$ are introduced, (thus the field $A_\mu(X)$ becomes an operator function of $X_\mu$), with the commutation relations

$$[p_\mu, X_\nu] = i\, g_{\mu\nu}, \quad [X_\mu, X_\nu] = 0, \quad [p_\mu, p_\nu] = 0. \tag{2.24}$$

The states $|x)$ are the eigenstates of the operator $X_\mu$, $X_\mu|x) = x_\mu|x)$.

Combining equations (2.22) and (2.23), we arrive at

$$T_{\mu\nu} = -\int dx\, e^{-iqx} (x|\bar{Q}(x)\Gamma_\mu \frac{1}{\slashed{\mathcal{P}} - m_q} \Gamma_\nu\, Q(X)|0). \tag{2.25}$$

As we have discussed above the operator $\mathcal{P}_\mu$ contains a large mechanical part $(P_0)_\mu = m_Q v_\mu$; the deviation from $P_0$ will be separated explicitly

$$\mathcal{P}_\mu = (P_0)_\mu + \pi_\mu \tag{2.26}$$

and we will expand in $\pi_\mu$. In this paper we will limit ourselves to the terms up to $\mathcal{O}(\pi^2)$ corresponding to $1/m_Q^2$ corrections. The master formulae to perform the expansion is

$$T_{\mu\nu} = -\int dx\, (x|\bar{Q}(X)\Gamma_\mu \frac{1}{\slashed{P_0} - \slashed{q} - m_q + \slashed{\pi}} \Gamma_\nu\, Q(X)|0). \tag{2.27}$$

There is a subtle point in the description of the formalism given above. Technically in the computation the $A_\mu(x)$ is assumed to be a c-number background field while in the final expression for local operators it should be understood as a second quantized operator. Since we are not considering any loop corrections this substitution is justified.

Let us now discuss the set of the operators relevant to the order $\mathcal{O}(m_Q^{-2})$. Without loss of generality we can work in the rest frame of the hadron $H_Q$, i.e. $v_\mu = (1, 0, 0, 0)$. Only those operators will be retained which produce non-vanishing results after being averaged over $H_Q$. The leading operator, as it was discussed above, is

$$\bar{Q}\gamma_0 Q, \tag{2.28}$$

its matrix element is fixed by the vector current conservation,

$$\frac{1}{2\, M_{H_Q}} \langle H_Q | \bar{Q}\gamma_0 Q | H_Q \rangle = 1. \tag{2.29}$$

Equation (2.29) is given in relativistic normalization we are using throughout this paper. In the non-relativistic normalization there is no need in the factor $1/2M_{H_Q}$ in the LHS.

As it has been noted in Ref. [18] there are no operators of dimension 4 in the problem at hand. The set includes two operators of dimension 5:

$$\mathcal{O}_G = \frac{i}{2}\bar{Q}\sigma^{\alpha\beta}G_{\alpha\beta}Q, \tag{2.30}$$

$$\mathcal{O}_\pi = -\bar{Q}\,\mathbf{D}^2 Q = \bar{Q}\,\boldsymbol{\pi}^2 Q, \tag{2.31}$$

where $\sigma^{\alpha\beta} = \frac{1}{2}(\gamma^\alpha\gamma^\beta - \gamma^\beta\gamma^\alpha)$, and $G_{\alpha\beta} = g\,G^q_{\alpha\beta}T^a$ is the gluon field strength tensor. The classification above takes into account the fact that the quark field $Q$ satisfies the equation of motion. In particular, it stems that the operator $\bar{Q}Q$ is not independent but is reducible to three operators (2.28), (2.30) and (2.31):

$$\begin{aligned}\bar{Q}\,Q &= \bar{Q}\,\gamma_0 Q - \frac{1}{2\,m_Q^2}\bar{Q}\,\boldsymbol{\pi}^2 Q + \\ &\quad \frac{i}{4\,m_Q^2}\bar{Q}\sigma^{\alpha\beta}G_{\alpha\beta}Q + \mathcal{O}(m_Q^{-3}).\end{aligned} \tag{2.32}$$

To get Eq. (2.32) we observe that the lower component of $Q$ is related to the upper one in the following way

$$\frac{1-\gamma_0}{2}Q = \frac{1}{2m_Q}\boldsymbol{\pi}\boldsymbol{\sigma}\frac{1+\gamma_0}{2}Q + \mathcal{O}(m_Q^{-2}), \tag{2.33}$$

and the difference between $\bar{Q}Q$ and $\bar{Q}\gamma_0 Q$ is due to the product of the lower components. (Here and below we will stick to the $H_Q$ rest frame.)

A few other useful relations which can be obtained in the same manner and are valid at the level $\mathcal{O}(m_Q^{-2})$ are:

$$\bar{Q}\boldsymbol{\gamma}\boldsymbol{\pi}Q = \frac{1}{m_Q}\bar{Q}(\boldsymbol{\pi}^2 - \frac{i}{2}\sigma G)Q + \mathcal{O}(m_Q^{-2}), \tag{2.34}$$

$$\bar{Q}\boldsymbol{\gamma}\boldsymbol{\pi}\gamma_0 Q = \mathcal{O}(m_Q^{-2}), \tag{2.35}$$

$$\bar{Q}\pi_0 Q = \frac{1}{2m_Q}\bar{Q}(\boldsymbol{\pi}^2 - \frac{i}{2}\sigma G)Q + \mathcal{O}(m_Q^{-2}). \tag{2.36}$$

A few comments are in order here concerning the actual technique of constructing the OPE. Since we work in the $H_Q$ rest frame it is convenient to compute different components of $T_{\mu\nu}$ separately, $T_{00}$, $T_{0i}$, $T_{i0}$ and $T_{ij}$. The calculation itself is a straightforward although rather tedious procedure of expanding the denominator in Eq. (2.27) in $/\!\!\!\pi$ using the properties of the $\gamma$ matrices, the commutation relation

$$[\pi_\mu, \pi_\nu] = i\,G_{\mu\nu} \tag{2.37}$$

and equations (2.34) - (2.36).

Notice that we must keep the terms of the first order in $\pi_0$ and of the second order in $\boldsymbol{\pi}$, since

$$\pi_0 Q = \frac{(\boldsymbol{\sigma}\boldsymbol{\pi})^2}{2m_Q}Q + \mathcal{O}(m_Q^{-2}). \tag{2.38}$$

Next, observe that the Green function in the background field can be written as follows:

$$\frac{1}{\not{\mathcal{P}} - \not{q} - m_q} = (\not{\mathcal{P}} - \not{q} + m_q)\frac{1}{(\mathcal{P}-q)^2 + (i/2)\sigma G - m_q^2}$$

$$\equiv (\not{\mathcal{P}} - \not{q} + m_q)\frac{1}{\Pi}. \tag{2.39}$$

To transpose $1/\Pi$ with $\Gamma_\nu$ it is convenient to use the identity

$$\frac{1}{\Pi}\Gamma_\nu = \Gamma_\nu \frac{1}{\Pi} + \frac{1}{\Pi}[\Gamma_\nu, \Pi]\frac{1}{\Pi}$$

$$= \Gamma_\nu \frac{1}{\Pi} + \frac{1}{\Pi}[\Gamma_\nu, \frac{i}{2}\sigma G]\frac{1}{\Pi}. \tag{2.40}$$

Acting on $Q$ and using the equations of motion we can now substitute $1/\Pi$ in both terms on the right-hand side by

$$\frac{1}{m_Q^2 - m_q^2 - 2\mathcal{P}q + q^2}, \tag{2.41}$$

provided that we limit ourselves to terms up to $\mathcal{O}(m_Q^{-2})$. The second term in (2.40) can be simplified even further since here we can additionally neglect $\pi$ in $\mathcal{P} = P_0 + \pi$.

We split the calculation into three parts: Vector×Vector, Axial×Axial and Vector× Axial in correspondence with the structure of $\Gamma_\mu$ as a sum of vector and axial vector, $\Gamma_\mu = \gamma_\mu + \gamma_\mu \gamma_5$. The full hadronic tensor $h_{\mu\nu}$ is given then by the following expression:

$$h_{\mu\nu} = h_{\mu\nu}^{VV} + h_{\mu\nu}^{AA} + h_{\mu\nu}^{AV} + h_{\mu\nu}^{VA} = \frac{1}{2\,M_{H_Q}}\langle H_Q|T_{\mu\nu}^{VV} + T_{\mu\nu}^{AA} + T_{\mu\nu}^{AV} + T_{\mu\nu}^{VA}|H_Q\rangle. \tag{2.42}$$

The complete expressions for the hadronic invariant functions are given in the Appendix B. In the order $\mathcal{O}(m_Q^{-2})$ they are defined by the matrix elements of operators $\mathcal{O}_G$, $\mathcal{O}_\pi$ given by eqs.(2.30), (2.31):

$$\frac{1}{2\,M_{H_Q}}\langle H_Q|\bar{Q}\,\frac{i}{2}\,\sigma_{\mu\nu}G^{\mu\nu}Q|H_Q\rangle = \frac{1}{2\,M_{H_Q}}\langle B|\bar{b}(\boldsymbol{\sigma}\cdot\boldsymbol{B})b|\rangle = \mu_G^2, \tag{2.43}$$

where $\boldsymbol{B}$ is the chromo-magnetic field operator, and

$$\frac{1}{2\,M_{H_Q}}\langle H_Q|\bar{Q}\,\pi^2 Q|H_Q\rangle = \mu_\pi^2. \tag{2.44}$$

The parameter $\mu_G^2$ coincides with $m_{\sigma H}^2$ introduced in [14]. For mesonic states it is expressible in terms of the quantity measured experimentally – the hyperfine mass splittings, and it has the zero value for baryonic states of the type of $\Lambda_Q$; the parameter $\mu_\pi^2$ has the meaning of the average square of spatial momentum of the heavy quark $Q$ in the hadronic state $H_Q$. The two quantities $\mu_G^2$ and $\mu_\pi^2$ often appear in the combination $\mu_\pi^2 - \mu_G^2$, cf. equation (2.32).

The value of $\mu_G^2$ is known from the hyperfine mass splitting between the members of the heavy quark symmetry doublet $B$ and $B^*$, for it one has the following expression

$$\mu_G^2 = \frac{3}{4}(M_{B^*}^2 - M_B^2) \approx 0.36 \text{ GeV}^2. \tag{2.45}$$

As for the value of $\mu_\pi^2$ it is not known that well. Some QCD sum rules estimates of $\mu_\pi^2$ were made in [4] Also the inequality $\mu_\pi^2 > \mu_G^2$ was proven in [52]. Let us say here, that $\mu_\pi^2$ is an important phenomenological parameter which enters different characteristics of the $B$ mesons and should be measured in experiments. One of the goals of this paper is to formulate constraints on the phenomenological structure functions which follow from the OPE sum rules and then fit them using the measured lepton energy spectrum. In this way we hope to be able to extract the QCD parameter $\mu_\pi^2$ from the experimentally measured lepton energy spectrum. Its value is expected to be around $0.5 \text{GeV}^2$.

Let us now introduce some definitions for the quark masses. In the works The heavy hadron mass is related to the heavy quark mass in the following way:

$$M_{H_Q} = m_Q + \bar{\Lambda} + \frac{\mu_\pi^2 - \mu_G^2}{2m_Q^2} + \mathcal{O}(\frac{1}{m_b^3}, \frac{1}{m_Q^3}), \tag{2.46}$$

where $Q$ is a heavy quark ( in this paper it is $b$ or $c$), $H_Q$ is the lowest lying meson containing $Q$, $\bar{\Lambda}$ is the so-called "mass" of the light degrees of freedom . The last expression is in fact an expansion of the hadron mass in $1/m_Q$. The $\mu_\pi^2/2m_Q^2$ part of the hadron mass is due to the motion of the heavy quark inside the hadron, $\mu_G^2/2m_Q^2$ is due to chromomagnetic interaction between the heavy quark and light degrees of freedom, $\bar{\Lambda}$ is the $m_Q$ independent part of the heavy meson mass.

Let us also introduce the following notation:

$$\delta_b \equiv M_B - m_b = \bar{\Lambda} + \frac{\mu_\pi^2 - \mu_G^2}{2m_b^2}, \quad \text{and} \quad \delta_c \equiv M_D - m_c = \bar{\Lambda} + \frac{\mu_\pi^2 - \mu_G^2}{2m_c^2}, \tag{2.47}$$

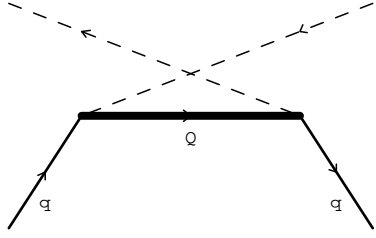

Figure 2.3: The tree diagram determining the operator without the heavy quark $Q$. Now the bold internal line describes the propagation of the heavy quark $Q$ and the solid external lines represent the quark $q$.

then
$$M_B - M_D = m_b - m_c + (\mu_\pi^2 - \mu_G^2)(\frac{1}{2m_b} - \frac{1}{2m_c}) + \mathcal{O}(\frac{1}{m_b^3}, \frac{1}{m_c^3}), \qquad (2.48)$$

$$M_B - M_{D^*} = m_b - m_c + (\mu_\pi^2 - \mu_G^2)(\frac{1}{2m_b} - \frac{1}{2m_c}) - \frac{2}{3}\frac{\mu_G^2}{m_c} + \mathcal{O}(\frac{1}{m_b^3}, \frac{1}{m_c^3}), \qquad (2.49)$$

and
$$\delta_b - \delta_c = (\mu_\pi^2 - \mu_G^2)(\frac{1}{2m_b} - \frac{1}{2m_c}) + \mathcal{O}(\frac{1}{m_b^3}, \frac{1}{m_c^3}). \qquad (2.50)$$

The last comment of this section is about the operators which are not bilinear in $\bar{Q}$, $Q$ fields. The simplest example of appearance of such operators is given by the diagram of Fig.2.3 where the heavy quark $Q$ propagates between the current vertices. This diagram is similar to the one of Fig. 2.2, and the corresponding operator follows from Eq.(2.21) by substitution $Q \Rightarrow q$, $m_q \Rightarrow m_Q$, $k_\mu \Rightarrow q_\mu$. The additional term in $T_{\mu\nu}$ has the form

$$\Delta T_{\mu\nu} = -\bar{q}\Gamma_\nu \frac{1}{\slashed{q} - m_Q}\Gamma_\mu q = -\frac{2}{(q^2 - m_Q^2)}[g_{\alpha\mu}q_\nu + g_{\alpha\nu}q_\mu - g_{\mu\nu}q_\alpha - i\epsilon_{\mu\nu\alpha\beta}q_\beta]\bar{q}\gamma^\alpha(1 + \gamma_5)q. \qquad (2.51)$$

The matrix element of the operator $\bar{q}\gamma^\alpha q$ over the $H_Q$ state counts the number of quarks $q$ and is not small in general. The operator coefficient given by Eq.(2.51) is particularly large when $q^2 \to m_Q^2$.

In terms of the intermediate hadronic states in the forward scattering off $H_Q$ this contribution is due to states in crossing channel containing two $Q$ quarks - the problem was pointed out in Ref. [18]. The cross-channel is not related to the weak inclusive decays under consideration. It is reasonable to accept the duality between the operators without heavy quark fields and the cross-channel contributions having nothing to do with heavy flavor decays. We rely on the assumption that we can consistently omit the crossing channel together with operators in $T_{\mu\nu}$ related to this channel.

## 2.4 Calculation of the differential distributions for massless lepton in the final state

The differential distributions we are interested in are determined by equation (2.8) containing three structure functions $w_1$, $w_2$ and $w_3$. They are obtained from the results for $h_i$ (see Appendix B) by taking the imaginary parts of the corresponding functions (see. Eq. (2.16)). The imaginary parts are due to the poles of $h_i$ and are obtained through the relations:

$$\text{Im} \frac{1}{z^n} = \pi \frac{(-1)^n}{(n-1)!} \frac{d^{n-1}}{dz^{n-1}} \delta(z), \qquad (2.52)$$

where $z$ is given by

$$z = m_Q^2 - 2\, m_Q\, q_0 + q^2 - m_q^2. \qquad (2.53)$$

We do not present here the expression for the triple differential distribution which can be easily obtained by combining the equations (2.8), (2.16), (2.52) and expressions for $h_i$ from the Appendix B.

Although the result is derived for the physical quantity $d^3\Gamma/dE_e dq^2 dq_0$, it can not be directly compared with the experimental data. An obvious signal for this is the presence of the delta function and its derivatives. It is not surprising because we are sitting now right on mass-shell of the $q$-quark. As we discussed in the introduction our results should be understood in the sense of duality: that is that the predictions should be smeared over certain duality interval. At the moment we have no purely theoretical tools to fix the size of the duality interval, therefore we are forced to rely on qualitative arguments and experimental data. For example the duality interval for $q_0$ can be inferred from the distribution in the invariant mass of the final hadronic states. Our $\delta$-functions reflect the resonance structure at low invariant masses. The smearing interval should be chosen in such a way as to cover the entire resonance domain up to the onset of the smooth behavior.

Instead of smearing of the distribution one can calculate the average characteristics like the total width $\Gamma$ or $\langle M_X^n \rangle$, where $M_X$ is the invariant mass of the final hadronic states. The power corrections we have calculated will enter in a specific way in each particular quantity.

Now let us proceed to the calculation of the double differential distribution $d^2\Gamma/dE_e\,dq^2$. To this end we must integrate over $q_0$, rather simple exercise with $\delta$-functions. However if one would perform the integration by merely substituting

$$q_0 \to q_0^* = \frac{m_Q^2 + q^2 - m_q^2}{2\,m_Q}, \tag{2.54}$$

and taking the derivatives in the case of $\delta'$ and $\delta''$, one would get the wrong answer. The point is that integration domain in $q_0$ has a boundary from below

$$q_0 \geq E_e + \frac{q^2}{4\,E_e}, \tag{2.55}$$

which corresponds to $4\,E_e\,E_\nu \geq q^2$. Therefore one should take into account the fact that $q_0$ can not cross the boundary (2.55). For that we introduce $\theta(q_0 - E_e - q^2/4E_e)$ into the integrand. The occurrence of the $\theta$-function is important for the integration of $\delta'(q_0 - q_0^*)$ and $\delta''(q_0 - q_0^*)$ which leads to appearance of $\delta(q_0^* - E_e - q^2/4E_e)$ and $\delta'(q_0^* - E_e - q^2/4E_e)$ in the double distribution $d^2\Gamma/dq^2\,dE_e$ because of differentiation of the $\theta$-function. The final formulae for the double differential distribution in the lepton energy $E_e$ and $q^2$ takes the form:

$$\frac{d^2\Gamma}{dx\,dt} = |V_{qQ}|^2 \frac{G_F^2\,m_Q^5}{96\,\pi^3}\,x^2\,\{\,6\,(1-t)(1-\rho-x+xt)+$$
$$G_Q\,[\,1 - 5\rho + 2t + 10\rho t + 10xt - 10xt^2 -$$
$$(-1 + 6\rho - 5\rho^2 + x - 5\rho x + t - 2\rho t + 5\rho^2 t + xt + 15\rho xt +$$
$$5x^2 t - 2xt^2 - 10\rho xt^2 - 10x^2 t^2 + 5x^2 t^3)\,\delta((1-t)(1-x)-\rho)\,] +$$
$$K_Q\,[-3 + 3\rho + 4t - 4\rho t - 6xt + 4xt^2 -$$
$$(1 - 2\rho + \rho^2 - 3x + 3\rho x - 3t + 2\rho t + \rho^2 t + 11xt - 3\rho xt -$$
$$3x^2 t - 6xt^2 - 2\rho xt^2 + 2x^2 t^2 + x^2 t^3)\,\delta((1-t)(1-x)-\rho) +$$
$$(1 - \rho - x + xt)(1 - t)(1 - 2\rho +$$
$$\rho^2 - 2xt - 2\rho xt + x^2 t^2)\,\delta'((1-t)(1-x)-\rho)\,]\,\}. \tag{2.56}$$

Here we have introduced the dimensionless variables

$$x = 2\,E_e/m_Q, \quad t = q^2/2m_Q E_e, \tag{2.57}$$

and the parameters

$$\rho = m_q^2/m_Q^2, \quad G_Q = \mu_G^2/m_Q^2, \quad K_Q = \mu_\pi^2/m_Q^2. \tag{2.58}$$

Let us emphasize that the scale $m_Q$ used in equation (2.57) is the heavy quark mass and does not coincide with $M_{H_Q}$ which is normally used in the experimental distributions.

The fact that OPE generates corrections only of the order of $\mathcal{O}(m_Q^{-2})$ (terms proportional to $K_Q$ and $G_Q$) is valid for the distributions only if we use $m_Q$ as a scale, i.e. in the variables $x,t$. Of course one can easily rescale them to $M_{H_Q}$; then the corrections of the order of $\mathcal{O}(m_Q^{-1})$ will show up for trivial kinematical reasons.

We can proceed further and obtain the energy spectrum by integrating over $q^2$. The range of integration is given by

$$0 \leq t \leq 1 - \frac{\rho}{1-x}. \tag{2.59}$$

The result for the energy spectrum coincides with that obtained in [7]. For the sake of completeness we present it here [2]:

$$\frac{d\Gamma}{dx} = \frac{|V_{qQ}|^2 G_F^2\, m_Q^5}{192\pi^3}\theta(1-x-\rho)2x^2\{(1-f)^2(1+2f)(2-x)+(1-f)^3(1-x)+$$
$$(1-f)[(1-f)(2+\frac{5}{3}x-2f+\frac{10}{3}fx)-\frac{f^2}{\rho}(2x+f(12-12x+5x^2))]G_Q -$$
$$[\frac{5}{3}(1-f)^2(1+2f)x+\frac{f^3}{\rho}(1-f)(10x-8x^2)+\frac{f^4}{\rho^2}(3-4f)(2x^2-x^3)\,]\,K_Q\}\tag{2.60}$$

where $f = \rho/(1-x)$. Finally, performing the last integration over $x$ in the domain

$$0 \leq x \leq 1 - \rho, \tag{2.61}$$

we arrive to the total width coinciding with that in [5]:

$$\Gamma = |V_{qQ}|^2 \frac{G_F^2\, m_Q^5}{192\pi^3}[\,z_0\,(1+\frac{1}{2}(G_Q - K_Q)) - 2\,z_1\,G_Q\,], \tag{2.62}$$

where $z_0 = 1 - 8\rho + 8\rho^3 - \rho^4 - 12\rho^2 \log \rho$ and $z_1 = (1-\rho)^4$.

---

[2] Let us draw the reader's attention to the difference of notation: $y$ in [7] is equal to our $x$.

Now let us discuss the characteristic features of the double distribution (2.56). The most striking one is the presence of the singular terms. The technical reason for occurrence of those terms was that we expanded the denominator of the pole expression (2.27) in $\pi$ and $\sigma G$. Physically this expansion reflects the shifts of the masses of particles due to the nonperturbative effects. As it was mentioned above these singularities reflect the structure of the resonance domain and the predictions suitable for comparison with the experimental data require smearing over the corresponding domain. To illustrate the most salient features of our prediction let us concentrate on the physically interesting case of the $b \to u$ transition.

For massless $u$ quark the kinematical region of $b$ quark semileptonic decay is shown on Fig. 2.4. It has the form of a square with the side equal to 1 in the plane ($x = 2E_e/m_b$, $t = q^2/2m_b E_e$). The right-hand side of the square corresponds to the maximal energy of electron $E_e = m_b/2$ while the upper side is a maximal energy of neutrino. In the real $B$ meson decay the kinematical region is certainly wider; if one neglects the pion mass the region is the square with the side $x_{max} = t_{max} = M_B/m_b$. The origin of this window is related to the motion of the heavy $b$ quark inside the $B$ meson. In our calculations we account for nonzero momentum of the $b$ quark in the form of expansion which produced singular $\delta$ and $\delta'$ terms on the boundary. It is possible to show (see refs.[7],[31]) that the expansion breaks down at distances $\sim (M_B - m_b)/m_b$ near the boundary, so we need to integrate our distributions over a range of the order of the window between quark and hadron boundaries. It is interesting to note that the distribution spreads off the distances of the order $(M_B - m_b)/m_b$ while the corrections to integrals are only of the second order in $1/m_b$.

Another effect we need to account for is the structure of the resonance region near the low end of the hadronic invariant masses. To imitate the effect let us imagine that this region corresponds to the $u$ quark fragmentation into the hadronic states with $s$ (the square of the invariant mass) from $s = 0$ to $s = s_0 = 2$ GeV$^2$. The curve corresponding to $s = s_0$ on Fig. 2.4 is given by the equation:

$$(1 - t)(1 - x) = s_0/m_b^2, \tag{2.63}$$

and the resonance region should be included as a whole into the process of integration; we can predict the integral but not the structure.

## 2.5 Application to the analysis of the experimental data

Our theoretical prediction (2.56) depends on the following parameters: $V_{qQ}$, $m_Q$, $m_q$, $K_Q$, $G_Q$. Let us remind that in this paper we do not consider perturbative in $\alpha_s$ corrections (see Ref. [17]), which, of course, should be added. The Cabibbo-Kobayashi-Maskawa matrix element $V_{qQ}$ does not effect the form of the differential distribution; the total semileptonic width is proportional to $|V_{qQ}|^2$. The quark masses enter at the level of leading approximation while $K_Q$ and $G_Q$ determine $1/m_Q^2$ corrections. It is important that our differential distributions by themselves could be used to fit these parameters. In particular it is a good place to extract the heavy quark mass.

Our purpose here is to give an idea of how important the $1/m_Q^2$ corrections are in the case of charmless $B$–meson decays ($b \to u$ transition). To this end we will use the approximate values for the parameters $m_B$, $K_b$ and $G_b$ obtained from other sources. First, we use $m_b \sim 4.8$ GeV as deduced from the QCD sum rules analysis of the Ypsilon system [51], and $m_u = 0$. Parameter $G_b$ can be extracted from the $B, B^*$ mass splitting [14]:

$$G_b = \frac{3}{4}(M^2(B^*) - M^2(B))/m_b^2 \sim 0.017. \qquad (2.64)$$

As a representative value we use for the parameter $K_b$ the value $\sim 0.02$. A close value was obtained in Ref. [4] from the QCD sum rules. Earlier QCD sum rule result [39] was a factor of two higher. Notice that the sensitivity of our results to the value of $K_b$ is essentially less than that to $G_b$. For example, Eq. (2.62) for $b \to u$ transition contains $G_b + \frac{1}{3}K_b$.

In accordance with the discussion at the end of the previous section the comparison with experiment should include integration of our distribution (2.56) over the domain which includes the area adjacent to the kinematical boundary. We will choose this area to be given by the resonance domain (see Eq. (2.63)) with $s_0 = 2$ GeV$^2$.

Let us introduce the quantity:

$$P(x_c, t_c) = \frac{1}{\Gamma_0} \int\int_{A(x_c, t_c)} dx\, dt\, \frac{d^2\Gamma}{dx\, dt}, \qquad (2.65)$$

where $x_c$, $t_c$ is the point in $(x, t)$ plane sitting not too close to the boundary (outside the resonance range), $\Gamma_0 = |V_{ub}|^2 G_F^2 m_b^5/192\pi^3$ and the area of integration $A(x_c, t_c)$ shown on Fig. 2.4 as shaded includes the resonance domain plus domain $x > x_c, t > t_c$. For the experimental distribution the range of integration should be extended to include the window between quark and hadron kinematical boundaries. Notice that in the limit of large $m_Q$ the size of the window $(M_{H_Q} - m_Q)/m_Q$ is parametrically larger then the resonance range $s_0/m_Q^2$. In the case of $b$–quark they are numerically close.

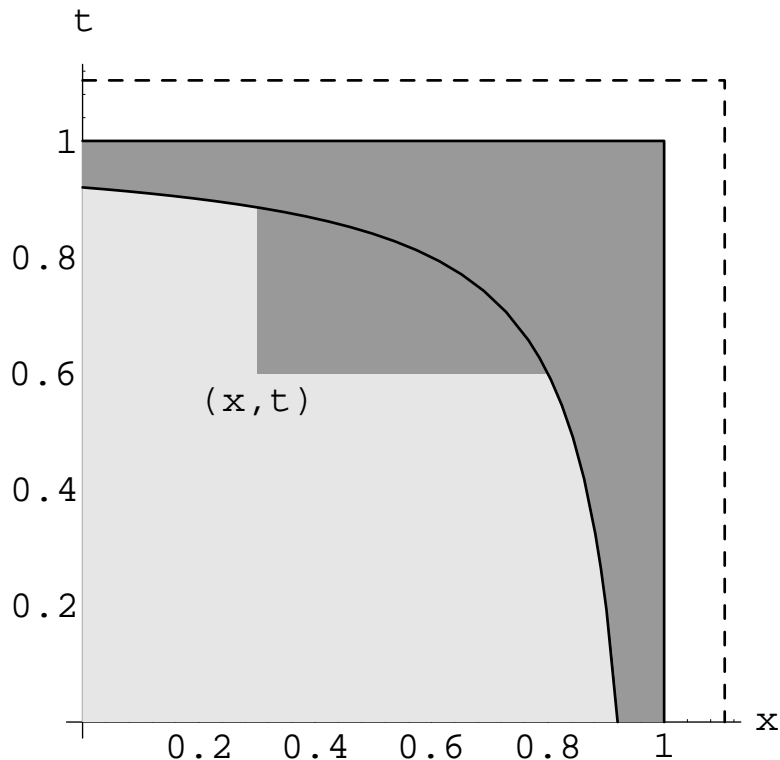

Figure 2.4: The kinematical region of the decay for $b \to u$ decays in coordinates $x = 2E_e/m_b$ and $t = q^2/2m_b E_e$. The solid lines are the kinematical boundary for the $b$ quark decay ($x_{max} = t_{max} = 1$) and the dashed lines are the boundary for $B$ meson decay ($x_{max} = t_{max} = M_B/m_b$). The area of integration for the distribution $P(x,t)$ is shaded. It includes integration over the resonance domain.

The function $P(x,t)$ is plotted as a function of $t$ on Fig. 2.5 for three values of $x$ equal to 0.3, 0.6, 0.8. The last value of $x$ is close to the border of the resonance region beyond which we cannot make reliable predictions for the distributions considered. The dashed lines on Fig. 2.5 describe the leading order distributions in $t$ while the solid lines include QCD corrections we have calculated. As we can see it from the curves, the corrections are negative and their relative magnitude is larger near the endpoints of the spectra.

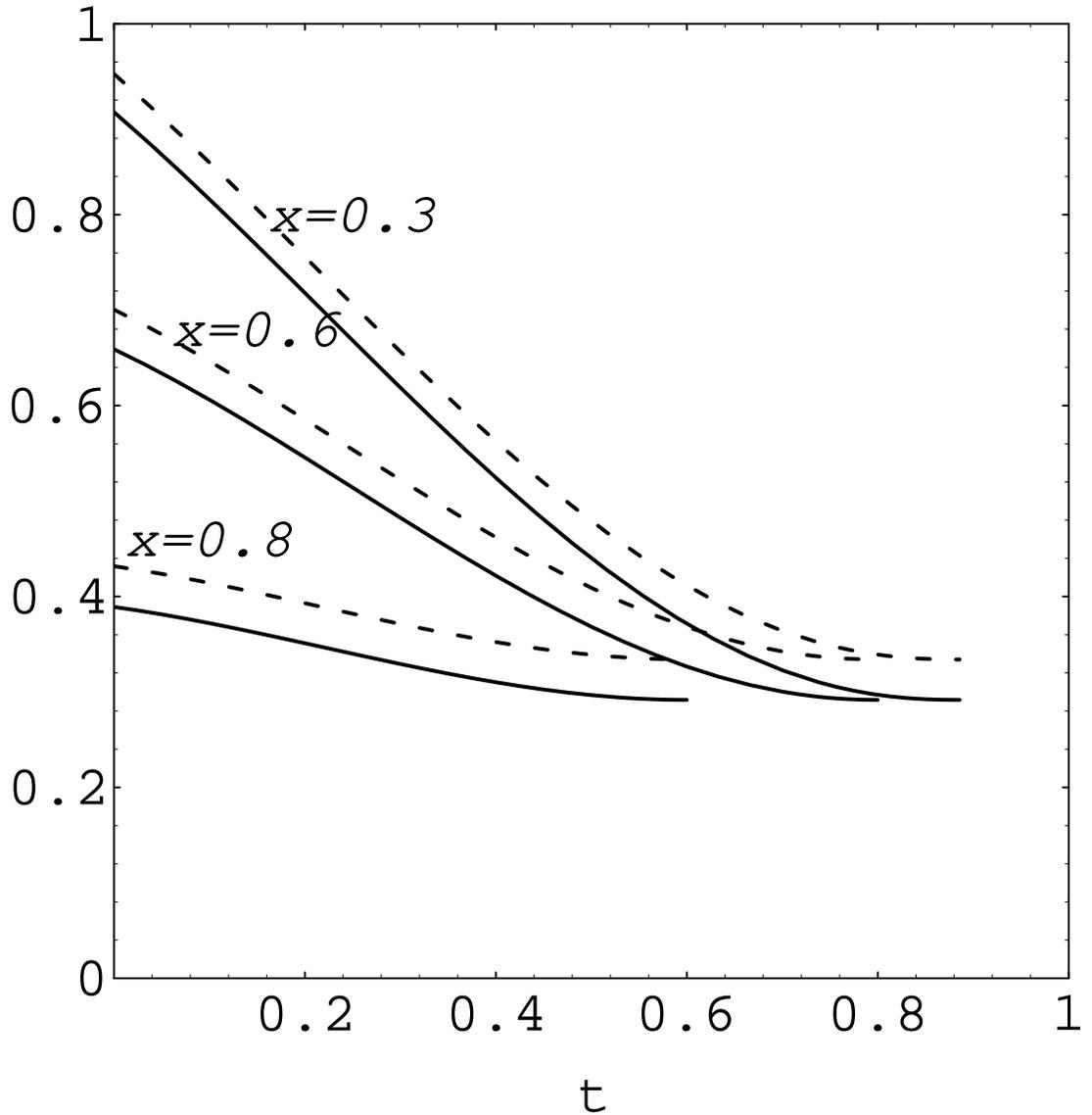

Figure 2.5: The integrated distribution $P(x,t)$ for the case $b \to u$ is plotted as function of $t = q^2/2m_b E_e$ for few values of $x = 2E_e/m_b$. The dashed lines correspond to the leading order distribution while the solid lines account for nonperturbative corrections (see Eq. (2.56)). The lines stop at the border of resonance region. It follows from the picture that the corrections are negative.

# Chapter 3

# Corrections to the heavy lepton energy distribution in the inclusive decays $H_b \to \tau \bar{\nu} X$

In this part of our work we investigate the power corrections to the heavy lepton energy distribution in the inclusive decays

$$H_b \to \tau \bar{\nu} X,$$

where $H_b$ is a hadron containing heavy quark $b$ and $\tau$ is the $\tau$-lepton. We include non-perturbative corrections up to the order $1/m_b^2$, and show that they cause the decay rate of $B$-mesons to decrease by 6 to 10 percent of its perturbative value depending on the mass of the quark in the final state. The lepton mass does not effect neither hadronic nor weak leptonic tensors. Therefore we can use expressions (2.2)–(2.8) and (B.7)–(B.11) for the hadronic invariant functions to get the corresponding matrix element. What differs the decay with a heavy lepton from the decay with a massless lepton is the phase space of the particles in the final state.

The first experimental observations of the decay were made recently using the missing energy tag [1],[44]. The missing energy was associated with two $\nu_\tau$ in the decay chain $b \to \tau^- \bar{\nu}_\tau X$, $\tau^- \to \nu_\tau X'$, which made it difficult to reconstruct. The branching ratio was found to be $4.08 \pm 0.76 \pm 0.62\%$ in [1] and $2.76 \pm 0.47 \pm 0.43\%$ in [44] (more recent analysis), which is compatible with the Standard Model. However because of the difficulties in identification of the decay mode $H_b \to \tau \bar{\nu} X$, the accuracy of the measurements is still insufficient for direct comparison of the results of this paper with the data.

## 3.1 Heavy lepton energy distributions

As was discussed earlier in the semileptonic inclusive decays there are three independent kinematical variables: $E_\tau$, $q_0$, and $q^2$, where $E_\tau$ is the energy of the emitted lepton ($\tau$ in this paper), $q_0$ is the energy and $q^2$ is the invariant mass of the lepton pair. The full differential distribution can be written as follows:

$$\frac{d\Gamma}{dE_\tau \, dq_0 \, dq^2} = \frac{1}{128 \, \pi^4 \, M_{H_b}} \, |\mathcal{M}(E_\tau, q_0, dq^2)|^2. \tag{3.1}$$

The matrix element of the process $\mathcal{M}(E_\tau, q_0, dq^2)$ involves the same hadronic and leptonic tensors as used in [13] and which are given by the expressions (2.2)–(2.8) and (B.7)–(B.11). The corresponding phase space and the boundaries for the kinematical variables for the massive lepton case are briefly discussed in Appendix A.

The double differential distribution in $E_\tau$ and $q^2$ could be obtained from (3.1) by integrating over $q_0$. The corresponding expressions are somewhat cumbersome because of the complicated limits of integration over $q_0$ in $(q^2, q_0^2)$ plane. Also there is little use of the double distribution because of the difficulties in observing the considered decays at present time. Therefore we will not write out the formulae for the double distributions in $(q^2, E_\tau)$, but will rather integrate Eq. (3.1) twice and get the energy distribution for the charged lepton in the final state. One more remark is in place here. Although we are looking at a hadron decay, in our dynamical approach we are considering a heavy quark decaying in the external field created by its interactions with the light degrees of freedom. That is why in the case at hand we have to use the quark kinematical boundaries rather then the hadronic ones. This means that in the formulas (A.7)–(A.10) we have to use $m_b$ and $m_q$ instead of $M_{H_b}$ and $M_D$ correspondingly. We will use such a kinematical boundary in the following calculations. Our results then should be understood in the sense of duality (see discussion in Chapter 2, [13]).

To get the energy distribution we integrate (3.1) over $dq^2$ and $dq_0$. In the dimensionless variables:

$$x = \frac{2\,E_\tau}{m_b}, \quad \rho_q = \frac{m_q^2}{m_b^2}, \quad \text{and} \quad \rho_\tau = \frac{m_\tau^2}{m_b^2}, \tag{3.2}$$

the result of the integration then takes the following form:

$$\begin{aligned}
\frac{1}{\Gamma_0}\frac{d\Gamma}{dx} &= 2\sqrt{x^2 - 4\rho_\tau}\,\{3\,(1 - \rho_q + \rho_\tau)\,x - 2\,x^2 - 4\,\rho_\tau + \rho_q\,(3 - \rho_q + 3\,\rho_\tau)\,f + \\
&\quad (-3 - \rho_q + 6\,\rho_\tau - \rho_q\,\rho_\tau - 3\,\rho_\tau^2)\,f^2 + 2\,(1 - \rho_\tau)^2\,f^3 + \\
&\quad K_b\,[\,-\frac{5\,x^2}{3} + \frac{14\,\rho_\tau}{3} - \frac{2}{3}\,\rho_q\,(3 - \rho_q + 3\,\rho_\tau)\,f + (1 + \frac{2}{3}\,\rho_q - 4\,\rho_\tau + \frac{2}{3}\,\rho_q\,\rho_\tau + \rho_\tau^2)\,f^2 + \\
&\quad \frac{2}{3}\,(6 + \rho_q - 6\,\rho_\tau + 4\,\rho_q\,\rho_\tau - 6\,\rho_\tau^2 + \rho_q\,\rho_\tau^2 + 6\,\rho_\tau^3)\,\frac{f^3}{\rho_q} + \\
&\quad 3\,(1 - \rho_\tau)^2\,(-1 - 2\,\rho_q + 2\,\rho_\tau - 2\,\rho_q\,\rho_\tau - \rho_\tau^2)\,\frac{f^4}{\rho_q^2} + 4\,(1 - \rho_\tau)^4\,\frac{f^5}{\rho_q^2}\,] + \\
&\quad G_b\,[\,2\,x + \frac{5}{3}\,x^2 + 4\,\rho_q - \frac{14}{3}\,\rho_\tau + (-2 + 3\,\rho_q - \frac{5}{3}\,\rho_q^2 - 2\,\rho_\tau + 5\,\rho_q\,\rho_\tau)\,f + \\
&\quad (-2 - \frac{5}{3}\,\rho_q^2 + 4\,\rho_\tau + 8\,\rho_q\,\rho_\tau - \frac{5}{3}\,\rho_q^2\,\rho_\tau - 2\,\rho_\tau^2 - 10\,\rho_q\,\rho_\tau^2)\,\frac{f^2}{\rho_q} + \\
&\quad (-1 + \rho_\tau)\,(3 + \frac{5}{3}\,\rho_q - 8\,\rho_\tau + \frac{25}{3}\,\rho_q\,\rho_\tau + 5\,\rho_\tau^2)\,\frac{f^3}{\rho_q} + 5\,(1 - \rho_\tau)^3\,\frac{f^4}{\rho_q}\,]\},
\end{aligned} \tag{3.3}$$

where

$$f = \frac{\rho_q}{1 + \rho_\tau - x}, \tag{3.4}$$

and

$$\Gamma_0 = |V_{qb}|^2\,\frac{G_F^2\,m_b^5}{192\,\pi^3}. \tag{3.5}$$

The quantities $K_b$ and $G_b$ are the hadronic matrix elements introduced in Eq. (2.44).

The energy distribution (3.3) spans in $x$ from $x = 2\sqrt{\rho_\tau}$ to $x = 1 + \rho_\tau - \rho_q$. It includes the nonperturbative corrections – terms proportional to $K_b$ and $G_b$. The part of Eq.(3.3) without corrections coincides with the electron spectrum in $\mu$ decay with a massive $\tau_\mu$ from Ref. [50]. In the limit $\rho_q \to 0$ we encounter the familiar end-point singularities of the lepton spectrum. To get the limit right we make the following substitutions:

$$\frac{f^n}{\rho_q} \Rightarrow \frac{\delta(1 + \rho_\tau - x)}{n - 1}, \quad n > 1, \tag{3.6}$$

and

$$\frac{f^n}{\rho_q^2} \Rightarrow \frac{\delta'(1 + \rho_\tau - x)}{(n - 1)(n - 2)}, \quad n > 2 \tag{3.7}$$

and only then take $\rho_q$ to zero. To see that this procedure gives the right limit, it is sufficient to compare the results of integration of the both sides of (3.6) and (3.7) in the limits $2\sqrt{\rho_\tau} \leq x \leq 1 + \rho_\tau - \rho_q$, multiplied by an arbitrary integrable function and check that the both sides are equal to each other. The distribution for $\rho_q \to 0$ takes the form:

$$\frac{1}{\Gamma_0}\frac{d\Gamma_{\rho_q \to 0}}{dx} = \sqrt{x^2 - 4\rho_\tau}\,\{(6 + 6\rho_\tau)\,x - 4\,x^2 - 8\rho_\tau +$$
$$K_b\,[\,\frac{28\,\rho_\tau}{3} + \delta'(1 + \rho_\tau - x)\,(-\frac{1}{3} + \frac{4\,\rho_\tau}{3} - 2\rho_\tau^2 + \frac{4\rho_\tau^3}{3} - \frac{\rho_\tau^4}{3}) - \frac{10\,x^2}{3}\,] +$$
$$G_b\,[-\frac{28\,\rho_\tau}{3} + \delta(1 + \rho_\tau - x)\,(-\frac{11}{3} + 9\rho_\tau - 7\rho_\tau^2 + \frac{5\rho_\tau^3}{3}) + 4\,x + \frac{10\,x^2}{3}\,]\}. \quad (3.8)$$

Now we can integrate distribution (3.3) to get the decay width including $1/m_b^2$ corrections:

$$\Gamma = |V_{qb}|^2\,\frac{G_F^2\,m_b^5}{192\,\pi^3}\{(1 + \frac{G_b - K_b}{2})\,z_0(\rho_q, \rho_\tau) - 2\,G_b\,z_1(\rho_q, \rho_\tau)\}, \quad (3.9)$$

where:

$$z_0(\rho_q, \rho_\tau) = \sqrt{\lambda}\,(1 - 7\rho_q - 7\rho_q^2 + \rho_q^3 - 7\rho_\tau - 7\rho_\tau^2 + \rho_\tau^3 + \rho_q\,\rho_\tau\,(12 - 7\rho_q - 7\rho_\tau)) +$$
$$12\,\rho_q^2\,(1 - \rho_\tau^2)\,\log\frac{(1 + v_q)}{(1 - v_q)} + 12\,\rho_\tau^2\,(1 - \rho_q^2)\,\log\frac{(1 + v_\tau)}{(1 - v_\tau)} \quad (3.10)$$

$$z_1(\rho_q, \rho_\tau) = \sqrt{\lambda}\,((1 - \rho_q)^3 + (1 - \rho_\tau)^3 - 1 - \rho_q\,\rho_\tau\,(4 - 7\rho_q - 7\rho_\tau)) +$$
$$12\,\rho_q^2\,\rho_\tau^2\,\log\frac{(1 + v_q)(1 + v_\tau)}{(1 - v_q)(1 - v_\tau)}, \quad (3.11)$$

$$\lambda = \lambda(1, \rho_q, \rho_\tau) = 1 + \rho_q^2 + \rho_\tau^2 - 2\rho_q - 2\rho_\tau - 2\rho_q\rho_\tau,$$

$v_q$ and $v_\tau$ are the maximal velocities of the quark and the $\tau$-lepton produced in the decay:

$$v_q = \frac{\sqrt{\lambda}}{1 + \rho_q - \rho_\tau}, \qquad v_\tau = \frac{\sqrt{\lambda}}{1 - \rho_q + \rho_\tau}. \quad (3.12)$$

There exists a simple relation between the two functions $z_0(\rho_q, \rho_\tau)$ and $z_1(\rho_q, \rho_\tau)$ [7]:

$$z_1(\rho_q, \rho_\tau) = -2\rho_q\,\frac{dz_0(\rho_q, \rho_\tau)}{d\rho_q} - 2\rho_\tau\,\frac{dz_0(\rho_q, \rho_\tau)}{d\rho_\tau} + 4\,z_0(\rho_q, \rho_\tau). \quad (3.13)$$

Width (3.9) is symmetrical function of $\rho_q$ and $\rho_\tau$ and therefore its limit when $\rho_q \to 0$

$$\Gamma_{\rho_q \to 0} = |V_{qb}|^2\,\frac{G_F^2\,m_b^5}{192\,\pi^3}\{(1 + \frac{G_b - K_b}{2})\,(1 - 8\rho_\tau + 8\rho_\tau^3 - \rho_\tau^4 - 12\rho_\tau^2\,\log\rho_\tau) - 2\,G_b\,(1 - \rho_\tau)^4\}$$
(3.14)

is the same as limit when $\rho_\tau \to 0$ with substitution $\rho_q \Rightarrow \rho_\tau$.

## 3.2 Numerical estimates and experimental predictions

To make numerical estimates in this section we will consider the $B$-meson decays $B \to \tau \bar\nu X$. We choose the following values for the parameters entering expressions (3.3), (3.8), (3.9) and (3.14) (see discussion in [7] and in chapter 2, [13]): $m_b = 4.8$ GeV, $m_c = 1.4$ GeV, $m_u = 0$, $m_\tau = 1.78$ GeV, $G_b = 0.02$ and $K_b = 0.017$ (note that for $\Lambda_b$ we have $G_b = 0$ and the corrections are much smaller).

For the width of the decay we see that the nonperturbative corrections are negative. For $b \to u\tau\bar\nu$ transitions they decrease the width by 6% of its perturbative value, while for the $b \to c\tau\bar\nu$ case they decrease it by 10% (note that for the massless lepton in the final state these numbers are 4% and 5% correspondingly). The quantity insensitive to the uncertainties of known values of $V_{qb}$ is $\Gamma(b \to \tau\nu X)/\Gamma(b \to e\nu X)$:

$$r_q = \frac{\Gamma(b \to \tau\nu X)}{\Gamma(b \to e\nu X)} = \frac{z_0(\rho_q, \rho_\tau)}{z_0(\rho_q, 0)} \left[ 1 - 2 G_b \left( \frac{z_1(\rho_q, \rho_\tau)}{z_0(\rho_q, \rho_\tau)} - \frac{z_1(\rho_q, 0)}{z_0(\rho_q, 0)} \right) \right]. \tag{3.15}$$

The first factor of last equation describes the phase space suppression while the second contains the nonperturbative corrections. The corrections reduce $r_c$ by 4% and $r_u$ by 2% of their perturbative values. Note that in the quantity $r_c$ the relative contribution of the corrections is almost independent of the uncertainties in the value of $c$-quark mass.

The obtained energy distributions (3.3) and (3.8) could be applied to the decays involving the transitions $b \to c$ and $b \to u$. The analysis of applicability of the distributions was made in Chapter 2 ([7] and [13]). There it was shown that the proper quantity to confront with experiment is

$$\gamma(x) = \int_x^{1+\rho_\tau-\rho_q} dx' \frac{1}{\Gamma_0} \frac{d\Gamma(x')}{dx'}, \quad 2\sqrt{\rho_\tau} \le x < 1 + \rho_\tau - \rho_q. \tag{3.16}$$

This quantity does not contain the end-point singularities and is suitable for direct comparison with the experimental data for $x$ not too close to its maximal value (so that the operator product expansion is still valid and we are not in the resonance region). As we mentioned above, the hadronic kinematical region is different from the quark one. Therefore to compare our results with experiment the range of integration of the experimental distribution should include the window between the quark and hadronic boundaries $1 + \rho_\tau - \rho_q \le x \le M_B/m_b + \rho_\tau - \rho_q$. Correspondingly, the reliable prediction can only be made for $2\sqrt{\rho_\tau} \leq x \leq x_{max}$, where $x_{max} = 1 + \rho_\tau - \rho_q - (M_B - m_b)/m_b$. For $u$-quark $x_{max} \sim 1.05$, for $c$-quark $x_{max} \sim 0.95$.

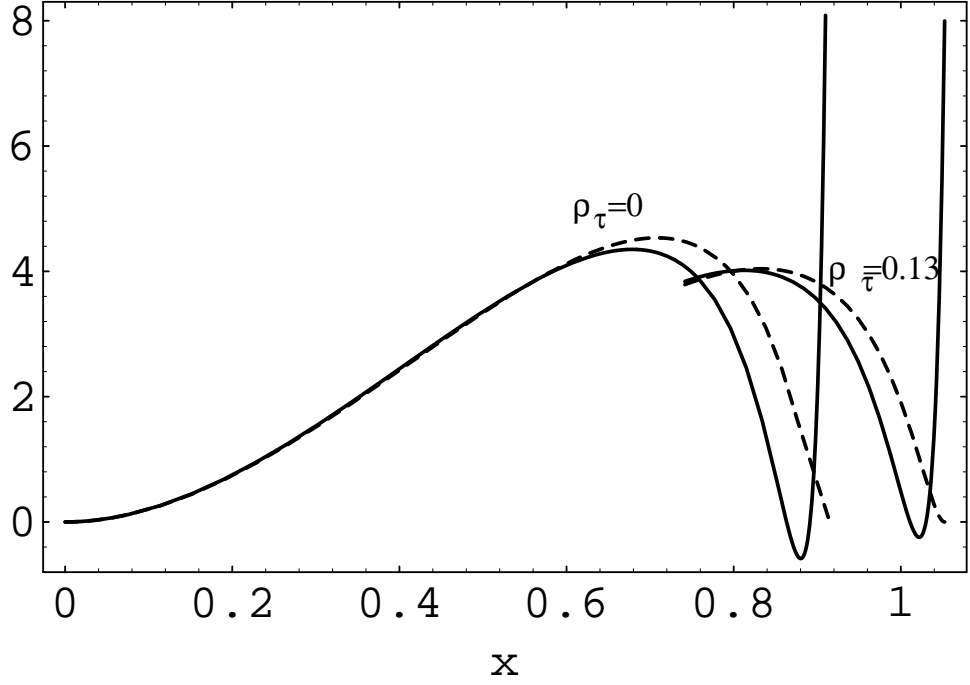

Figure 3.1: The energy spectrum of $\tau$ is plotted for $b \to c\tau\bar{\nu}$ transitions. The solid line shows the distribution with the nonperturbative corrections, while the dashed line – without them. For comparison, on the same plot we show the electron energy distribution for $b \to ce\bar{\nu}$ transitions ($\rho_\tau = 0$). The graph can only be trusted for $x < x_{max} \sim 0.95$.

The lepton energy spectrum is plotted on the Fig. 3.1 for $b \to c\tau\bar{\nu}$ and on the Fig. 3.2 for $b \to u\tau\bar{\nu}$. The delta-functions of Eq.(3.8) are not shown on the graph. For comparison, on the same plots we show the energy distributions for electrons in $b \to ce\bar{\nu}$ and $b \to ue\bar{\nu}$ transitions correspondingly ($\rho_\tau = 0$).

The function $\gamma(x)$ is plotted on Fig. 3.3 for the case $b \to c\tau\bar{\nu}$ and on Fig. 3.4 for $b \to u\tau\bar{\nu}$. The solid line shows $\gamma(x)$ with the nonperturbative corrections while the dashed

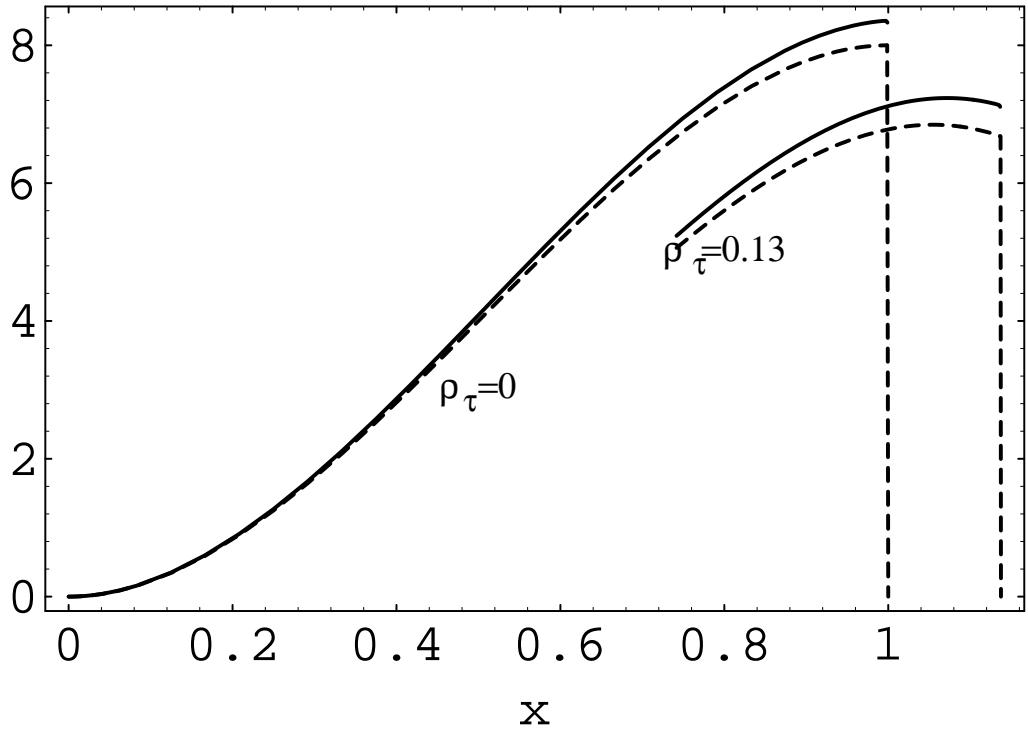

Figure 3.2: The energy spectrum (3.8) of $\tau$ is plotted for $b \to u\tau\bar{\nu}$ transitions ($\rho_q = 0$). The solid line shows the distribution with the nonperturbative corrections, while the dashed line – without them. For comparison, on the same plot we show the electron energy distribution for $b \to ue\bar{\nu}$ transitions ($\rho_\tau = 0$). The graph can only be trusted for $x < x_{max} \sim 1.05$.

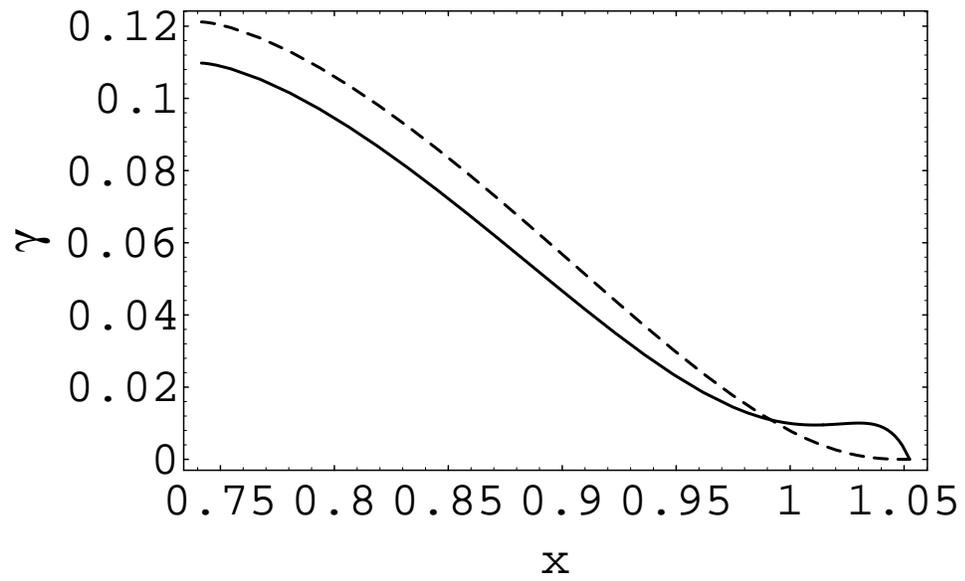

Figure 3.3: The function $\gamma(x)$ plotted for the case $b \to c\tau\bar{\nu}$. The solid line shows $\gamma(x)$ with the nonperturbative corrections while the dashed line – without them. The graph can only be trusted for $x < x_{max} \sim 0.95$.

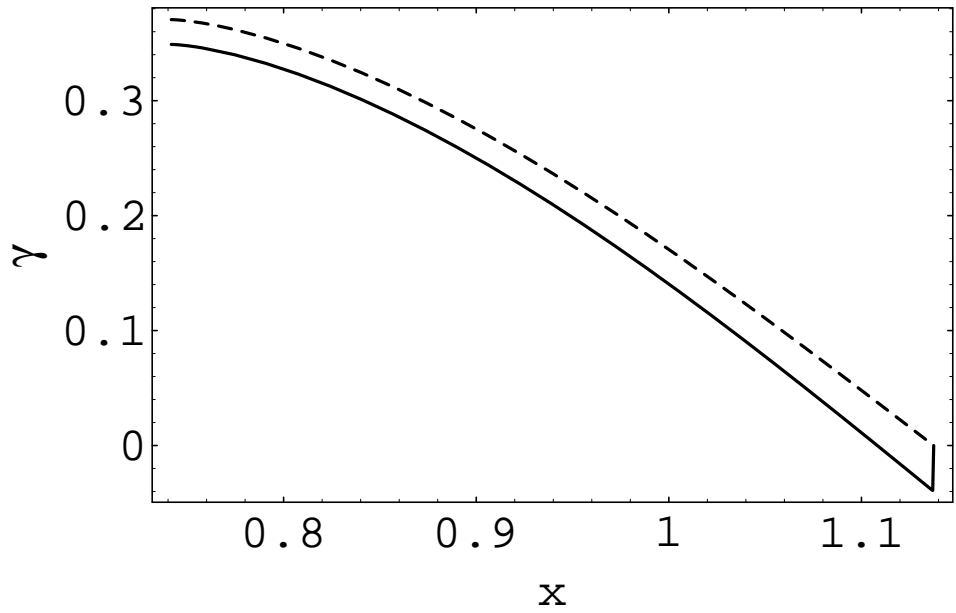

Figure 3.4: The function $\gamma(x)$ for the case $b \to u\tau\bar{\nu}$ ($\rho_q = 0$). The solid line shows $\gamma(x)$ with the nonperturbative corrections while the dashed line – without them. The graph can only be trusted for $x < x_{max} \sim 1.05$.

line – without them.

Although the functions are plotted for the whole range of $x$, $2\sqrt{\rho_\tau} \leq x \leq 1 + \rho_\tau - \rho_q$, we can only trust the graphs for $x < x_{max}$.

## 3.3 Summary of results of chapters 2 and 3

Let us now summarize our results. Model independent approach to nonperturbative effects $(1/m_Q)^n$ is used for calculations of differential distributions. The effects are most pronounced near the endpoints of the spectra. We discussed how the comparison with experiment should be formulated accounting for the boundary effects. Somewhat disappointing is that we cannot use our results to improve an extraction of $V_{ub}$ by the consideration of $q^2$ dependence. Indeed, experimentally the signal of $b \to u$ is due to the range of electron energy $E_e$ near the upper end where $b \to c$ is absent. However as it follows from Fig.2.4 the distribution in $q^2$ at such energies is concentrated in the resonance domain, and no model-independent prediction emerges.

We also investigated the nonperturbative corrections up to order $1/m_b^2$ to the decays $H_b \to \tau\bar{\nu}X$ and found their contribution to the $\tau$-lepton energy spectrum and the width of the decay. In the case of $B$-meson decays the corrections could be up to 10% of the width. Unfortunately at present time the experimental measurements are not accurate enough in order to compare our results with the data.

# Chapter 4

# Sum rules for heavy flavor transitions

## 4.1 Cuts and sums. OPE sum rules and predictability of moments and inclusive distributions

One of the goals of this work is to formulate a model for the semileptonic decays of heavy flavors which could describe transitions into individual hadronic final states. We will see that QCD provides sufficient number of constraints on the formfactors of such transitions. Those constraints are imposed in the form of sum rules which follow from the OPE analysis of the inclusive decays developed in the previous chapters. We consider these sum rules to be the QCD conditions on the structure functions of different final state hadronic resonances which should be satisfied by any phenomenological model. This is in fact what we will call the QCD constrained model of the semileptonic decay of the $B$ mesons. So, we define our model by the requirement that it satisfies these sum rules.

The sum rules for semileptonic heavy flavor transitions were considered in the works [9],[48], [10],[11],[35],[52]. These sum rules are based on the dispersion relations which follow from analyticity and unitarity of the scattering matrix. This approach follows many ideas of the classical works [49]. Despite the fact that we are not going to use Borel transform in our analysis in this work, it is clear that it also could be utilized and we are going to do it somewhere.

In this chapter we are going to briefly review and summarize the results for the OPE sum rules derived in the above cited papers and provide the full list of the sum rules for heavy flavor transitions up to the order $1/m_Q^2$. This is the highest order we can go based on the structure functions calculations done in Chapter 2 and listed in the Appendix B (see also [13] and [38]).

To be specific, we will be talking about $b \to c$ transitions, however no sum rules will contain the $c$–quark mass in the denominator, thus they could be used for $b \to u$ transitions as well. As we will see, it is only $E_c$ (or $E_u$, correspondingly) that should remain large parameter for the sum rules to be valid.

The hadronic invariant functions introduced in the previous section satisfy the following dispersion relations

$$h_j(q_0, \mathbf{q}^2) = \frac{1}{2\pi} \int_{cut} \frac{w_j(\tilde{q}_0, \mathbf{q}^2)}{\tilde{q}_0 - q_0} d\tilde{q}_0. \tag{4.1}$$

Here $cut$ means the physical cut in the $q_0$ complex plain. $q_0$ is some point away from the cut. We will distinguish two kinds of cuts: hadronic for which we keep notation $cut$ and quark, which we will denote as $qcut$.

In Chapter 2 (see also [13] and [38]) the tensor $h_{\mu\nu}$ was calculated by means of the operator product expansion (OPE) for the correlator of currents in (2.13). Let us call $h_{\mu\nu}$ calculated in this way $h_{\mu\nu}^{\text{theor}}$, which means that these functions vave been calculated using theoretical metods. The expressions for the functions $h_j^{\text{theor}}$ obtained by means of the OPE are given in the Appendix B (see also [13]). For them the dispersion relation (4.1) is only valid for the values of $q_0$ for which the OPE is valid: far from the cuts.

On the other hand, we can represent the hadronic tensor $W_{\mu\nu}$ by inserting the complete set of physical states in the Eq. (2.4). We will call the tensor calculated in such a way $W_{\mu\nu}^{\text{phen}}$, phenomenological. In the points where the OPE is valid, the $h_{\mu\nu}^{\text{phen}}(q_0, \mathbf{q}^2)$ calculated using (4.1) for $w_j^{\text{phen}}$ and the $h_{\mu\nu}^{\text{theor}}(q_0, \mathbf{q}^2)$ should be the same with the accuracy with which the OPE was calculated. Then in a remote from the cuts point $q_0$ one obtains the following equation:

$$h_j^{\text{theor}}(q_0, \mathbf{q}^2) \approx \frac{1}{2\pi} \int_{cut} \frac{w_j^{\text{phen}}(\tilde{q}_0, \mathbf{q}^2)}{\tilde{q}_0 - q_0} d\tilde{q}_0. \tag{4.2}$$

At the same time, we can also write:

$$h_j^{\text{theor}}(q_0, \mathbf{q}^2) = \frac{1}{2\pi} \int_{qcut} \frac{w_j^{\text{theor}}(\tilde{q}_0, \mathbf{q}^2)}{\tilde{q}_0 - q_0} d\tilde{q}_0. \tag{4.3}$$

Now equating RHS's of (4.2) and (4.3) we have:

$$\frac{1}{2\pi} \int_{cut} \frac{w_j^{\text{phen}}(\tilde{q}_0, \mathbf{q}^2)}{\tilde{q}_0 - q_0} d\tilde{q}_0 \approx \frac{1}{2\pi} \int_{qcut} \frac{w_j^{\text{theor}}(\tilde{q}_0, \mathbf{q}^2)}{\tilde{q}_0 - q_0} d\tilde{q}_0. \tag{4.4}$$

The point $q_0$ in which the comparison of the theoretical and phenomenological invariant functions is performed will be called reference pount.

Let us describe the analytical structure of the hadronic tensor $h_{\mu\nu}(q_0, \mathbf{q}^2)$ in the complex plane of $q_0$ at fixed $\mathbf{q}^2$. This tensor has discontinuities on the physical cuts, which represent possible physical processes. The semileptonic decay cut spans from $q_0 = 0$ to $q_0 = M_B - \sqrt{M_D^2 + \mathbf{q}^2}$, we will call it $cut_1$. Another cut corresponding to fusion of $B$ and $\bar{D}$ mesons, goes from $q_0 = M_B + \sqrt{M_D^2 + \mathbf{q}^2}$ to $q_0 = \infty$, we will call it $cut_2$. There is the third cut going from $q_0 = -\infty$ to $q_0 = 0$, which corresponds to the process of $\bar{D}$ absorbing a $W$-boson and going into $\bar{B}$. This cuts structure is shown on Fig. 4.1.

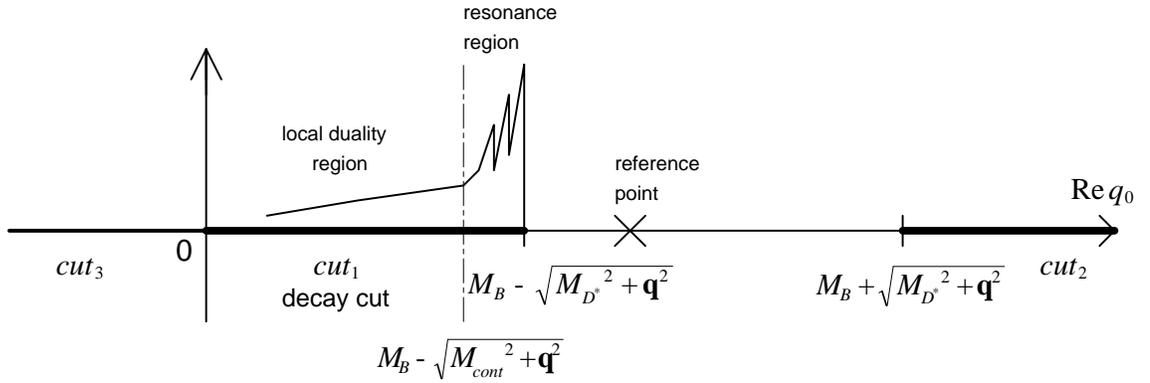

Figure 4.1: Cuts of the forward scattering amplitude in the complex plane $q_0$ and position of the resonance region and continuum spectrum.

Let us introduce a new variable

$$\epsilon = M_B - \sqrt{M_{D^*}^2 + \mathbf{q}^2} - q_0, \tag{4.5}$$

which describes the distance from the right end of the decay cut to the reference point $q_0$. Then $\tilde{\epsilon} = M_B - \sqrt{M_{D^*}^2 + \mathbf{q}^2} - \tilde{q}_0$ is the final state excitation energy counted from the $E_{D^*}$.

Note, that in the definition of $\epsilon$ we used the hadronic masses, and not the quark ones. Then for hadronic invariant functions we have the following expression:

$$h_j^{\text{theor}}(\epsilon, \mathbf{q}^2) \approx \frac{1}{2\pi}\int_{cut_1} \frac{w_j^{\text{phen}}(\tilde{\epsilon}, \mathbf{q}^2)}{\tilde{\epsilon}-\epsilon}d\tilde{\epsilon} + \frac{1}{2\pi}\int_{cut_2+cut_3} \frac{w_j^{\text{phen}}(\tilde{\epsilon}, \mathbf{q}^2)}{\tilde{\epsilon}-\epsilon}d\tilde{\epsilon}. \qquad (4.6)$$

In the Eq. (4.6) $\tilde{\epsilon}$ is in fact excitation energy of the final hadronic state counted from the $D^*$ energy. The equation (4.6) in fact is the starting point for the derivation of the sum rules. [49],see [10], [11],[9],[48],[52],[35]. Despite of the different ways used for deriving the sum rules in the original papers, they all could be obtained in a systematic way from the equation (4.6) (see [10]).

The choice of the reference point $\epsilon$ is based on the following consideration. The only experimentally observed processes lie on the decay part of the physical cut of the forward scattering amplitude. Therefore, the integral in the RHS of the Eq. (4.2) is known only for the part of the cut corresponding to the $B$ meson decays. This dictates the choice of the reference point close to the decay part of the cut (which is $cut_1$) and as far from the other cuts as possible. At the same time $\epsilon$ can not be too close to $cut_1$ for the OPE to be valid. Therefore we choose the reference point in between $cut_1$ and $cut_2$ but as close to $cut_1$ as possible. This will lead to the fact that in Eq. (4.6) only integral over $cut_1$ will be relevant for derivation of the sum rules. Note, that the value of $\epsilon$ at the reference point is negative.

One of important assumtions we are going to make is the assumption of local duality starting at the scale of $M_{\text{dual}} - M_{D^*}$. In practical terms this duality means that hadronic tensor $w_j^{\text{phen}}(q_0, \mathbf{q}^2)$ is the same as the theoretical one $w_j^{\text{theor}}(q_0, \mathbf{q}^2)$.

The choice of the reference point close to $cut_1$ enables us to leave only integrations over $cut_1$ and neglect contributions of other cuts. Then with the help of the local duality we can write:

$$\frac{1}{2\pi}\int_{E_D-E_{D^*}}^{\epsilon_{\text{dual}}} \frac{w_j^{\text{phen}}(\tilde{\epsilon}, \mathbf{q}^2)}{\tilde{\epsilon}-\epsilon}d\tilde{\epsilon} \approx \frac{1}{2\pi}\int_{(M_B-m_b)-(E_{D^*}-E_c)}^{\epsilon_{\text{dual}}} \frac{w_j^{\text{theor}}(\tilde{\epsilon}, \mathbf{q}^2)}{\tilde{\epsilon}-\epsilon}d\tilde{\epsilon} \qquad (4.7)$$

where $\epsilon_{\text{dual}} = M_B - \sqrt{M_{\text{dual}}^2 + \mathbf{q}^2}$. The unusually looking lower limit of integration in the RHS is due to the fact that integration in quark $q_0$ goes from 0 to $m_b - E_c$, and, therefore, in $\epsilon$ it starts from $(M_B - m_b) - (E_{D^*} - E_c)$.

For $\tilde{\epsilon}$ lying on $cut_1$ where $|\tilde{\epsilon}| \ll |\epsilon|$, $\epsilon < 0$ let us expand the denominators of the integrals over the $cut_1$ in the following series:

$$\frac{1}{\tilde{q}_0 - q_0} = \frac{1}{\epsilon - \tilde{\epsilon}} = \sum_{k=0}^{\infty} \frac{(-\tilde{\epsilon})^k}{\epsilon^{k+1}}, \qquad (4.8)$$

For the other cuts, on the contrary, we have different condition satisfied: $|\epsilon| \ll |\tilde{\epsilon}|$, and the expansion goes in the positive powers of $\epsilon$ and not in negative.

Now by equating the coefficients in front of $1/\epsilon^k$, we arrive at the OPE sum rules:

$$\frac{1}{2\pi} \int_{M_B - E_{D^*}}^{\epsilon_{\text{dual}}} w_j^{\text{phen}}(\epsilon, \mathbf{q}^2) \epsilon^k d\epsilon = \frac{1}{2\pi} \int_{(M_B - m_b) - (E_{D^*} - E_c)}^{\epsilon_{\text{dual}}} w_j^{\text{theor}}(\epsilon, \mathbf{q}^2) \epsilon^k d\epsilon. \qquad (4.9)$$

It is clear now that the operator product expansion approach provides answers for the moments of physical structure functions, but not for the functions themselves. This is sufficient for calculation of *inclusive* quantities such as inclusive differential distributions and total widths. In the next chapters however, it will be shown that if one consideres the moments (or the corresponding sum rules) as constraints on a model of the decay, then determination of *exclusive* quantities also becomes possible.

Let us notice that the first three sum rules corresponding to $k = 0, 1, 2$ could be calculated using the results for $h_i$'s derived in the work [13] (see Appendix B), since in that work $h_j$'s were calculated only to the order $1/m_Q^2$. The last term on the RHS in fact contains gluon corrections to the forward scattering amplitude and therefore all corrections of order $\alpha_s$ due to virtual and real gluons should be included. The virtual gluon corrections have been calculated in the previous works [40]. Corrections due to emission of a real gluon will be calculated later in this work.

## 4.2 Complete list of the OPE sum rules up to order $1/m_Q^2$

The RHS of the sum rules (4.9) for $k = 0, 1, 2$ for $j = 1, 2, 3$ were calculated in the work [10]. Here we reproduce the results of that work and add the results of calculations for $j = 4, 5$.

As it was explained in the previous section, for the moments

$$I_j^{J_1 J_2, k} = \frac{1}{2\pi} \int_{(M_B - m_b) - (E_{D^*} - E_c)}^{\epsilon_{\text{dual}}} w_j^{\text{phen}}(\epsilon, \mathbf{q}^2) \epsilon^k d\epsilon$$

we have theoretical predictions. Here $I_j^{J_1 J_2, k}$ is the moment corresponding to the meson average of $T\{J_1 J_2\}$. As it was explained in the previous section, for these quantities we have theoretical predictions. In the derivation of the hadronic structure functions the expressions for $AA$ current could be obtained from the expressions for $VV$ current by substitution $m_c \to -m_c$. For the moments of the structure functions however this is not true because of the shift of variable $q_0$ to $\epsilon$ (4.5). The practical way of calculation was expansion of the corresponding hadronic invariant functions in $1/\epsilon$ (see [10] for details). In this derivation we used mass relations (2.47-2.50). Here we provide here the complete list of moments of all structure functions without perturbative corrections. The perturbative corrections to the sum rules will be calculated in the next chapter.

For the $VV$ currents:

$$\begin{aligned} I_1^{VV,0} &= \frac{1}{2} - \frac{m_c}{2\,E_c} + \left(\frac{m_c}{12\,E_c{}^3} - \frac{m_c}{4E_c\,m_b{}^2} + \frac{m_c{}^2}{6\,E_c{}^3\,m_b}\right)\mu_G^2 \\ &+ \left(\frac{m_c}{4\,E_c\,m_b{}^2} - \frac{m_c{}^2}{6\,E_c{}^3\,m_b} + \frac{m_c{}^3}{4\,E_c{}^5}\right)\mu_\pi^2, \end{aligned} \quad (4.10)$$

$$\begin{aligned} I_1^{VV,1} &= \frac{\bar{\Lambda}}{2}\left(1 - \frac{m_c}{E_c}\right)^2 - \frac{\bar{\Lambda}^2}{4E_c}\left(1 - \frac{m_c^2}{E_c^2}\right)\left(1 - \frac{m_c}{E_c}\right) + \left(1 - \frac{m_c{}^2}{E_c{}^2}\right)\frac{\mu_G^2}{6m_b} \\ &+ \left(\frac{-1}{12\,E_c} - \frac{1}{6\,m_b} + \frac{m_c}{4\,E_c{}^2}\right)(1 - \frac{m_c^2}{E_c^2})\mu_\pi^2, \end{aligned} \quad (4.11)$$

$$I_1^{VV,2} = \frac{\bar{\Lambda}^2}{2}\left(1 - \frac{m_c}{E_c}\right)^3 + \left(1 - \frac{m_c}{E_c}\right)\left(1 - \frac{m_c{}^2}{E_c{}^2}\right)\frac{\mu_\pi^2}{6}, \quad (4.12)$$

$$I_2^{VV,0} = \frac{m_b}{E_c} + \left(\frac{-5}{6\,E_c\,m_b} + \frac{m_b}{2\,E_c{}^3} + \frac{m_c}{3\,E_c{}^3}\right)\mu_G^2 + \left(\frac{5}{6\,E_c\,m_b} - \frac{2\,m_b}{3\,E_c{}^3} - \frac{m_b\,m_c{}^2}{2\,E_c{}^5}\right)\mu_\pi^2, \quad (4.13)$$

$$\begin{aligned} I_2^{VV,1} &= \bar{\Lambda}\frac{m_b}{E_c}\left(1 - \frac{m_c}{E_c}\right) - \bar{\Lambda}^2\frac{m_b}{2E_c^2}\left(1 - \frac{m_c{}^2}{E_c^2}\right) + \\ &\left(1 - \frac{m_c}{E_c} - \frac{2\,m_b}{E_c}\right)\frac{\mu_G^2}{3E_c} + \left(\frac{-2}{3} + \frac{m_b}{6\,E_c} + \frac{m_b\,m_c{}^2}{2\,E_c{}^3}\right)\frac{\mu_\pi^2}{E_c}, \end{aligned} \quad (4.14)$$

$$I_2^{VV,2} = \bar{\Lambda}^2 \frac{m_b}{E_c} \left(1 - \frac{m_c}{E_c}\right)^2 + \left(1 - \frac{m_c^2}{E_c^2}\right) \frac{m_b}{3\,E_c} \mu_\pi^2, \tag{4.15}$$

$$I_4^{VV,0} = I_4^{AA,0} = \frac{\mu_G^2}{3\,E_c^3\,m_b} - \frac{\mu_\pi^2}{3\,E_c^3\,m_b}, \tag{4.16}$$

$$I_4^{VV,1} = I_4^{AA,1} = \frac{-\mu_G^2}{3\,E_c^2\,m_b} + \frac{\mu_\pi^2}{3\,E_c^2\,m_b}, \tag{4.17}$$

$$I_4^{VV,2} = I_4^{AA,2} = 0 \tag{4.18}$$

$$I_5^{VV,0} = I_5^{AA,0} = \frac{-1}{2\,E_c} - \frac{5\,\mu_G^2}{12\,E_c^3} + \left(1 + \frac{m_c^2}{2\,E_c^2}\right) \frac{\mu_\pi^2}{2 E_c^3}, \tag{4.19}$$

$$\begin{aligned} I_5^{VV,1} = I_5^{AA,1} = & -\frac{\bar{\Lambda}}{2E_c} \left(1 - \frac{m_c}{E_c}\right) + \frac{\bar{\Lambda}^2}{4 E_c^2} \left(1 - \frac{m_c^2}{E_c^2}\right) + \\ & \left(\frac{1}{2} - \frac{E_c}{6\,m_b}\right) \frac{\mu_G^2}{E_c^2} + \left(\frac{-1}{4} + \frac{E_c}{6\,m_b} - \frac{m_c^2}{4\,E_c^2}\right) \frac{\mu_\pi^2}{E_c^2}, \end{aligned} \tag{4.20}$$

$$I_5^{VV,2} = I_5^{AA,2} = -\frac{\bar{\Lambda}^2}{2E_c} \left(1 - \frac{m_c}{E_c}\right)^2 - \left(1 - \frac{m_c^2}{E_c^2}\right) \frac{\mu_\pi^2}{6 E_c}, \tag{4.21}$$

For the $AA$ currents:

$$\begin{aligned} I_1^{AA,0} = & \frac{1}{2} + \frac{m_c}{2\,E_c} + \left(\frac{-m_c}{12\,E_c^3} + \frac{m_c}{4\,E_c\,m_b^2} + \frac{m_c^2}{6\,E_c^3\,m_b}\right) \mu_G^2 + \\ & \left(\frac{-m_c}{4\,E_c\,m_b^2} - \frac{m_c^2}{6\,E_c^3\,m_b} - \frac{m_c^3}{4\,E_c^5}\right) \mu_\pi^2, \end{aligned} \tag{4.22}$$

$$\begin{aligned} I_1^{AA,1} = & \frac{\bar{\Lambda}}{2}\left(1 - \frac{m_c^2}{E_c^2}\right) - \frac{\bar{\Lambda}^2}{4 E_c} \left(1 + \frac{m_c}{E_c}\right)\left(1 - \frac{m_c^2}{E_c^2}\right) + \left(1 - \frac{m_c^2}{E_c^2}\right) \frac{\mu_G^2}{6 m_b} - \\ & \left(\frac{1}{12\,E_c} + \frac{1}{6\,m_b} + \frac{m_c}{4\,E_c^2}\right)(1 - \frac{m_c^2}{E_c^2})\mu_\pi^2, \end{aligned} \tag{4.23}$$

$$I_1^{AA,2} = \bar{\Lambda}^2 \left(\frac{1}{2} - \frac{m_c}{2\,E_c} - \frac{m_c^2}{2\,E_c^2} + \frac{m_c^3}{2\,E_c^3}\right) + \left(\frac{1}{6} + \frac{m_c}{6\,E_c} - \frac{m_c^2}{6\,E_c^2} - \frac{m_c^3}{6\,E_c^3}\right) \mu_\pi^2, \tag{4.24}$$

$$I_2^{AA,0} = \frac{m_b}{E_c} + \left(\frac{-5}{6\,m_b} + \frac{m_b}{2\,E_c^2} - \frac{m_c}{3\,E_c^2}\right)\frac{\mu_G^2}{E_c} + \left(\frac{5}{6\,m_b} - \frac{2\,m_b}{3\,E_c^2} - \frac{m_b\,m_c^2}{2\,E_c^4}\right)\frac{\mu_\pi^2}{E_c}, \qquad (4.25)$$

$$\begin{aligned}I_2^{AA,1} &= \bar{\Lambda}\frac{m_b}{E_c}\left(1 - \frac{m_c}{E_c}\right) - \bar{\Lambda}^2\frac{m_b}{2\,E_c^2}\left(1 - \frac{m_c^2}{E_c^2}\right) + \\ &\quad \left(1 - \frac{2\,m_b}{E_c} + \frac{m_c}{E_c}\right)\frac{\mu_G^2}{3E_c} + \left(-\frac{2}{3} + \frac{m_b}{6\,E_c} + \frac{m_b\,m_c^2}{2\,E_c^3}\right)\frac{\mu_\pi^2}{E_c},\end{aligned} \qquad (4.26)$$

$$I_2^{AA,2} = \bar{\Lambda}^2\frac{m_b}{E_c}\left(1 - \frac{m_c}{E_c}\right)^2 + \left(1 - \frac{m_c^2}{E_c^2}\right)\frac{m_b}{3\,E_c}\mu_\pi^2, \qquad (4.27)$$

For the $VA$ currents:

$$I_3^{VA,0} = \frac{1}{2\,E_c} - \frac{\mu_G^2}{12\,E_c^3} - \left(\frac{1}{6} + \frac{m_c^2}{4\,E_c^2}\right)\frac{\mu_\pi^2}{E_c^3}, \qquad (4.28)$$

$$\begin{aligned}I_3^{VA,1} &= \frac{\bar{\Lambda}}{2\,E_c}\left(1 - \frac{m_c}{E_c}\right) - \frac{\bar{\Lambda}^2}{4\,E_c^2}\left(1 - \frac{m_c^2}{E_c^2}\right) + \frac{\mu_G^2}{6\,E_c\,m_b} \\ &\quad + \left(\frac{-1}{12} - \frac{E_c}{6\,m_b} + \frac{m_c^2}{4\,E_c^2}\right)\frac{\mu_\pi^2}{E_c^2},\end{aligned} \qquad (4.29)$$

$$I_3^{VA,2} = \frac{\bar{\Lambda}^2}{2\,E_c}\left(1 - \frac{m_c}{E_c}\right)^2 + \frac{\mu_\pi^2}{6\,E_c}\left(1 - \frac{m_c^2}{E_c^2}\right). \qquad (4.30)$$

All moments that are not written are equal to zero. Let us notice that all second moments are related to each other by the heavy quark symmetry, this happens because these moments are only known in the leading order and nonperturbative corrections to them have not been calculated.

For the $V - A$ current we have

$$I_j^{V-A,k} = I_j^{VV,k} - 2I_j^{VA,k} + I_j^{AA,k}.$$

For the reader's convenience let us reproduce them here:

$$I_1^{V-A,0} = 1 + \frac{m_c^2(\mu_G^2 - \mu_\pi^2)}{3E_c^3 m_b}, \qquad (4.31)$$

$$\begin{aligned}I_1^{V-A,1} &= \bar{\Lambda}\left(1 - \frac{m_c}{E_c}\right) - \frac{\bar{\Lambda}^2}{2E_c}\left(1 - \frac{m_c^2}{E_c^2}\right) + \\ &\quad \left(1 - \frac{m_c^2}{E_c^2}\right)\frac{\mu_G^2}{3m_b} - \left(1 - \frac{m_c^2}{E_c^2}\right)\left(\frac{1}{6E_c} + \frac{1}{3m_b}\right)\mu_\pi^2,\end{aligned} \qquad (4.32)$$

$$I_1^{V-A,2} = \bar{\Lambda}^2 \left(1 - \frac{m_c}{E_c}\right)^2 + \left(1 - \frac{m_c^2}{E_c^2}\right) \frac{\mu_\pi^2}{3}, \tag{4.33}$$

$$I_2^{V-A,0} = \frac{2m_b}{E_c} + \left(\frac{-5}{3m_b} + \frac{m_b}{E_c^2}\right) \frac{\mu_\pi^2}{E_c} + \left(\frac{5}{3m_b} - \frac{4m_b}{3E_c^2} - \frac{m_b m_c^2}{E_c^4}\right) \frac{\mu_\pi^2}{E_c}, \tag{4.34}$$

$$\begin{aligned}I_2^{V-A,1} &= \bar{\Lambda}\frac{2m_b}{E_c}\left(1 - \frac{m_c}{E_c}\right) - \bar{\Lambda}^2 \frac{m_b}{E_c^2}\left(1 - \frac{m_c^2}{E_c^2}\right) + \\ &\quad \left(1 - \frac{2m_b}{E_c}\right)\frac{2\mu_G^2}{3E_c} + \left(\frac{-4}{3} + \frac{m_b}{3E_c} + \frac{m_b m_c^2}{E_c^3}\right)\frac{\mu_\pi^2}{E_c},\end{aligned} \tag{4.35}$$

$$I_2^{V-A,2} = \bar{\Lambda}^2 \frac{2m_b}{E_c}\left(1 - \frac{m_c}{E_c}\right)^2 + \frac{2m_b}{3E_c}\left(1 - \frac{m_c^2}{E_c^2}\right)\mu_\pi^2, \tag{4.36}$$

$$I_3^{V-A,0} = \frac{1}{E_c} - \frac{\mu_G^2}{6E_c^3} - \left(\frac{1}{3} + \frac{m_c^2}{2E_c^2}\right)\frac{\mu_\pi^2}{E_c^3}, \tag{4.37}$$

$$\begin{aligned}I_3^{V-A,1} &= \frac{\bar{\Lambda}}{E_c}\left(1 - \frac{m_c}{E_c}\right) - \frac{\bar{\Lambda}^2}{2E_c^2}\left(1 - \frac{m_c^2}{E_c^2}\right) + \\ &\quad \frac{\mu_G^2}{3E_c m_b} + \left(\frac{-1}{6} - \frac{E_c}{3m_b} + \frac{m_c^2}{2E_c^2}\right)\frac{\mu_\pi^2}{E_c^2},\end{aligned} \tag{4.38,4.39}$$

$$I_3^{V-A,2} = \frac{\bar{\Lambda}^2}{E_c}\left(1 - \frac{m_c}{E_c^2}\right)^2 + \left(1 - \frac{m_c^2}{E_c^2}\right)\frac{\mu_\pi^2}{3E_c}, \tag{4.40}$$

$$I_4^{V-A,0} = 2I_4^{VV,0}, \tag{4.41}$$

$$I_4^{V-A,1} = 2I_4^{VV,1}, \tag{4.42}$$

$$I_4^{V-A,2} = 0, \tag{4.43}$$

$$I_5^{V-A,0} = 2I_5^{VV,0}, \tag{4.44}$$

$$I_5^{V-A,1} = 2I_5^{VV,1}, \tag{4.45}$$

$$I_5^{V-A,2} = 2I_5^{VV,2}. \tag{4.46}$$

The listed moments could be viewed as expanded in the inverse heavy quark mass, they contain leading order contributions plus nonperturbative corrections (except second moments for which only the leading order is calculated). Radiative corrections to the sum rules will be calculated in Chapter 5.

# Chapter 5

# Perturbative corrections to the sum rules

## 5.1 Introduction

In the discussion of the operator product expansion we did not take into account any perturbative corrections to the forward scattering amplitude. This also means that we have neglected contributions of perturbative corrections in the theoretical part of the sum rules.

There are two places the perturbative corrections are coming from: virtual gluon exchange and real gluon emission. Therefore one can write for the structure functions:

$$w_j^{\text{theor}}(\epsilon, \mathbf{q}^2) = w_j^{(0)}(\epsilon, \mathbf{q}^2) + w_j^{\text{virt}}(\epsilon, \mathbf{q}^2) + w_j^{\text{real}}(\epsilon, \mathbf{q}^2),$$

where $w_j^{(0)}(\epsilon, \mathbf{q}^2)$ is the leading order in $\alpha_s$ result including nonperturbative corrections obtained in Chapter 2, $w_j^{\text{virt}}(\epsilon, \mathbf{q}^2)$ and $w_j^{\text{real}}(\epsilon, \mathbf{q}^2)$ are perturbative corrections coming from virtual and real gluons correspondingly. Now expresstions for each moment of each structure function will contain two additional parts, due to virtual and real gluons.

The contribution of virtual gluons to the hadronic tensor could be found in the work [40], for readers convenience and to make this text self contained, we review the results of that work in the following Section 5.2. The real gluon contributions to the sum rules are calculated in Section 5.3.

## 5.2 Virtual gluons contributions

In this section the results of Ref. [40] of calculations of the virtual gluon contributions to the hadronic vertex are used to calculate the corresponding contributions to the hadronic tensor and the OPE sum rules.

The virtual gluon exchange results in the modification of the interaction vertex [40] (see also the review paper [41]):

$$\Gamma_{\text{QCD}}^{V,\mu} = (1 + a^V)\gamma^\mu + b^V v^\mu + c^V v'^\mu, \tag{5.1}$$

$$\Gamma_{\text{QCD}}^{A,\mu} = (1 + a^A)\gamma^\mu\gamma^5 + b^A v^\mu\gamma^5 + c^A v'^\mu\gamma^5, \tag{5.2}$$

where

$$a^{V,A} = \frac{\alpha_s}{\pi}[\ln\frac{m_b}{m_c} - \frac{\gamma_0^{hh}(w)}{4}\ln\frac{m_c}{\lambda} + \frac{2}{3}(f(w) \pm r(w) + g(z,w))], \tag{5.3}$$

$$b^{V,A} = -\frac{2\alpha_s}{3\pi}[2r(w) \mp 1 + h_1^{V,A}(z,w)], \tag{5.4}$$

$$c^{V,A} = \mp\frac{2\alpha_s}{3\pi}h_2^{V,A}(z,w), \tag{5.5}$$

where $z = m_c/m_b$, $w = v \cdot v' = E_c/m_c = \sqrt{1 + \mathbf{q}^2/m_c^2}$, and $v'$ is the velocity of the final $c$-quark, $\lambda$ is the gluon mass which is needed to regularize the infrared divergency. This divergency is to be cancelled in the sum rules by the corresponding divergency in the gluon emission part of perturbative corrections. The upper signs refer to the vector current, whereas the lower signs refer to the axial current. Expressions for the functions $f(w)$, $r(w)$, $g(z,w)$, $h_1^{V,A}(z,w)$, and $h_2^{(A)}(z,w)$ are given in the works [40], [41], we cite them here for the reader's convenience.

$$w_\pm = w \pm \sqrt{w^2 - 1}, \tag{5.6}$$

$$\gamma_0^{hh} = \frac{16}{3}[wr(w) - 1], \tag{5.7}$$

$$r(w) = \frac{1}{\sqrt{w^2 - 1}}\ln(w + \sqrt{w^2 - 1}), \tag{5.8}$$

$$f(w) = wr(w) - 2 - \frac{w}{\sqrt{w^2 - 1}}[L_2(1 - w_-^2) + (w^2 - 1)r^2(w)], \tag{5.9}$$

$$g(w) = \frac{w}{\sqrt{w^2 - 1}}[L_2(1 - zw_-^2) - L_2(1 - zw_+^2)] - \frac{z}{1 - 2wz + z^2}[(w^2 - 1)r(w) + (w - z)\ln z], \tag{5.10}$$

$$L_2(x) = -\int_0^x dt \frac{\ln(1-t)}{t}, \tag{5.11}$$

$$h_2^{V,A}(z,w) = \frac{z}{(1-2wz+z^2)^2}\{2(w \mp 1)z(1 \pm z)\ln z - $$
$$[(w \pm 1) - 2w(2w \pm 1)z + (5w \pm 2w^2 \mp 1)z^2 - 2z^3]r(w)\} - $$
$$\frac{z}{1-2wz+z^2}[\ln z - 1 \pm z], \tag{5.12}$$

$$h_1^{V,A}(z,w) = h_2^{V,A}(z^{-1},w) - 2r(w) \pm 1. \tag{5.13}$$

After plugging this vertex into the expressions for the hadronic tensor, one can read off the corrections to the corresponding hadronic structure functions. The corrected expressions look as follows (nonperturbative corrections are not included):

$$w_1^{VV} = 2\pi(1+2a^V)(m_b - m_c - q_0)\delta(m_b^2 - 2m_b q_0 + q_0^2 - \mathbf{q}^2 - m_c^2), \tag{5.14}$$

$$w_2^{VV} = 4\pi(m_b + 2a^V m_b + b^V(m_b + m_c) + $$
$$c^V(\frac{m_b^2}{m_c} + m_b))\delta(m_b^2 - 2m_b q_0 + q_0^2 - \mathbf{q}^2 - m_c^2), \tag{5.15}$$

$$w_3^{VV} = 0, \tag{5.16}$$

$$w_4^{VV} = 4\pi\frac{c^V}{m_c}\delta(m_b^2 - 2m_b q_0 + q_0^2 - \mathbf{q}^2 - m_c^2), \tag{5.17}$$

$$w_5^{VV} = -2\pi((1+b^V+c^V) + 2(a^V + c^V\frac{m_b}{m_c}))\delta(m_b^2 - 2m_b q_0 + q_0^2 - \mathbf{q}^2 - m_c^2), \tag{5.18}$$

and for the $VA$ currents only $h_3^{VA}$ is not equal to zero,

$$w_3^{VA} = 2\pi(1+a^V+a^A)\delta(m_b^2 - 2m_b q_0 + q_0^2 - \mathbf{q}^2 - m_c^2). \tag{5.19}$$

For the $AA$ currents expressions are obtained from the ones for $VV$ currents by replacing $m_c \to -m_c$.

In the last expressions we can rearrange the argument of the delta function, since in the sum rules we integrate over $\epsilon = M_B - E_{D^*} - q_0$ and only over the decay cut:

$$\delta(m_b^2 - 2m_b q_0 + q_0^2 - \mathbf{q}^2 - m_c^2) \to \frac{1}{2E_c}\delta(m_b - E_c - q_0) = \frac{1}{2E_c}\delta(\epsilon - (M_B - m_b) + (E_{D^*} - E_c)),$$

where $E_i = \sqrt{M_i^2 + \mathbf{q}^2}$. The corrected sum rules are now obtained by multiplying Eq. (5.14)-(5.19) by the appropriate power of $\epsilon$, dividing by $2\pi$ and by $2E_c$, dropping the delta functions and substituting $q_0 \to m_b - E_c$. Let us note that these corrections contain the infrared logarithmic divergencies which are to be cancelled by the corresponding real gluon contributions. Let us call perturbative corrections to the moments including only virtual gluon contributions $S_j^k$. Then for the corrections to the moments we get:

$$S_1^{VV,k} = \frac{a^V}{E_c}(E_c - m_c)((M_B - m_b) - (E_{D^*} - E_c))^k, \tag{5.20}$$

$$S_2^{VV,k} = \frac{1}{E_c}[2a^V m_b + b^V(m_b + m_c) +$$

$$c^V(\frac{m_b^2}{m_c} + m_b)]((M_B - m_b) - (E_{D^*} - E_c))^k, \tag{5.21}$$

$$S_3^{VV,k} = 0, \tag{5.22}$$

$$S_4^{VV,k} = \frac{c^V}{m_c E_c}((M_B - m_b) - (E_{D^*} - E_c))^k, \tag{5.23}$$

$$S_5^{VV,k} = -\frac{1}{2E_c}[(b^V + c^V) + 2(a^V + c^V \frac{m_b}{m_c})]((M_B - m_b) - (E_{D^*} - E_c))^k, \tag{5.24}$$

$$S_3^{VA,k} = \frac{1}{2E_c}(a^V + a^A)((M_B - m_b) - (E_{D^*} - E_c))^k. \tag{5.25}$$

As it was already mentioned, $S_j^{AA,k}$ could be obtained from equations (5.20)-(5.24) by substituting $m_c \to -m_c$.

## 5.3 Real gluons contributions

The corrections to the hadronic tensor coming from the real gluon emission have also been calculated, they could be read off from different works (see for example [17],[26],[32], [20]-[22],[27]). The corresponding expressions are however pretty lengthy and we are not going to use them in our calculation of the corrections to the sum rules. Instead we are going to derive approximate formulas for the moments of the structure functions, utilizing the fact that the energy release in the decay is of the order of $m_b - m_c \gg M_B - M_{\text{dual}} \sim \Lambda_{QCD}$, where $M_{\text{dual}}$ is the hadron invariant mass scale at which the local duality starts. In fact, $M_B - M_{\text{dual}}$ is the size of the resonance domain, in which we are going to expand (speaking more presicely, we are going to expand in $(M_B - M_{\text{dual}})/E_c$).

Let us first define the quantity we are going to calculate. The contribution of real gluons to the hadronic tensor $W_{\mu\nu}^{\text{virt}}$ in the leading order in $1/m_b$ is defined by the following expression, containing $c$ quark and a gluon $g$ in the intermediate state:

$$W_{\mu\nu}^{\text{virt}} = \sum_{X=cg} (2\pi)^4 \delta^4(p_Q - q - p_X) \frac{1}{2m_b} \langle b|j_\mu^\dagger(0)|X\rangle\langle X|j_\nu(0)|b\rangle$$

$$= \frac{1}{2m_b(2\pi)^2} \int \frac{d^3k}{2\omega_k} \frac{d^3p_c}{2E_c} \delta^4(p_b - q - p_c - k) \sum_{\text{pol.}} \langle b|j_\mu^\dagger(0)|cg\rangle\langle cg|j_\nu(0)|b\rangle, \quad (5.26)$$

where $k$ is the gluon momentum and the sum runs over gluon polarizations. The matrix elements $\langle cg|j_\nu(0)|b\rangle$ are given by the diagrams on Fig. 5.1.

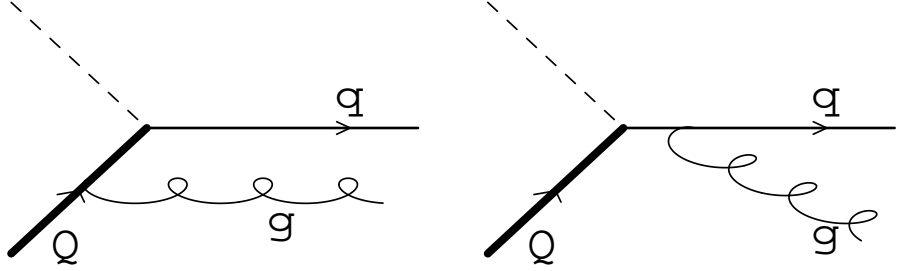

Figure 5.1: Feynman diagrams for the real gluon contributions to the hadronic tensor.

Functions $w_j(\epsilon, \mathbf{q}^2)$ satisfy the following sum rules:

$$\frac{1}{2\pi} \int_{M_B-E_{D^*}}^{\epsilon_{\text{dual}}} w_j^{\text{phen}}(\epsilon, \mathbf{q}^2) \epsilon^k d\epsilon = \frac{1}{2\pi} \int_{(M_B-m_b)-(E_{D^*}-E_c)}^{\epsilon_{\text{dual}}} w_j^{\text{theor}}(\epsilon, \mathbf{q}^2) \epsilon^k d\epsilon.. \quad (5.27)$$

Let us denote

$$\frac{1}{2\pi} \int_{(M_B-m_b)-(E_{D^*}-E_c)}^{\epsilon_{\text{dual}}} w_j^{\text{theor}}(\epsilon, \mathbf{q}^2) \epsilon^k d\epsilon. = I_j^k + I_j^{k,\text{virt}} + I_j^{k,\text{real}}, \quad (5.28)$$

where $I_j^n$ are the leading order in $\alpha_s$ moments,

$$I_j^{k,\text{virt}} = \frac{1}{2\pi} \int_{(M_B-m_b)-(E_{D^*}-E_c)+\lambda}^{\epsilon_{\text{dual}}} d\epsilon\, \epsilon^k\, w_j^{\text{virt}},$$

are the contributions of virtual gluons, and

$$I_j^{k,\text{real}} = \frac{1}{2\pi} \int_{(M_B-m_b)-(E_{D^*}-E_c)+\lambda}^{\epsilon_{\text{dual}}} d\epsilon\, \epsilon^k\, w_j^{\text{real}},$$

are the contributions of the real gluons, $E_c = \sqrt{m_c^2 + \mathbf{q}^2}$, $\epsilon_{\text{dual}} = \sqrt{M_{\text{dual}}^2 + \mathbf{q}^2} - \sqrt{M_{D^*}^2 + \mathbf{q}^2}$ is the excitation energy where the continuum spectrum of hadrons starts. The lower limits of integrations correspond to $0 \ll q_0 \ll m_b - E_c - \lambda$, with $\lambda$ being the gluon mass needed for regularization of the infrared divergency.

We now need to calculate the right hand sides of Eq. (5.27). To be able to perform the calculation we will make one more approximation. In fact in Eq. (5.27) the integration goes over the resonance domain of the final hadron energies. We will assume that this resonance domain is narrow compared to the maximal energy release in the process. Then $\epsilon_{\text{dual}}$ can be considered as a small parameter along with all gluon energies in the integration domain and we can expand in it. (Note, that in fact we need to calculate moments of different structure functions, and not the functions themselves. This allows us to change the order of integration and first expand and integrate in $\epsilon$, which makes the calculation easier to do). Let us denote

$$R_j^{JJ,k} = \frac{1}{2\pi} \int_\lambda^{\epsilon_{\text{dual}}} d\epsilon\, \epsilon^k\, w_j^{JJ,\text{real}}.$$

Let us introduce for convenience the variable $\mathbf{v}' = -\mathbf{q}/E_c$, which is the spatial velocity of the final quark. Then the results of the calculations look as follows.

For $VV$ currents:

$$\begin{aligned}
R_1^{VV,0} &= \frac{\alpha_s}{3\pi m_b E_c^3}\{\epsilon_{\text{dual}}[\,E_c\,(4\,m_b\,E_c - 2\,m_b m_c - 2E_c^2) + \\
&\qquad (m_c^2 + m_b\,(-2E_c + m_c(1+\mathbf{v}'^2)))\frac{E_c}{|\mathbf{v}'|}\ln(\frac{1+|\mathbf{v}'|}{1-|\mathbf{v}'|})] + \\
&\qquad 2m_b E_c^2(E_c - m_c)E_c(-2+\frac{1}{|\mathbf{v}'|})\ln(\frac{1+|\mathbf{v}'|}{1-|\mathbf{v}'|}))\ln(\frac{\epsilon_{\text{dual}}}{\lambda})\}, \qquad (5.29)
\end{aligned}$$

$$R_1^{VV,1} = \frac{\alpha_s \epsilon_{\text{dual}}}{3\pi}\,(1 - \frac{m_c}{E_c})[\,2 - \frac{m_c^2}{|\mathbf{v}'|E_c^2}\ln(\frac{1+|\mathbf{v}'|}{1-|\mathbf{v}'|})\,], \qquad (5.30)$$

$$R_1^{VV,2} = \frac{2\,\alpha_s\,\mathbf{v}'^2\,(E_c - m_c)\,\epsilon_{\text{dual}}^2}{9\pi\,E_c}, \qquad (5.31)$$

$$R_2^{VV,0} = \frac{2\alpha_s}{3\pi E_c^3}\{\frac{\epsilon_{\text{dual}}}{\mathbf{v}'^2 E_c^2}\,[\,(m_b\,E_c\,(4\,m_c^2 + 6\,\mathbf{v}'^2 E_c^2) - 4\,E_c^4) +$$

$$(2\,m_c{}^2 E_c^3 - m_b\,(3\,m_c{}^2\mathbf{v'}^2 E_c^2 + 2\,(m_c{}^4 + \mathbf{v'}^4 E_c^4)))\frac{1}{|\mathbf{v'}|E_c}\ln(\frac{1+|\mathbf{v'}|}{1-|\mathbf{v'}|})\,]+$$

$$2m_b E_c^2\,(-2 + \frac{1}{|\mathbf{v'}|}\ln(\frac{1+|\mathbf{v'}|}{1-|\mathbf{v'}|}))\ln(\frac{\epsilon_{\text{dual}}}{\lambda})\}, \tag{5.32}$$

$$R_2^{VV,1} = \frac{2\,\alpha_s m_b}{3\pi\,E_c^2}\,\epsilon_{\text{dual}}\,[\,2\,E_c - \frac{m_c{}^2}{|\mathbf{v'}|E_c}\ln(\frac{1+|\mathbf{v'}|}{1-|\mathbf{v'}|})\,], \tag{5.33}$$

$$R_2^{VV,2} = \frac{4\,\alpha_s\,m_b\mathbf{v'}^2\,\epsilon_{\text{dual}}^2}{9\pi\,E_c}, \tag{5.34}$$

$$R_4^{VV,0} = \frac{2\,\alpha_s \epsilon_{\text{dual}}}{3\pi\,m_b\mathbf{v'}^2 E_c^2}\,[\,2 - \frac{m_c{}^2}{|\mathbf{v'}|E_c^2}\ln(\frac{1+|\mathbf{v'}|}{1-|\mathbf{v'}|})\,], \tag{5.35}$$

$$R_5^{VV,0} = \frac{\alpha_s}{3\pi m_b E_c^3}\{\frac{\epsilon_{\text{dual}}}{\mathbf{v'}^2 E_c^2}\,[\,2E_c(-m_b(3\,m_c{}^2 + 4\mathbf{v'}^2 E_c^2) + E_c{}^3)+$$

$$(-m_c^2\,E_c^3 + m_b\,(2\,\mathbf{v'}^4 E_c^4 + m_c{}^2\,(3\,m_c{}^2 + 4\,\mathbf{v'}^2 E_c^2)\,))\frac{1}{|\mathbf{v'}|E_c}\ln(\frac{1+|\mathbf{v'}|}{1-|\mathbf{v'}|})\,]+$$

$$2m_b E_c^2\,(2 - \frac{1}{|\mathbf{v'}|}\ln(\frac{1+|\mathbf{v'}|}{1-|\mathbf{v'}|}))\ln(\frac{\epsilon_{\text{dual}}}{\lambda})\}, \tag{5.36}$$

$$R_5^{VV,1} = \frac{\alpha_s\,\epsilon_{\text{dual}}}{3\pi\,E_c^2}\,[-2\,E_c + \frac{m_c{}^2}{|\mathbf{v'}|E_c}\ln(\frac{1+|\mathbf{v'}|}{1-|\mathbf{v'}|})\,], \tag{5.37}$$

$$R_5^{VV,2} = \frac{-2\alpha_s\mathbf{v'}^2\,\epsilon_{\text{dual}}^2}{9\pi\,E_c}. \tag{5.38}$$

For $VA$ currents:

$$R_3^{VA,0} = \frac{\alpha_s}{3\pi m_b\,E_c^3}\{\frac{\epsilon_{\text{dual}}}{\mathbf{v'}^2 E_c^2}\,[\,2E_c(m_b\,(\,m_c{}^2 + 2\,\mathbf{v'}^2 E_c^2) - E_c{}^3) +$$

$$(m_c{}^2 E_c^3 - m_b\,(m_c{}^4 + 2\,\mathbf{v'}^2 E_c^4\,))\frac{1}{|\mathbf{v'}|E_c}\ln(\frac{1+|\mathbf{v'}|}{1-|\mathbf{v'}|})\,]+$$

$$2m_b\,E_c^2\,(-2 + \frac{1}{|\mathbf{v'}|}\ln(\frac{1+|\mathbf{v'}|}{1-|\mathbf{v'}|})\,)\ln(\frac{\epsilon_{\text{dual}}}{\lambda})\}, \tag{5.39}$$

$$R_3^{VA,1} = \frac{\alpha_s\,\epsilon_{\text{dual}}}{3\pi E_c^2}\,[\,2\,E_c - \frac{m_c{}^2}{|\mathbf{v'}|E_c}\ln(\frac{1+|\mathbf{v'}|}{1-|\mathbf{v'}|})\,], \tag{5.40}$$

$$R_3^{VA,2} = \frac{2\,\alpha_s \mathbf{v}'^2 \epsilon_{\text{dual}}^2}{9\pi\, E_c}. \tag{5.41}$$

For $AA$ currents expressions are obtained from the ones for $VV$ currents by replacing $m_c \to -m_c$.

In these calculations we did not retain terms of the order $(\alpha_s/\pi) \times$ *(nonperturbative corrections)* to the corresponding moments, because these terms have extra $\alpha_s/\pi$ and are expected to be suppressed in respect to the nonperturbative corrections. In other words, here we are not considering perturbative corrections to nonperturbative corrections.

Note, that first and second moments of different structure functions are simply related to each other, which again is a reflection of the heavy quark symmetry:

$$(E_c - m_c)R_1^1 = 2m_b R_2^1 = -R_5^1 = R_3^1, \tag{5.42}$$

and

$$(E_c - m_c)R_1^2 = 2m_b R_2^2 = -R_5^2 = R_3^2. \tag{5.43}$$

Terms violating the heavy quark symmetry show up only in zeroth moments.

In this section we used notation slightly different from [41] and from the previous section, but the relation between the variables is clear: $w = E_c/m_c = 1/\sqrt{1-\mathbf{v}'^2}$, $\sqrt{w^2-1} = |\mathbf{v}'|/\sqrt{1-\mathbf{v}'^2}$. It is easy to see that the calculated expressions are not singular as $\mathbf{v}' \to 0$.

An explicit check shows that the infrared divergency is cancelled. To do this it is sufficient to see that in the virtual gluons part the only coefficient that has infrared divergent terms is $a^{V,A}$:

$$\begin{aligned} a^{V,A} &= \frac{\alpha_s}{\pi} \frac{\gamma_0^{hh}(w)}{4} \ln\frac{m_c}{\lambda} + ... = -\frac{4}{3}\frac{\alpha_s}{\pi}[\frac{w}{\sqrt{w^2-1}}\ln(w+\sqrt{w^2-1}) - 1]\ln\frac{m_c}{\lambda} + ... \\ &= -\frac{4}{3}\frac{\alpha_s}{\pi}[\frac{m_c}{2|\mathbf{v}'|E_c}\ln\frac{w+\sqrt{w^2-1}}{w-\sqrt{w^2-1}} - 1]\ln\frac{m_c}{\lambda} + ... \\ &= -\frac{2}{3}\frac{\alpha_s}{\pi}[-2 + \frac{1}{|\mathbf{v}'|}\ln\frac{1+|\mathbf{v}'|}{1-|\mathbf{v}'|}]\ln\frac{m_c}{\lambda} + ... \end{aligned} \tag{5.44}$$

where we only kept logarithmically divergent terms.

# Chapter 6

# The QCD constrained model of semileptonic decays of the $B$ mesons in the heavy quark limit

## 6.1 Introduction. Description of the model

In this chapter we introduce a model of semileptonic decays of $B$ meson into charmed final states $B \to l\bar{\nu}X_c$, based on QCD. In this model we will use experimental information and heavy quark symmetry along with the constituent quark model spectroscopy in order to identify possible final states and to parametrize the structure functions of the decay. We use these states to saturate the phenomenological part of the OPE sum rules (4.9) and then use the sum rules to fix the hadronic structure functions.

Let us start with the description of what is known about the final hadronic states of the decay. The lowest energy final hadronic states seen in experiments with $B$ mesons are (all $D$ mesons electric charges are zeros, like in decays $B^- \to X_c^0 l\bar{\nu})$ : $D(1870)$, $D^*(2007)$, $D_1(2420)$ and $D_2^*(2460)$ [45]. In this chapter we use the notion of the heavy quark spin symmetry [29],[30] and consider final states as doublets of that symmetry. Therefore in our model we will consider $B$ meson decaying into leptons and hadronic resonances: pseudoscalar $D$ and vector $D^*$, axial-vector and tensor $D_1$, $D_2^*$, scalar and axial-vector $D^{**}$, $D_1^{**}$ and pseudoscalar and vector $D^{***}$, $D_1^{***}$. In the constituent quark models the last doublet represents the first radial excitation and for some quark-antiquark potentials is degenerate with the third one. Anything lying above will be considered continuum states. We consider the members of the heavy quark doublets to have degenerate masses. We hope that this picture provides a good approximation for the final states of the decay and that these final states will saturate the OPE sum rules. In the present work we will limit the discrete states by the final states listed above, however one can of course add more final states such as the first radial excitations, etc. We note, however that the structure functions for the case of three heavy doublets in the final state are completely fixed by the first three OPE sum rules and therefore could be considered most suitable as the model with the next to leading order accuracy. The results obtained in the present section will be used as zero approximation for the model including next to leading order corrections to the structure functions (see Section 7).

The resonances listed above could be viewed in the following way in terms of the constituent quark model [29],[30]. Since the $c$ quark is heavy, in the first approximation we can neglect the interaction of its spin with the light degrees of freedom. Then we can classify the states of the light subsystem as $1^1S_{1/2}$, $1^1P_{1/2}$, $1^1P_{3/2}$ and $2^1S_{1/2}$, where the usual spectroscopic notation $n^{2s+1}l_j$ was employed: $n$ is the radial quantum number, $s$ is the spin of the light quark, $l$ is the orbital momentum of the light quark and $j$ is the total momentum of light degrees of freedom. The last state is the first radial excitation. Now we add the spin of the $c$ quark. Each of the listed states gets doubled thus forming the heavy quark doublets and we get the following classification [30] (now in the $n^{2S+1}L_J$ notation, $S$ is the combined spin of the two quarks inside the meson, $L$ is the orbital momentum of the light quark, and $J$ is the spin of the meson or total angular momentum of the $b\bar{q}$ system):

pseudoscalar $D \to |1^1S_0\rangle$ and vector $D^* \to |1^3S_1\rangle$;

axial-vector $D_1 \to \sqrt{\frac{3}{2}}|1^1P_1\rangle + \sqrt{\frac{1}{2}}|1^3P_1\rangle$ and tensor $D_2^* \to |1^3P_2\rangle$ ;

scalar $D^{**} \to |1^3P_0\rangle$ and axial-vector $D_1^{**} \to \sqrt{\frac{3}{2}}|1^1P_1\rangle - \sqrt{\frac{1}{2}}|1^3P_1\rangle$;

pseudoscalar $D^{***} \to |2^1S_0\rangle$ and vector $D_1^{***} \to |2^3S_1\rangle$,

The notation for the last two doublets is in no way standard, we use it here in order to merely have notation for these particles.

Since, as it was already mentioned, the last two doublets are usually degenerate in quark potential models, we will consider them as one and the same doublet number three. Therefore, contribution of the last two doublets in the heavy quark symmetry limit will always look like a contribution of a combined doublet.

As a consequence of the heavy quark mass being large one also can expect that the mass splitting inside heavy quark doublets (which is in fact the hyperfine splitting in the quark model of hadrons) is smaller then that between different heavy quark doublets (the latter is due to the $LS$ coupling in this model).

In the limit of the exact heavy quark symmetry the hadronic tensor is described by just one structure function $w(q_0, q^2)$ and has the following form (same as for the free quark decay)

$$W_{\mu\nu} = 4\pi w(q_0, q^2) \left(-(M_B - q_0) g_{\mu\nu} + 2 M_B v_\mu v_\nu + i\epsilon_{\mu\nu\alpha\beta} v^\alpha q^\beta - (v_\mu q_\nu + v_\nu q_\mu)\right), \quad (6.1)$$

where we stick to the convention $\epsilon^{0123} = -\epsilon_{0123} = 1$. Note that $w_4$ (coefficient in front of the structure $q_\mu q_\nu$) is zero in this approximation. Therefore in this limit it is sufficient to write the sum rules for the structure function $w_1$

$$w_1 = 4\pi (M_B - q_0) w(q_0, q^2).$$

Let us note that for a free quark decay $w(q_0, q^2) = 1 \cdot \delta(M_B^2 - 2M_B q_0 + q_0^2 - \mathbf{q}^2 - M_i^2)$.

As it was described above, in the sum rules we fix $\mathbf{q}^2$ and calculate the following moments for the $V - A$ current:

$$I_{0,1}^k(\mathbf{q}^2) = \frac{1}{2\pi} \int_0^{\epsilon_{max}} \epsilon^k [w_1(q_0, \mathbf{q}^2)]_0 \, d\epsilon, \quad (6.2)$$

where subscript 0 denotes the leading order in $1/m_Q$ expansion and

$$\epsilon = M_B - q_0 - E_{D^*} = E_X - E_{D^*} = \sqrt{M_X^2 + \mathbf{q}^2} - \sqrt{M_{D^*}^2 + \mathbf{q}^2}, \quad (6.3)$$

$$\epsilon_{max} = M_B - E_D^*. \quad (6.4)$$

The variable $\epsilon$ is just the energy of the final hadron counted from the energy of the $D^*$. According to the sum rules, these moments of the phenomenological structure functions should be equated to the corresponding moments of the theoretically calculated function $w_1^{\text{theor}}$. The number of moments which we are able to calculate is limited by the accuracy of the operator product expansion (see discussion in [10] and in Section 4.1).

If one tries to approximate the final hadronic state by a sum over heavy quark doublets and neglects the contribution of continuum states, then for $w(q_0, \mathbf{q}^2)$ one can write:

$$\begin{aligned} w_1(q_0, \mathbf{q}^2) &= 4\pi \sum_i (M_B - q_0) w^i(\mathbf{q}^2) \delta(M_B^2 - 2M_B q_0 + q_0^2 - \mathbf{q}^2 - M_i^2) \\ &= 2\pi \sum_i w^i(\mathbf{q}^2) \delta(M_B - E_i - q_0), \end{aligned} \quad (6.5)$$

where $M_i$ is mass and $E_i = \sqrt{M_i^2 + \mathbf{q}^2}$ is energy of $i$'s heavy quark doublet and the sum runs over the heavy doublets. Note that the second $\delta$-function $\delta(M_B + E_i - q_0)$ was dropped, since it does not contribute to the sum rules, as it was explained in Section 4.1.

Let us say that both nonperturbative and radiative corrections violate the heavy quark symmetry, for them the equation (6.1) does not hold. We will not consider them in this section but will take them into account later as small perturbations of the sum rules.

The function $w$ satisfies the following leading order sum rules:

$$\sum_i w^i = 1, \quad (6.6)$$

$$\sum_i (E_i - E_{D^*}) w^i = \bar{\Lambda} \left(1 - \frac{m_c}{E_c}\right), \quad (6.7)$$

$$\sum_i (E_i - E_{D^*})^2 w^i = \bar{\Lambda}^2 \left(1 - \frac{m_c}{E_c}\right)^2 + \frac{\mu_\pi^2}{3} \left(1 - \frac{m_c^2}{E_c^2}\right) \quad (6.8)$$

As we can see, these equations express phenomenological structure functions through quark masses, $\bar{\Lambda}$ and $\mu_\pi^2$. If we could solve these equations for $w$'s and express $w$'s through $\bar{\Lambda}$ and $\mu_\pi^2$, then we could fit $\bar{\Lambda}$ and $\mu_\pi^2$ by comparing decay distributions (2.11) with data.

For the differential distribution in variables $E_e$, $q^2$, $q_0$ we have for the case at hand:

$$\frac{d^3\Gamma}{dE_e\, dq^2 dq_0} = |V_{cb}|^2 \frac{G_F^2}{2\pi^3} w\, (q_0 - E_e)(2 E_e M_B - q^2). \quad (6.9)$$

Now integrating over $q_0$ we obtain the expression for the double distribution in $q^2$ and $E_e$:

$$\frac{d^2\Gamma}{dE_e\, dq^2} = |V_{cb}|^2 \frac{G_F^2}{4\pi^3} (2 E_e\, M_B - q^2) \sum_i \frac{w^i(\mathbf{q}^2)}{M_B} (M_B - E_i - E_e). \quad (6.10)$$

In the last equation $\mathbf{q}^2$ is not independent:

$$\mathbf{q}^2 = q_0^2 - q^2 = \frac{M_B^4 + M_i^4 + q^4 - 2M_B M_i - 2M_B q_0 - 2M_i q_0}{4 M_B^2}, \qquad (6.11)$$

and

$$E_i = \frac{M_B^2 + M_i^2 - q^2}{2 M_B}. \qquad (6.12)$$

For each resonance the variable $q^2$ changes between the following limits:

$$0 < q^2 < \frac{2E_e(M_B^2 - 2M_B E_e - M_i^2)}{M_B - 2E_e}. \qquad (6.13)$$

The distribution in electron energy is now obtained by integrating the double distribution over $q^2$:

$$\frac{d\Gamma}{dE_e} = \int_0^{\frac{2E_e(M_B^2 - 2M_B E_e - M_D^2)}{M_B - 2E_e}} \frac{d^2\Gamma}{dE_e\, dq^2}.$$

For each resonance in the final state the electron energy lies within the limits

$$0 < E_e < \frac{M_B}{2}(1 - \frac{M_i^2}{M_B^2}). \qquad (6.14)$$

The general case including perturbative and nonperturbative corrections will be considered in Section 7.

We are now going to consider two simple models for the function $w(q_0, \mathbf{q}^2)$, namely models with two and three heavy hadronic doublets in the final state of the decay. For these models equations (6.6)-(6.8) could be solved and solutions could be used for measuring $\mu_\pi^2$.

## 6.2 Two doublets model

Let us now try to satisfy these sum rules in the following simple way: assume that there exist only two heavy quark doublets with masses $M_{D^*}$ and $M_{D_2^*}$.

Let us denote for convenience:

$$\Delta = M_{D_2^*} - M_{D^*}, \qquad (6.15)$$

$$\tilde{\Delta}^2 = M_{D_2^*}^2 - M_{D^*}^2. \qquad (6.16)$$

Note, that in this chapter we do not distinguish between masses of particles within the same heavy quark doublet.

Now the sum rules (6.6)-(6.8) take the form:

$$w^1 + w^2 = 1 \tag{6.17}$$

$$w^2 \, \tilde{\Delta}^2 = \bar{\Lambda}(1 - \frac{m_c}{E_c})(E_{D_2^*} + E_{D^*}), \tag{6.18}$$

$$w^2 \, (\tilde{\Delta}^2)^2 = (\bar{\Lambda}^2 (1 - \frac{m_c}{E_c})^2 + \frac{\mu_\pi^2}{3}(1 - \frac{m_c^2}{E_c^2}))(E_{D_2^*} + E_{D^*})^2. \tag{6.19}$$

We see that the system of equations is overdetermined and there is no way to satisfy these sum rules at arbitrary $\mathbf{q}^2$.

However let us expand the RHS in

$$\mathbf{v}^2 = \frac{\mathbf{q}^2}{E_c^2} \approx \frac{\mathbf{q}^2}{E_{D^*}^2}, \quad and \quad \mathbf{v}_i^2 = \frac{\mathbf{q}^2}{E_i^2}, \tag{6.20}$$

where

$$E_c^2 = m_c^2 + \mathbf{q}^2. \tag{6.21}$$

Then for small velocities of the particles in the final states we can cast the sum rules into the form:

$$w^1 + w^2 = 1 \tag{6.22}$$

$$w^2 \, \tilde{\Delta}^2 = \bar{\Lambda}\frac{\mathbf{q}^2}{2\, E_{D^*}^2}(M_{D_2^*} + M_{D^*}), \tag{6.23}$$

$$w^2 \, \tilde{\Delta}^4 = \frac{\mu_\pi^2}{3} \frac{\mathbf{q}^2}{E_{D^*}^2}(M_{D_2^*} + M_{D^*})^2, \tag{6.24}$$

The last three equations can be solved:

$$\mu_\pi^2 = \frac{3}{2}\bar{\Lambda}\Delta, \tag{6.25}$$

$$w^2 = \frac{\bar{\Lambda}}{2\, \Delta}\frac{\mathbf{q}^2}{E_{D^*}^2}, \tag{6.26}$$

$$w^1 = 1 - \frac{\bar{\Lambda}}{2\, \Delta}\frac{\mathbf{q}^2}{E_{D^*}^2}. \tag{6.27}$$

The Isgur-Wise $\rho$ parameter is defined as the slope of the universal Isgur-Wise formfactor [30], which is related to the slope of $w^1(\mathbf{q}^2)$: $w^1 = 1 - (\rho - \frac{1}{4})(\mathbf{v}')^2$, where $\mathbf{v}' = \mathbf{q}/\sqrt{M_{D^*}^2 + \mathbf{q}^2}$. From the equation (6.27) we see that $\rho^2$ is related to $\Lambda$ and $\Delta$ in a very simple way:

$$\rho^2 = \frac{\bar{\Lambda}}{2\, \Delta} + \frac{1}{4}. \tag{6.28}$$

Finally,
$$w^1 = 1 - (\rho^2 - \frac{1}{4})\frac{\mathbf{q}^2}{M_{D^*}^2 + \mathbf{q}^2}. \tag{6.29}$$

$$w^2 = (\rho^2 - \frac{1}{4})\frac{\mathbf{q}^2}{M_{D^*}^2 + \mathbf{q}^2}. \tag{6.30}$$

Now, with $\bar{\Lambda} = 0.55\text{GeV}$, $M_{D^*} = 2.009\text{GeV}$, $M_{D_2^*} = 2.462\text{GeV}$ we obtain:

$$\mu_\pi^2 = \frac{3}{2}\bar{\Lambda}(M_{D_2^*} - M_{D^*}) \simeq 0.37\text{GeV}, \tag{6.31}$$

$$\rho^2 = \frac{\bar{\Lambda}}{2(M_{D_2^*} - M_{D^*})} \simeq 0.86, \tag{6.32}$$

which is in agreement with the other estimates and experimental data.

Summarizing, we can say that the first (Bjorken) sum rule along with the second (Voloshin) and the third (Bigi-Uraltsev-Vainshtein-Shifman) sum rules relate $\bar{\Lambda}$ and the mass difference $\Delta$ with $\mu_\pi^2$ and the hadronic structure function $w$. Now using Eq. (6.10)-(6.14) one can fit $\mu_\pi^2$ in $d\Gamma/dE_e$ from experimental data on the decay.

## 6.3 Three doublets model

It was shown in the previous secsion that in the case of two doublets in the final state the sum rules equations could not be solved exactly at any $\mathbf{q}^2$. In fact, we did not have enough degrees of freedom to satisfy the sum rules. However if one adds the third doublet to the model then the sum rules could be solved exactly. In this section we solve the equations for the structure functions in the case of three heavy quark doublets in the final state. In fact, if we consider masses of the third and fourth doublets degenerate (which happens in potential quark models, as it was mentioned in Section 6.1), the structure function $w^3$ could be considered as a sum of contributions of those two doublets. In this sense the proposed model in the heavy quark limit takes care of the fourth doblet. But since it is described by just tree independend functions, we will still call it three doublets model.

Let us first introduce the notation. We will call the three doublets $d_1 = \{D, D^*\}$, $d_2 = \{D_2^*, D_1\}$, and $d_3 = \{D^{**}, D_1^{**}\}$ and use the following definitions:

$$\begin{aligned}\Delta_2 &= E_{d_2} - E_{d_1} = \sqrt{M_{d_2}^2 + \mathbf{q}^2} - \sqrt{M_{d_1}^2 + \mathbf{q}^2}, \\ \Delta_3 &= E_{d_3} - E_{d_1} = \sqrt{M_{d_3}^2 + \mathbf{q}^2} - \sqrt{M_{d_1}^2 + \mathbf{q}^2}.\end{aligned} \tag{6.33}$$

In this model in the heavy quark limit we use only three doublets, bearing in mind that there is a fourth one, which is degenerate with the third, then we do not separate their contributions and use generic notation $d_3$ for both of them.

Now the sum rules for the three particles take the following form.

$$\begin{aligned}
w^1 + w^2 + w^3 &= 1, \\
w^2 \Delta_2 + w^3 \Delta_3 &= \bar{\Lambda}\,(1 - \frac{m_c}{E_c}), \\
w^2 \Delta_2^2 + w^3 \Delta_3^2 &= \bar{\Lambda}^2\,(1 - \frac{m_c}{E_c})^2 + \frac{\mu_\pi^2}{3}\,(1 - \frac{m_c^2}{E_c^2}).
\end{aligned} \quad (6.34)$$

The last three equations could be solved:

$$\begin{aligned}
w^1 &= 1 - \frac{1}{\Delta_2\,\Delta_3}\,(\bar{\Lambda}\,(\Delta_3 + \Delta_2) - \bar{\Lambda}^2\,(1 - \frac{m_c}{E_c}) - \frac{\mu_\pi^2}{3}(1 + \frac{m_c}{E_c}))(1 - \frac{m_c}{E_c}), \\
w^2 &= \frac{1}{\Delta_2\,(\Delta_3 - \Delta_2)}(\Delta_3\bar{\Lambda} - \bar{\Lambda}^2\,(1 - \frac{m_c}{E_c}) - \frac{\mu_\pi^2}{3}(1 + \frac{m_c}{E_c}))\,(1 - \frac{m_c}{E_c}), \\
w^3 &= \frac{1}{\Delta_3\,(\Delta_3 - \Delta_2)}(\bar{\Lambda}^2\,(1 - \frac{m_c}{E_c}) - \Delta_2\bar{\Lambda} + \frac{\mu_\pi^2}{3}(1 + \frac{m_c}{E_c}))\,(1 - \frac{m_c}{E_c}).
\end{aligned} \quad (6.35)$$

The three functions $w^1$, $w^2$, $w^3$ must be positive, which leads to the following inequality:

$$\Delta_2\bar{\Lambda} \leq \bar{\Lambda}^2\,(1 - \frac{m_c}{E_c}) + \frac{\mu_\pi^2}{3}(1 + \frac{m_c}{E_c}) \leq \Delta_3\bar{\Lambda}. \quad (6.36)$$

Of course, the last inequality is specific for the three resonance doublets case of our model, but it still gives some physical limitations on the quantities involved. Let us note also that in the limit $\mathbf{q}^2 \to 0$ it looks as follows:

$$\Delta m_2 \bar{\Lambda} \leq \frac{2}{3}\mu_\pi^2 \leq \Delta m_3 \bar{\Lambda},$$

which, in turn, could be viewed as limitations on the values of $\mu_\pi^2$, $\Delta m_3$ and $\bar{\Lambda}$. For example, if one takes $\Delta m_2 \simeq 0.6\text{GeV}$, $\bar{\Lambda} \simeq 0.5\text{GeV}$, then one gets:

$$\begin{aligned}
0.45\text{GeV}^2 &\leq \mu_\pi^2, \\
0.6\text{GeV} &\leq \Delta m_3,
\end{aligned}$$

which is in accordance with the estimates for $\mu_\pi^2$ of the works [10],[9],[48],[53] (in these works the inequality $\mu_\pi^2 \leq \mu_G^2$ was proven; while $\mu_G^2 \simeq 0.36\text{GeV}^2$). In fact, however, these inequalities do not take into account perturbative corrections to the sum rules, which could somewhat alter them.

Expanding $w^1$ at small $\mathbf{v}^2$ we find

$$w^1 = 1 - \mathbf{v}_c^2 \left( \frac{\bar{\Lambda}(\Delta_3 + \Delta_2)}{2\,\Delta_3\Delta_2} - \frac{\mu_\pi^2}{3\,\Delta_3\Delta_2} \right), \tag{6.37}$$

where $\mathbf{v}_c^2 = \mathbf{q}^2/(m_c^2 + \mathbf{q}^2)$. From the last expression we can see that the Isgur-Wise $\rho$-parameter in this case is

$$\rho^2 - \frac{1}{4} = \frac{m_c^2}{M_{D^*}^2} \left( \frac{\bar{\Lambda}(\Delta m_3 + \Delta m_2)}{2\,\Delta m_3 \Delta m_2} - \frac{\mu_\pi^2}{3\,\Delta m_3 \Delta m_2} \right), \tag{6.38}$$

where $\Delta m_2 = M_{d_2} - M_{d_1}$ and $\Delta m_3 = M_{d_3} - M_{d_1}$ and we have taken into account the fact that the $\rho$-parameter is defined for hadrons and not for quarks.

For the differential distributions one can again use formulas (6.9)-(6.14).

# Chapter 7

# Model of the decay including perturbative and nonperturbative corrections to the sum rules

## 7.1 Description of the model.

In the previous chapters we have considered the model of semileptonic decays of heavy flavors in the heavy quark limit. We considered the final states resonances to be heavy quark doublets and also neglected terms breaking the heavy quark symmetry in the sum rules. In this chapter we are going to take into account the fact that the resonance doublets masses are split and that the hadronic tensor and, therefore, the sum rules, contain terms violating the heavy quark symmetry. We are going to use the previous chapters results as a zero approximation model and consider perturbative and nonperturbative corrections as small perturbations to it.

Let us consider a model with $n$ heavy quark doublets. To accommodate deviations from the heavy quark symmetry we now need to consider mass splitting within the heavy quark doublets as well as independence of different structure functions in the hadronic tensor (they were not independent in the heavy quark limit, see Eq. (6.1)).

Now we have $2n$ resonances with different masses, we will consider mass splittings of the doublets to be of order $1/m_Q^2$.

For the masses of the first four resonances experimental data is available, and therefore we do not need to consider them as parameters to be fixed from the sum rules [45]. For higher resonances, however, no data exists, but as numerical analysis of the model will show, they are not going to play a major role for their contribution is going to be small anyway.

The hadronic tensor now has all structure functions independent from each other. Three of these functions contribute to semileptonic decays. The operator product expansion was calculated only with the accuracy $1/m_Q^2$. This limits the order of the sum rules we can use to restrict the structure functions by three. It is valid to include $1/m_Q$ corrections to the first two sum rules while in order to calculate them for the third one we would have to know the higher order terms of the OPE.

First let us introduce the notation. Since now we are considering deviations from the heavy quark symmetry, we assign each resonance its own distinctive mass $M_i$, where for odd $i$'s $M_i, M_{i+1}$ come from the same heavy quark doublet. Let us also denote $w_j^i, w_j^{i+1}$ the corresponding structure functions, where $j$ stands for the structure function number and $i, i+1$ denote the functions coming from the same doublet for odd $i$'s. Note that functions $w_j^i$ refer to the separate resonances within the heavy quark doublets, while the previously defined functions $w^i$ which did not have subscripts referred to the contributions of the whole doublets. Let us also denote the leading order moments as $I_{0,j}^k$. In this section we only consider $V - A$ current.

Our model is now formulated as follows:

$$\begin{aligned} w_j^{\text{phen}}(q_0, \mathbf{q}^2) &= 2\pi \sum_i w_j^i(\mathbf{q}^2)\, \delta(M_B - \sqrt{M_i^2 + \mathbf{q}^2} - q_0) + \\ &\quad w_j^{\text{cont}}(q_0, \mathbf{q}^2)\, \theta(M_B - \sqrt{M_{\text{cont}}^2 + \mathbf{q}^2} - q_0), \end{aligned} \quad (7.1)$$

where $M_{\text{cont}}$ is the hadronic invariant mass at which the continuum spectrum starts.

In our model we will assume that $M_{\text{cont}} = M_{\text{dual}}$, other words, we assume that local duality starts at the same scale as continuum spectrum. Then in the LHS of the sum rules only resonance contributions should be counted.

The first three sum rules now read:

$$\sum_{i=1,3,5,\ldots} (w_j^i + w_j^{i+1}) = \tilde{I}_j^0, \qquad (7.2)$$

$$\sum_{i=1,3,5,\ldots} w_j^i(E_i - E_{D^*}) + w_j^{i+1}(E_{i+1} - E_{D^*}) = \tilde{I}_j^1, \qquad (7.3)$$

$$\sum_{i=1,3,5,\ldots} w_j^i(E_i - E_{D^*})^2 + w_j^{i+1}(E_{i+1} - E_{D^*})^2 = \tilde{I}_j^2. \qquad (7.4)$$

In the last equations $\tilde{I}_j^k$ are the moments including all perturbative and nonperturbative corrections found in previous sections.

We are interested in the lowest order in $1/m_Q$ corrections to the sum rules. Let us expand energies of the final particles in heavy mass splittings:

$$E_i \approx E_{i+1} - \frac{M_{i+1}^2 - M_i^2}{2E_{i+1}} = E_{i+1} - \frac{\delta M_i^2}{2E_{i+1}} = E_{i+1} - \frac{M_{i+1}\delta M_i}{E_{i+1}}. \qquad (7.5)$$

Then we have for the sum rules:

$$\sum_{i=1,3,5,\ldots} (w_j^i + w_j^{i+1})(E_{i+1} - E_{D^*})^k - w_j^i \frac{kM_{i+1}\delta M_i}{E_{i+1}}(E_{i+1} - E_{D^*})^{k-1} = \tilde{I}_j^k. \qquad (7.6)$$

For $w_j^i + w_j^{i+1}$ we can write

$$w_j^i + w_j^{i+1} = (w_j^i + w_j^{i+1})_0 + \delta w_j^i + \delta w_j^{i+1}, \qquad (7.7)$$

where $(w_j^i)_0$ are the leading order structure functions which are determined from the sum rules in the heavy quark limit (6.6)-(6.8), and $\delta w_j^i$ are corresponding corrections. In fact, by comparing (7.1) and (6.5) one finds:

$$w^i = (w_1^i + w_1^{i+1})_0, \ \ i = 1, 3, 5.$$

Let us take into account the fact that $\delta M_i \sim \Lambda_{QCD}^2/m_Q$ and therefore $w_j^{i+1}$ in the second term of Eq. (7.6) could be taken in the leading approximation. By the virtue of the heavy quark symmetry both $(w_j^i)_0$ and $(w_j^{i+1})_0$ could be expressed through the universal Isgur-Wise formfactors [30]. The explicit expressions for these functions are given in the appendix C.

Since it is only the sum of corrections for the members of a heavy quark doublet which enters the sum rules and the expressions for lepton energy distributions, we will introduce a special notation for it:

$$\delta W_j^i = \delta w_j^i + \delta w_j^{i+1}, \quad i = 1, 3, 5.$$

Finally corrections to the sum rules can be written in the form:

$$\sum_{i=1,3,5,\ldots} \delta W_j^i (E_{i+1} - E_{D^*})^k = \tilde{I}_j^k - I_{0,j}^k + \sum_{i=1,3,5,\ldots} (w_j^i)_0 \frac{k M_{i+1} \delta M_i}{E_{i+1}} (E_{i+1} - E_{D^*})^{k-1}, \quad (7.8)$$

where $I_{0,j}^k$ are defined in Eq. (6.2).

This system of equations relates corrections to the structure functions to the mass splittings inside the heavy doublets. For the model to be self consistent in the zeroth approximation we take resonance doublets masses equal to the masses of heavier particles of the doublets. This does not matter in the zeroth approximation itself (heavy quark symmetry limit) but does matter when we consider nonperturbative corrections to the sum rules coming from the mass splittings within the doublets. Again, for the first two doublets the mass splittings are known experimentally [45].

We see that the OPE sum rules provide the system of equations for the functions $\delta W_j^i$, which is determined to the same degree as the system of equations for the functions $(w_j^i)_0$. In fact for $V - A$ current we have three systems of equations, independent for each structure function number.

## 7.2 Three doublets model

For three heavy doublets in the final state the system of equations (7.8) can be easily solved, for it has the following generic form:

$$\begin{aligned} x + y + z &= a, \\ y \Delta_3 + z \Delta_5 &= b, \\ y \Delta_3^2 + z \Delta_5^2 &= c. \end{aligned}$$

The solution is:

$$x = \frac{c - b(\Delta_3 + \Delta_5) + a \Delta_3 \Delta_5}{\Delta_3 \Delta_5},$$

$$y = \frac{b\Delta_5 - c}{\Delta_3\Delta_5 - \Delta_3^2},$$
$$z = \frac{b\Delta_3 - c}{\Delta_3\Delta_5 - \Delta_5^2}.$$

Let us denote:

$$\Delta_i = E_i - E_{D^*},$$
$$\delta I_j^k = \tilde{I}_j^k - I_{0,j}^k - \sum_{i=1,3,5,\ldots} (w_j^{i+1})_0 \frac{kM_i \delta M_i}{E_i} \Delta_i^{k-1}.$$

Then for each structure function the equations are:

$$\delta W_j^1 + \delta W_j^3 + \delta W_j^5 = \delta I_j^0,$$
$$\delta W_j^3 \Delta_3 + \delta W_j^5 \Delta_5 = \delta I_j^1,$$
$$\delta W_j^3 \Delta_3^2 + \delta W_j^5 \Delta_5^2 = \delta I_j^2,$$

and the solution is given by:

$$\delta W_j^1 = \frac{\delta I_j^2 - \delta I_j^1 (\Delta_3 + \Delta_5) + \delta I_j^0 \Delta_3 \Delta_5}{\Delta_3 \Delta_5},$$
$$\delta W_j^3 = \frac{\delta I_j^1 \Delta_5 - \delta I_j^2}{\Delta_3 \Delta_5 - \Delta_3^2},$$
$$\delta W_j^5 = \frac{\delta I_j^1 \Delta_3 - \delta I_j^2}{\Delta_3 \Delta_5 - \Delta_5^2}. \tag{7.9}$$

## 7.3 Differential distributions for the model including perturbative and nonperturbative corrections

One of possible uses of the proposed model is to describe the lepton energy and invariant mass distribution in the semileptonic decays of heavy mesons. In this section we will show that the information about structure functions derived from the corrected sum rules is sufficient in order to write expressions for the differential distributions. Unlike the pure OPE results, this model predicts smooth distributions, which do not have anything like $\delta$-functions at the ends, and still is totally based on OPE.

The hadronic structure functions of the model are given by the following expression

$$\begin{aligned}
w_j^{phenom} &= 2\pi \sum_i w_j^i(\mathbf{q}^2)\delta(M_B - q_0 - \sqrt{M_i^2 + \mathbf{q}^2}) + \\
&\quad w_j^{\text{cont}}(q_0, \mathbf{q}^2)\theta(M_B^2 - 2M_B q_0 + q^2 - M_{\text{cont}}^2) \\
&= 2\pi \sum_i w_j^i(q^2) 2E_i \delta(M_B^2 - 2M_B q_0 + q^2 - M_i^2) + \\
&\quad w_j^{\text{cont}}(q_0, q^2)\theta(M_B^2 - 2M_B q_0 + q^2 - M_{\text{cont}}^2). \quad (7.10)
\end{aligned}$$

In the last equations we have taken into account the relation

$$\int_{decay\ phase\ space} dq^2 dq_0 f(q_0, q^2) 2E_i \delta(M_B^2 - 2M_B q_0 + q^2 - M_i^2)$$
$$= \int_{decay\ phase\ space} d\mathbf{q}^2 dq_0 f(q_0, q_0^2 - \mathbf{q}^2)\delta(M_B - q_0 - \sqrt{M_i^2 + \mathbf{q}^2}), \quad (7.11)$$

with corresponding limits of integration. Because, it is the double distribution in $E_e, q^2$, that we studied in the previous chapters and experimentalists are going to measure, in what follows we will work in variables $(q_0, q^2)$. The form of the equations in the variables $(q_0, \mathbf{q}^2)$ could be readily restored.

The full differential distribution is given by the Eq. (2.9), which we rewrite here showing explicitly contributions of different states:

$$\begin{aligned}
\frac{d^3\Gamma}{dE_e dq^2 dq_0} &= |V_{cb}|^2 \frac{G_F^2}{32\,\pi^4}\{ \\
&\quad 2\pi \sum_i [2q^2 w_1^i + [4\,E_e(q_0 - E_e) - q^2]\,w_2^i + 2\,q^2(2\,E_e - q_0)\,w_3^i] \times \\
&\quad 2E_i\delta(M_B^2 - 2M_B q_0 + q^2 - M_i^2) + \\
&\quad [2q^2 w_1^{\text{cont}} + [4\,E_e(q_0 - E_e) - q^2]\,w_2^{\text{cont}} + 2\,q^2(2\,E_e - q_0)\,w_3^{\text{cont}}] \times \\
&\quad \theta(M_B^2 - 2M_B q_0 + q^2 - M_{\text{cont}}^2)\}. \quad (7.12)
\end{aligned}$$

The total width is then given by the formulae:

$$\begin{aligned}
\Gamma &= \int_\Phi dE_e dq^2 dq_0 \frac{d^3\Gamma}{dE_e dq^2 dq_0} \\
&= \frac{|V_{cb}|^2 G_F^2}{32\,\pi^4}\{2\pi \sum_i \int_{\Phi_i} dE_e dq^2 \frac{E_i}{M_B}[2q^2 w_1^i + [4\,E_e(M_B - E_i - E_e) - q^2]\,w_2^i + \\
&\quad 2\,q^2(2\,E_e - M_B + E_i)\,w_3^i] + \\
&\quad \int_\Phi d\Phi [2q^2 w_1^{\text{cont}} + [4\,E_e(q_0 - E_e) - q^2]\,w_2^{\text{cont}} + 2\,q^2(2\,E_e - q_0)\,w_3^{\text{cont}}] \times \\
&\quad \theta(M_B^2 - 2M_B q_0 + q^2 - M_{\text{cont}}^2)\}. \quad (7.13)
\end{aligned}$$

where $\Phi, \Phi_i$'s are volumes in the phase space in corresponding variables available to the particles in the final state, and the sum runs over all resonances. Note that the phase spaces available to heavier particles in the final state are smaller then those available to the lighter ones, therefore $\Phi_{i+1} > \Phi_i$, $i = 1, 3, 5$.

Let us examine the resonance contributions to the total width. First let us denote:

$$f^i = \frac{E_i}{M_B}[2q^2 w_1^i + [4 E_e(M_B - E_i - E_e) - q^2] w_2^i + 2 q^2 (2 E_e - M_B + E_i) w_3^i], \qquad (7.14)$$

where

$$E_i = \frac{M_B^2 + M_i^2 - q^2}{2 M_B}.$$

For a contribution of a heavy doublet, up to the first order in the heavy doublet mass splitting, we have:

$$\begin{aligned}
&\int_{\Phi_i} dE_e dq^2 f^i + \int_{\Phi_{i+1}} dE_e dq^2 f^{i+1} \\
&\approx \int_{\Phi_i} dE_e dq^2 (f^i)_0 + \int_{\Phi_i} dE_e dq^2 \delta f^i + \int_{\Phi_{i+1}} dE_e dq^2 (f^{i+1})_0 + \int_{\Phi_{i+1}} dE_e dq^2 \delta f^{i+1} \\
&\approx \int_{\Phi_{i+1}} dE_e dq^2 (f^i)_0 + \int_{\Phi_{i+1}} dE_e dq^2 \delta f^i + \int_{\Phi_{i+1}} dE_e dq^2 (f^{i+1})_0 + \\
&\quad \int_{\Phi_i - \Phi_{i+1}} dE_e dq^2 (f^i)_0 + \int_{\Phi_{i+1}} dE_e dq^2 \delta f^{i+1} \\
&\approx \int_{\Phi_{i+1}} dE_e dq^2 [(f^i)_0 + (f^{i+1})_0 + (\delta f^i + \delta f^{i+1})] + \int_{\Phi_i - \Phi_{i+1}} dE_e dq^2 (f^i)_0, \qquad (7.15)
\end{aligned}$$

where $(f^i)_0$ are taken in the leading order (heavy quark symmetry limit) and $\delta f^i$ are correction to them. From the last equation it follows that there are regions of the decay phase space where only one resonance of a doublet is present. Therefore unlike heavy quark symmetry case, now we need to know not only leading order structure functions describing resonance doublets, but also the leading order contributions of individual resonances. These contributions could be expressed through the universal Isgur-Wise form factors [29],[30], which is done in Appendix C.

In the three doublets model, all the integrands in the last line of Eq. (7.15) are known. To find $(f^i)_0 + (f^{i+1})_0$ we can use solutions (6.35) of the model without corrections, and to find $(\delta f^i + \delta f^{i+1})$ we can use (7.9):

$$\delta f^i + \delta f^{i+1} = \frac{E_i}{M_B}[2q^2 w_1^i + (4 E_e(M_B - E_i - E_e) - q^2) w_2^i + 2 q^2(2 E_e - M_B + E_i) w_3^i] -$$

$$\{\frac{E_i}{M_B}[2q^2 w_1^i + (4 E_e(M_B - E_i - E_e) - q^2) w_2^i + 2 q^2(2 E_e - M_B + E_i) w_3^i]\}_0 +$$

$$\frac{E_{i+1}}{M_B}[2q^2 w_1^{i+1} + (4 E_e(M_B - E_{i+1} - E_e) - q^2) w_2^{i+1} + 2 q^2(2 E_e - M_B + E_{i+1}) w_3^{i+1}] -$$

$$\{\frac{E_{i+1}}{M_B}[2q^2 w_1^{i+1} + (4 E_e(M_B - E_{i+1} - E_e) - q^2) w_2^{i+1} + 2 q^2(2 E_e - M_B + E_{i+1}) w_3^{i+1}]\}_0$$

$$= \frac{E_i}{M_B}[2q^2 \delta w_1^i + (4 E_e(M_B - E_i - E_e) - q^2) \delta w_2^i + 2 q^2(2 E_e - M_B + E_i) \delta w_3^i] +$$

$$\frac{E_{i+1}}{M_B}[2q^2 \delta w_1^{i+1} + (4 E_e(M_B - E_{i+1} - E_e) - q^2) \delta w_2^{i+1} + 2 q^2(2 E_e - M_B + E_{i+1}) \delta w_3^{i+1}]$$

$$\approx \frac{E_i}{M_B}[2q^2 (\delta w_1^i + \delta w_1^{i+1}) + (4 E_e(M_B - E_i - E_e) - q^2)(\delta w_2^i + \delta w_2^{i+1}) +$$

$$2 q^2(2 E_e - M_B + E_i)(\delta w_3^i + \delta w_3^{i+1})]$$

$$= \frac{E_i}{M_B}[2q^2 \delta W_1^i + (4 E_e(M_B - E_i - E_e) - q^2) \delta W_2^i + 2 q^2(2 E_e - M_B + E_i) \delta W_3^i]. \quad (7.16)$$

Finally, we get for the total width:

$$\Gamma = |V_{cb}|^2 \frac{G_F^2}{32 \pi^4} \{ \sum_{i=1,3,5} 2\pi \int_{\Phi_{i+1}} dE_e dq^2 \{ 4(2 E_e M_B - q^2) \frac{1}{M_B}(M_B - E_i - E_e) w^i$$

$$+ \frac{E_i}{M_B}[2q^2 \delta W_1^i + (4 E_e(M_B - E_i - E_e) - q^2) \delta W_2^i + 2 q^2(2 E_e - M_B + E_i) \delta W_3^i]\}$$

$$+ 2\pi \sum_{i=1,3,5} \int_{\Phi_i - \Phi_{i+1}} dE_e dq^2 [2q^2 (w_1^i)_0 + (4 E_e(M_B - E_i - E_e) - q^2)(w_2^i)_0$$

$$+ 2 q^2(2 E_e - M_B + E_i)(w_3^i)_0] \frac{E_i}{M_B}$$

$$+ \int_\Phi dE_e dq^2 dq_0 2\pi [2q^2 w_1^{\text{cont}} + (4 E_e(q_0 - E_e) - q^2) w_2^{\text{cont}} + 2 q^2(2 E_e - q_0) w_3^{\text{cont}}]$$

$$\times \theta(M_B^2 - 2M_B q_0 + q^2 - M_{\text{cont}}^2) \}.$$

In the last equation the first term is the leading order contribution, it contains functions $w^i$ defined in Eq. (5.27)) which are found by solving leading order sum rules equations (6.35), second and third terms are the nonperturbative corrections and the fourth is perturbative corrections to the total width. The functions $(w_j^i)_0$ are read off from the equations (C.2), (C.11), (C.20), $q_0 = (M_B^2 + M_X^2 - q^2)/2M_B$, where $M_X$ is the mass of the corresponding particle. Numerical analysis shows however, that contributions of the phase space regions $\Phi_i - \Phi_{i+1}$ are negligibly small.

The energy distribution is the integrand of the integral over $E_e$, and the double distribution in $E_e$ and $q^2$ is given by the integrand of the integral over $dE_e dq^2$. For each resonance the variables change in the following limits:

$$\begin{aligned} 0 \quad &< \quad q^2 \quad < \quad \frac{2E_e(M_B^2 - 2M_B E_e - M_i^2)}{M_B - 2E_e}, \\ 0 \quad &< \quad E_e \quad < \quad \frac{M_B}{2}(1 - \frac{M_i^2}{M_B^2}). \end{aligned} \qquad (7.17)$$

As for the contribution of continuum states, we recall that in our model it was chosen to be the same as radiative gluon emission correction to the free quark decay starting at some invariant hadronic mass $M_{\min}$. These corrections have been calculated many times before (see for example [17],[32],[20]-[22], [26],[27]), and we are not going to reproduce them here. The range of the variables for the continuum is as follows:

$$\begin{aligned} E_e + \frac{q^2}{4E_e} \quad &< \quad q_0 < \frac{m_b^2 + q^2 - M_{\min}^2}{2m_b} \\ 0 \quad &< \quad q^2 \quad < \quad \frac{2E_e(m_b^2 - 2m_b E_e - M_{\min}^2)}{m_b - 2E_e}, \qquad (7.18) \\ 0 \quad &< \quad E_e \quad < \quad \frac{m_b}{2}(1 - \frac{M_{\min}^2}{m_b^2}). \qquad (7.19) \end{aligned}$$

Use of the quark variables is dictated by the nature of the calculation in which one considers decay of a quark into a quark and gluon. However use of hadronic variables (which is effectively use of $M_B$ instead of $m_b$) is also permissible within our accuracy since we did not take into account corrections of the order $\alpha_s \Lambda_{QCD}^2/m_b^2$.

This effectively concludes formulation of our model.

## 7.4 Numerical analysis. Importance of the nonperturbative corrections to the sum rules.

In this section we are going to investigate how important the nonperturbative corrections to the sum rules are from the point view of affecting predictions of our model. This section is mostly illustrative. More work is needed in order to make reliable connections between the proposed model and real experiments.

Let us make some assumptions about different quantities that enter the model. First about hadron masses. For them we take the following values [45]. For $B$-meson mass $M_B = 5.2\text{GeV}$. For the first resonance doublet: $M_D = 1.87\text{GeV}$, $M_{D^*} = 2.01\text{GeV}$; for the second: $M_{D_2^*} = 2.42\text{GeV}$, $M_{D_1} = 2.46\text{GeV}$. About the third resonance doublet experimental information is not available and for testing of our model we take the following values: $M_{D^{**}} = 3.8\text{GeV}$, $M_{D_1^{**}} = 3.83\text{GeV}$. For the matrix elements of the operators in the OPE we can use the following information. First, as we mentioned earlier, $\mu_G^2 = .36\text{GeV}^2$ is known from the hyperfine mass splitting between $B, B^*$. For $\mu_\pi^2$ there is no direct measurements, but we can use the inequality $\mu_\pi^2 \geq \mu_G^2$ to get an idea about its possible range. For our estimates we take $\mu_\pi^2 = .4\text{GeV}^2$, which is consistent with the other people's estimates (see, for example [23]).

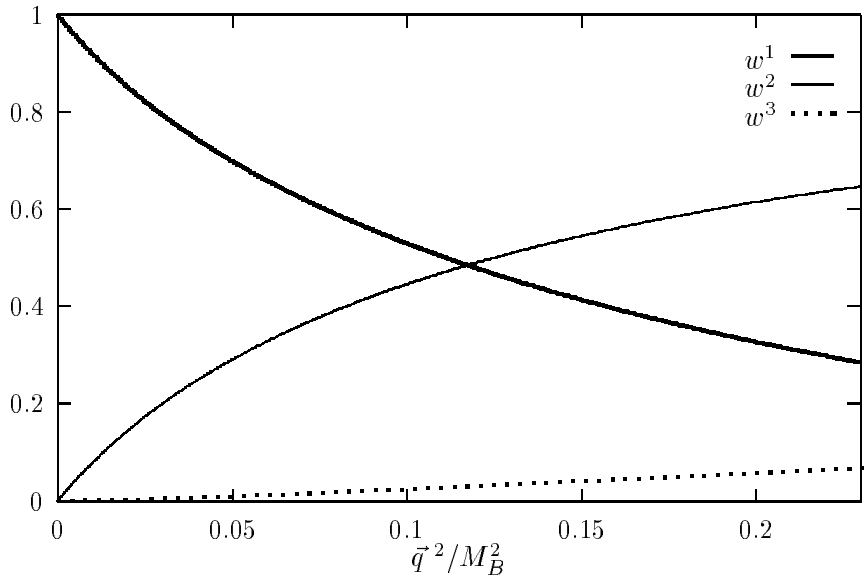

Figure 7.1: Hadronic structure functions $w^i(\mathbf{q}^2)$, $i = 1, 2, 3$, for the model without nonperturbative corrections to the sum rules.

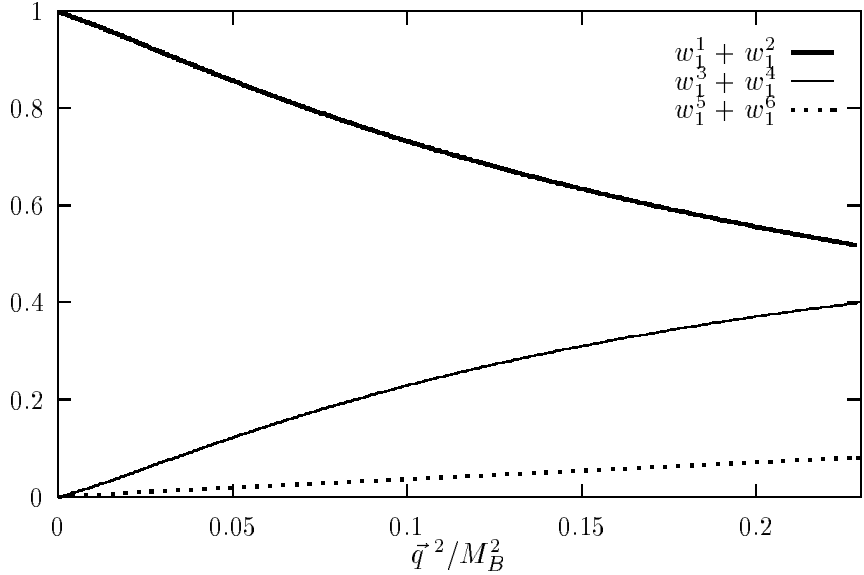

Figure 7.2: Hadronic structure functions $(w_1^i + w_1^{i+1})(\mathbf{q}^2)$ for the model including nonperturbative corrections to the sum rules. These should be compared directly with the functions on Fig. 7.1

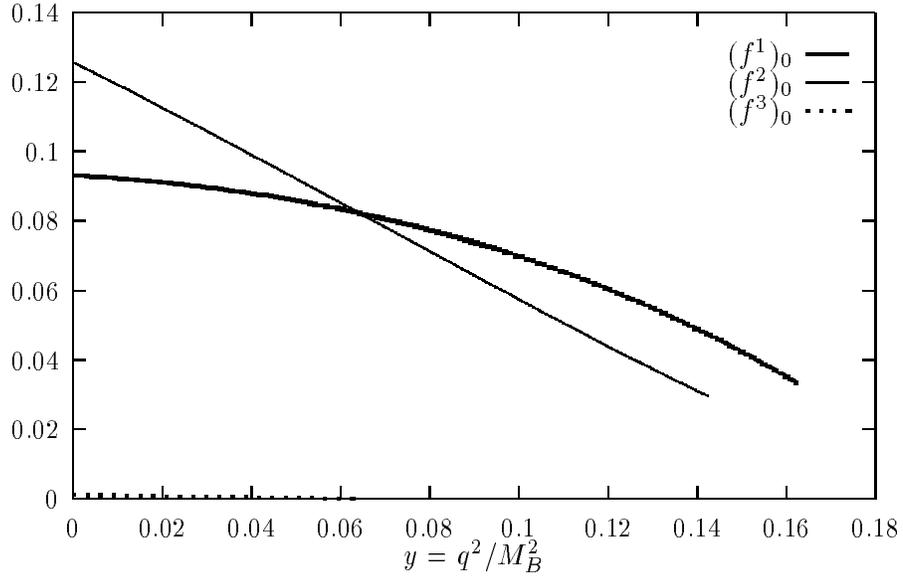

Figure 7.3: Functions $f^i(E_e, q^2)_0$ of the resonance doublets for the model without nonperturbative corrections to the sum rules for $x = 2E_e/M_B = 0.2$.

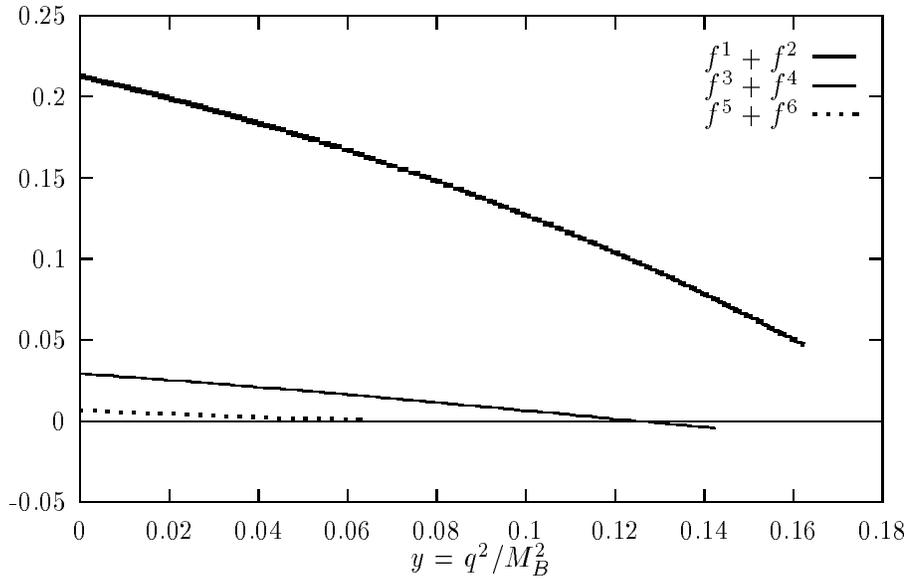

Figure 7.4: Functions $(f^i + f^{i+1})(E_e, q^2)$ for the model including nonperturbative corrections to the sum rules for $x = 2E_e/M_B = 0.2$.

We did calculations of our model both in the heavy quark symmetry limit (without nonperturbative corrections to the sum rules), and without heavy quark symmetry (nonperturbative corrections to the sum rules taken into account). We calculated hadronic structure functions, which in fact determine the corresponding hadronic formfactors, functions $f^i$'s (7.14), which represent contributions of different resonances to the differential distributions, and electron energy distributions.

The hadronic structure functions $w^i(\mathbf{q}^2)$ and $w_1^i(\mathbf{q}^2)$ are plotted on Fig. 7.1 and Fig. 7.2 correspondingly. One can see that the functions are significantly different, which signals that nonperturbative corrections to the sumrules are relatively large and important. Same conclusion follows from the analysis of plots of functions $f^i$'s on Fig. 7.3 and Fig. 7.4. The most different between symmetric and non-symmetric cases is behavior of contributions of the first and second resonance doublets, while the contributions of the third one in both cases is relatively insignificant. This could be an indication of the fact that two doublets approximation of Section 6.2 is sufficient (with nonperturbative and perturbative corrections to the sum rules taken into account).

On Fig. 7.5three normalized electron energy distributions are plotted. Two of them – for the cases of the model with and without nonperturbative corrections to the sum rules taken into account. The third curve shows the electron energy distribution for the free quark decay. We plotted all of them on the same graph in order to demonstrate differences in the electron spectrums. We see that the model predicted spectrum is different from the free quark one. Let us also mention, that despite the fact that the electron spectrums for the models with and without the corrections seem close, the determination of phenomenological parameters from experimental data based on the model with and without the corrections of course will be different.

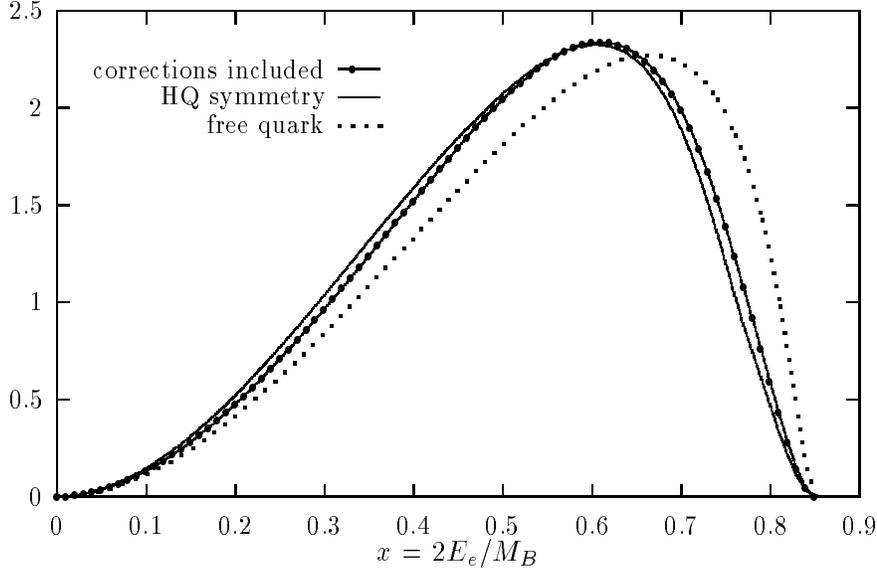

Figure 7.5: Normalized electron spectrum plotted for the model with and without non-perturbative corrections. For comparison, electron spectrum for a free quark decay is also show.

## 7.5 Summary and outlook

The proposed model of the semileptonic decays $B \to X_c l \nu$ is based on some general heavy quark spectroscopy arguments and the operator product expansion sum rules. In this sense the model is so to speak "model-independent", for it does not utilize any quark-antiquark potential nor it introduces any phenomenological parameters like "Fermi momentum" which are not coming from the QCD based analysis. All the parameters of the model have a clear meaning in QCD, like hadronic matrix elements $\mu_\pi^2$, $\mu_G^2$, $\bar{\Lambda}$, $etc$ .

There are still some unanswered questions about the model, however. One of them is about the fourth resonance doublet, which in potential quark models is degenerate with the third one. As it was shown, this is not a problem for the heavy quark symmetry case, but the issue has to be addressed in the full model as well. One of the possible solutions could be considering more sum rules, not just for the $V - A$ current but also for $V$ and $A$ currents separately. This would provide more sum rules and could describe the structure of the final state in more details. Another approach could be not using the notion of the third and fourth resonances at all, but instead use the sum rules to restrain part of the continuum spectrum in the domain where local duality does not work (one can think about it as of some broad resonance merging with the continuum spectrum).

The other direction of improving the model and making it possibly more accurate, is to use Borel transformed sum rules. This would effectively suppress contributions of the higher states along with continuum states, and at the same time better describe the contributions of the lowest resonances.

Numerical analysis of the effects of perturbative corrections also needs to be done.

The main question about the model has to do with the accuracy of the operator product exopansion itself. We have seen that the nonerturbative corrections of order $1/m_Q^2$ to the OPE sum rules strongly affect predictions for the structure functions of the resonances. What would happen if the next order corrections were taken into account? How this would affect determination of the value of $\mu_\pi^2$ from experiment? The $1/m_Q^3$ corrections have been considered in recent works [12],[25], but there is still no conclusive numerical estimates for hasronic invariant functions, which could be used in the considered model.

# Chapter 8

# Conclusions.

In the first part of the work a model independent operator product expansion approach was used to calculate differential distributions in semileptonic decays of heavy flavors in QCD. Double distributions in lepton energy and invariant mass of the lepton pair for the massless lepton (electron) in the final state and the energy spectrum for massive $\tau$-lepton have been calculated.

Based on expressions for hadronic invariant functions, the full set of sum rules for heavy flavor transitions up to order $1/m_Q^2$ was calculated along with perturbative corrections to them up to order $\alpha_s \Lambda_{\rm QCD}/m_Q$.

A new model of semileptonic decays of $B$-mesons was proposed, which is based on the model-independent OPE sum rules approach. It was shown that the OPE sum rules provide sufficient number of restrictions to fully fix contributions of different resonances to the hadronic tensor. In fact, in this approach it is possible to calculate different hadronic structure functions for the transitions induced by the weak currents. Important here is that the proposed model is so to speak "model independent". It does not use any potential quark model calculations and is entirely based on very general symmetry and spectroscopic considerations.

The proposed model could have several uses. On one hand it can realistically describe the end point (maximal lepton energy) part of the lepton spectrum as well as double differential distributions in lepton energy and combined invariant mass of the lepton pair. On the other hand, it provides relations between the lepton spectrum and important phe-nomenological parameters describing strong interactions in heavy mesons, which are given by the matrix elements of operators in the operator product expansion. This should enable experimentalists to extract the OPE matrix elements from the experimentally measured lepton spectrum.

# Appendix A

# The phase space of the decay

In this Appendix we briefly outline the derivation of the Lorentz Invariant Phase Space (LIPS) for the case of the inclusive semileptonic decay $H_b \to \tau \bar{\nu} X$. Let us introduce the invariant kinematical variables in the following way. Let $P$ be the 4-momentum of the decaying particle, so that $M_{H_b}^2 = P^2$; $q = p_\tau + p_{\bar{\nu}}$ is the 4-momentum of the lepton pair, $p_\tau$ and $p_{\bar{\nu}}$ are the momentums of the emitted charged lepton and neutrino correspondingly. We introduce $m_X^2 = p_X^2$, – the invariant mass squared of the born hadronic state and $m_{X\bar{\nu}}^2 = p_{X\bar{\nu}}^2$, – the combined mass of the hadrons and $\bar{\nu}$. We have the following relations: $p_X = P - q$ and $p_{X\bar{\nu}} = P - p_\tau$. The introduced invariant variables are related to the ones used in [13] in the following way:

$$m_X^2 = M_{H_b}^2 + q^2 - 2 M_{H_b} q_0, \tag{A.1}$$

$$m_{X\bar{\nu}}^2 = M_{H_b}^2 + m_\tau^2 - 2 M_{H_b} E_\tau. \tag{A.2}$$

The inclusive decay rate in the normalization of [13] is given by the following expression:

$$\Gamma = \frac{1}{128\pi^6 M_{H_b}} \int_{m_{X\,min}^2}^{m_{X\,max}^2} dm_X^2 \, |\mathcal{M}(m_{X\bar{\nu}}^2, m_X^2, q^2)|^2 \, d\Phi(P, p_X, p_\tau, p_{\bar{\nu}}), \tag{A.3}$$

where $|\mathcal{M}(m_{X\bar{\nu}}^2, m_X^2, q^2)|^2$ is the matrix element which describes the transition into the hadronic states with mass $m_X^2$, it is given by the expressions (2)–(8) in [13]. The lower boundary of $m_X^2$ is given by the mass squared of the lightest final hadronic state possible in the decay ($D$-meson mass for $b \to c$ and pion mass for $b \to u$ transitions ), while the upper boundary of $m_X^2$ is $(M_{H_b} - m_\tau)^2$. The $d\Phi(P, p_X, p_\tau, p_{\bar\nu})$ is the three particle LIPS for a particle with the 4-momentum $P$ going into three particles with 4-momentums $p_X, p_\tau$ and $p_{\bar\nu}$. It is given by the formulae:

$$d\Phi(P, p_X, p_\tau, p_{\bar\nu}) = (2\pi)^4 \delta^4(P - p_X - p_\tau - p_{\bar\nu})\theta(p_X)\theta(p_\tau)\theta(p_{\bar\nu}) \frac{d^4 p_X}{(2\pi)^3} \frac{d^4 p_{\bar\nu}}{(2\pi)^3} \frac{d^4 p_\tau}{(2\pi)^3}, \quad (A.4)$$

where $\theta(a)$ denotes $\theta(a_0)$, and $a_0$ is the zero component of the 4-vector $a$. For the three-particle phase space one can write:

$$d\Phi(P, p_X, p_\tau, p_{\bar\nu}) = \int \frac{dm_{X\bar\nu}^2}{2\pi} d\Phi(P, p_{X\bar\nu}, p_\tau) d\Phi(p_{X\bar\nu}, p_X, p_{\bar\nu}) \delta(m_{X\bar\nu}^2 - p_{X\bar\nu}^2) \quad (A.5)$$

where $d\Phi(a, b, c)$ is a two particle LIPS for a particle with 4-momentum $a$ going into particles with 4-momentums $b$ and $c$. The equation (A.5) gives the decomposition of a 3-particle LIPS into 2-particle LIPSes.

In order to introduce a new variable $q^2$ into the phase space integral one can multiply the equation (A.5) by "1":

$$\int dq^2 \, \delta(q^2 - (p_{\bar\nu} + p_\tau)^2) \theta(q_0) = 1. \quad (A.6)$$

Performing the integrations over all variables except for $m_X^2, q^2$ and $m_{X\bar\nu}^2$, which amounts to integrating the delta functions in LIPSes along with determining the conditions for the delta functions to be non-zeros, we arrive to the following formulae:

$$\Gamma = \frac{1}{512\pi^4 M_{H_b}^3} \int_{M_D^2}^{(M_{H_b}-m_\tau)^2} dm_{X\bar\nu}^2 \int_{M_D^2}^{m_{X\bar\nu}^2} dm_X^2 \int_{q_{min}^2(m_X^2, m_{X\bar\nu}^2)}^{q_{max}^2(m_X^2, m_{X\bar\nu}^2)} dq^2 \, |\mathcal{M}(m_{X\bar\nu}^2, m_X^2, q^2)|^2, \quad (A.7)$$

where

$$q_{min}^2(m_X^2, m_{X\bar\nu}^2) = \frac{1}{2}[M_{H_b}^2 + m_X^2 - m_{X\bar\nu}^2 + m_\tau^2 - \frac{M_{H_b}^2 - m_\tau^2}{m_{X\bar\nu}^2} m_X^2 \quad (A.8)$$

$$- \frac{\sqrt{\lambda(M_{H_b}^2, m_\tau^2, m_{X\bar\nu}^2)}(m_{X\bar\nu}^2 - m_X^2)}{m_{X\bar\nu}^2}], \quad (A.9)$$

$$q_{max}^2(m_X^2, m_{X\bar\nu}^2) = \frac{1}{2}[M_{H_b}^2 + m_X^2 - m_{X\bar\nu}^2 + m_\tau^2 - \frac{M_{H_b}^2 - m_\tau^2}{m_{X\bar\nu}^2} m_X^2 \quad (A.10)$$

$$+ \frac{\sqrt{\lambda(M_{H_b}^2, m_\tau^2, m_{X\bar\nu}^2)}(m_{X\bar\nu}^2 - m_X^2)}{m_{X\bar\nu}^2}], \quad (A.11)$$

and
$$\lambda(a, b, c) = a^2 + b^2 + c^2 - 2\,a\,b - 2\,a\,c - 2\,b\,c. \tag{A.12}$$

The physical meaning of the function $\lambda(a, b, c)$ lies in its relation to the square of the spatial momentum $\mathbf{p}^{*2}$ of particles born in a two - body decay in the center of mass reference frame, for example:

$$\mathbf{p}_{X\bar{\nu}}^{*2} = \mathbf{p}_{\tau}^{*2} = \frac{\lambda(M_{H_b}^2, m_{\tau}^2, m_{X\bar{\nu}}^2)}{4\,M_{H_b}^2}. \tag{A.13}$$

The fully differential distribution (3.1) in the invariant variables looks as follows:

$$\frac{d\Gamma}{dm_X^2\,dm_{X\bar{\nu}}^2\,dq^2} = \frac{1}{512\,\pi^4\,M_{H_b}^3}\,|\mathcal{M}(m_X^2, m_{X\bar{\nu}}^2, dq^2)|^2. \tag{A.14}$$

# Appendix B

# Hadronic invariant functions

Here we present the results of calculations of different hadronic invariant functions $h_i$, introduced by eqs.(2.15), (2.42). The structure functions $w_i$ are simply related to $h_i$ by eqs.(2.16) and (2.52). We use the following notation: $q_0 = q \cdot v$, $^2 = q_0^2 - \mathbf{q}^2$ and $z = m_Q^2 - 2\, m_Q\, q_0 + q^2 - m_q^2$.

For the Vector×Vector functions we have:

$$h_1^{VV} = -\,[\,(\,m_Q - m_q - q_0\,) - (\mu_G^2 - \mu_\pi^2\,)\frac{1}{2m_Q}(\frac{1}{3} + \frac{m_q}{m_Q}\,)\,]\,\frac{1}{z} -$$
$$\frac{1}{m_Q}\,[\,\frac{1}{3}\,\mu_G^2\,(\,(4\,m_Q - 3q_0)(\,m_Q - m_q - q_0) + 2\,\mathbf{q}^2\,) +$$
$$\mu_\pi^2\,(q_0\,(m_Q - m_q - q_0) - \frac{2}{3}\mathbf{q}^2\,)\,]\,\frac{1}{z^2} - \frac{4}{3}\,\mu_\pi^2\,\mathbf{q}^2\,(m_Q - m_q - q_0)\,\frac{1}{z^3}\,, \qquad \text{(B.1)}$$

$$h_2^{VV} = -\,[\,2\,m_Q - \frac{5}{3m_Q}\,(\mu_G^2 - \mu_\pi^2)\,]\,\frac{1}{z} -$$
$$\frac{2}{3}\,[\,2\mu_G^2\,(m_Q - m_q) - 5\,\mu_G^2\,q_0 + 7\,\mu_\pi^2\,q_0]\,\frac{1}{z^2} - \frac{8}{3}\,m_Q\,\mu_\pi^2\,\mathbf{q}^2\,\frac{1}{z^3}, \qquad \text{(B.2)}$$

$$h_3^{VV} = 0\,, \qquad \text{(B.3)}$$

$$h_4^{VV} = -\frac{4}{3m_Q}\,(\mu_\pi^2 - \mu_G^2)\,\frac{1}{z^2}\,, \qquad \text{(B.4)}$$

$$h_5^{VV} = \frac{1}{z} - \frac{1}{3}\,[\,5\,\frac{q_0}{m_Q}(\mu_G^2 - \mu_\pi^2\,) - 4\,\mu_\pi^2)\,]\,\frac{1}{z^2} + \frac{4}{3}\,\mu_\pi^2\,\mathbf{q}^2\,\frac{1}{z^3}\,. \qquad \text{(B.5)}$$

To get the functions $h_i^{AA}$ for Axial×Axial tensor from $h_i^{VV}$ one should substitute $m_q$ by $(-m_q)$ in eqs.(B.1 - B.5).

For the Axial×Vector tensor only one invariant structure survives:

$$h_3^{VA} = \frac{1}{z} + [\, 2\,\mu_G^2 + \frac{5}{3}\,(\mu_\pi^2 - \mu_G^2)\,\frac{q_0}{m_Q}\,]\,\frac{1}{z^2} + \frac{4}{3}\,\mu_\pi^{2\,2}\,\frac{1}{z^3}\,. \tag{B.6}$$

Summing up we get the result for the full hadronic tensor $h_{\mu\nu}$.

$$h_1 = -\,[\, 2\,(m_Q - q_0) - \frac{1}{3 m_Q}\,(\mu_G^2 - \mu_\pi^2)\,]\,\frac{1}{z} -$$
$$[\,\frac{2}{3 m_Q}\,\mu_G^2\,(4\,m_Q^2 + 2\,\mathbf{q}^2 - 7\,m_Q\,q_0 + 3\,q_0^2) + \frac{\mu_\pi^2}{2 m_Q}\,(4\,q_0\,(m_Q - q_0) - \frac{8}{3}\,\mathbf{q}^2\,)\,]\,\frac{1}{z^2} -$$
$$\frac{8}{3}\,\mu_\pi^2\,\mathbf{q}^2\,(m_Q - q_0)\,\frac{1}{z^3}\,, \tag{B.7}$$

$$h_2 \;=\; -\,[\, 4\,m_Q + \frac{10}{3 m_Q}\,(\mu_\pi^2 - \mu_G^2)\,]\,\frac{1}{z} - [\,\frac{28}{3}\,\mu_\pi^2\,q_0 + \mu_G^2\,(\frac{8}{3}\,m_Q - \frac{20}{3}\,q_0)\,]\,\frac{1}{z^2} -$$
$$\frac{16}{3}\,\mu_\pi^2\,m_Q\,\mathbf{q}^2\,\frac{1}{z^3}\,, \tag{B.8}$$

$$h_3 = -\,2\,\frac{1}{z} - [\,4\,\mu_G^2 + \frac{10}{3}\,(\mu_\pi^2 - \mu_G^2)\,\frac{q_0}{m_Q}\,]\,\frac{1}{z^2} - \frac{8}{3}\,\mu_\pi^2\,\mathbf{q}^2\,\frac{1}{z^3}\,, \tag{B.9}$$

$$h_4 = -\,\frac{8}{3 m_Q}\,(\mu_\pi^2 - \mu_G^2)\,\frac{1}{z^2}\,, \tag{B.10}$$

$$h_5 = 2\,\frac{1}{z} - \frac{2}{3}\,[\,5\,(\mu_G^2 - \mu_\pi^2)\,\frac{q_0}{m_Q} - 4\,\mu_\pi^2)\,]\,\frac{1}{z^2} + \frac{8}{3}\,\mu_\pi^2\,\mathbf{q}^2\,\frac{1}{z^3}\,. \tag{B.11}$$

# Appendix C

# Structure functions of the resonances in the heavy quark limit

In this appendix we are going to write down expressions for the contributions of separate resonances to the structure functions in the leading order in the heavy quark mass. We need these expressions for the formulation of the full model of the decays which takes into account independence of different structure functions and mass differences between particles in the same heavy quark doublets. Due to these mass differences there are regions in the phase space of the decay, where only one (the heavier) resonance of the doublet contributes.

These formulas could be obtained using the results of the original works of Isgur and Wise [29],[30].

Now we need to write down parts of the hadronic tensor coming from each resonance in the heavy quark limit. For our purposes we only need to know the structure functions for the $V - A$ currents.

$$
\begin{aligned}
W_{\mu\nu} &= \sum_X (2\pi)^4 \delta^4(p_B - q - p_X) \frac{1}{2M_B} \langle B|j_\mu^\dagger(0)|P_X,\lambda\rangle\langle P_X,\lambda|j_\nu(0)|B\rangle \\
&= 2\pi \sum_i \delta(M_B - q_0 - \sqrt{M_i^2 + \mathbf{q}^2}) \frac{1}{2M_B} \sum_\lambda \{\langle B|V_\mu^\dagger(0)|i,\mathbf{q},\lambda\rangle\langle i,\mathbf{q},\lambda|V_\nu(0)|B\rangle \\
&\quad - \langle B|V_\mu^\dagger(0)|i,\mathbf{q},\lambda\rangle\langle i,\mathbf{q},\lambda|A_\nu(0)|B\rangle - \langle B|A_\mu^\dagger(0)|i,\mathbf{q},\lambda\rangle\langle i,\mathbf{q},\lambda|V_\nu(0)|B\rangle \\
&\quad + \langle B|A_\mu^\dagger(0)|i,\mathbf{q},\lambda\rangle\langle i,\mathbf{q},\lambda|A_\nu(0)|B\rangle\} \\
&= 2\pi \sum_i w_{\mu\nu}(i) \delta(M_B - q_0 - \sqrt{M_i^2 + \mathbf{q}^2}),
\end{aligned}
$$

where $i$ runs over all resonances of our model (not just the resonance doublets), $\lambda$ runs over all polarizations of the corresponding particle and

$$w_{\mu\nu}(i) = \frac{1}{2M_B}\sum_\lambda \{\langle B|V_\mu^\dagger(0)|i,\mathbf{q},\lambda\rangle\langle i,\mathbf{q},\lambda|V_\nu(0)|B\rangle - \langle B|V_\mu^\dagger(0)|i,\mathbf{q},\lambda\rangle\langle i,\mathbf{q},\lambda|A_\nu(0)|B\rangle$$
$$-\langle B|A_\mu^\dagger(0)|i,\mathbf{q},\lambda\rangle\langle i,\mathbf{q},\lambda|V_\nu(0)|B\rangle + \langle B|A_\mu^\dagger(0)|i,\mathbf{q},\lambda\rangle\langle i,\mathbf{q},\lambda|A_\nu(0)|B\rangle\}.$$

Now we define the structure functions of the resonances in the following way:

$$w_{\mu\nu}(i) = -w_1^i g_{\mu\nu} + w_2^i v_\mu v_\nu - i w_3^i \epsilon_{\mu\nu\alpha\beta} v^\alpha q^\beta + w_4^i q_\mu q_\nu + w_5^i (q_\nu v_\mu + q_\mu v_\nu),$$

so that

$$w_j(q_0,\mathbf{q}^2) = 2\pi \sum_i w_j^i(\mathbf{q}^2)\delta(M_B - \sqrt{M_i + \mathbf{q}^2} - q_0).$$

To calculate the hadronic tensor we will also need formulas for sums over spins of the vector and tensor particles in the final states:

$$\sum_\lambda \epsilon_\mu(v',\lambda)\epsilon_\nu(v',\lambda) = -g_{\mu\nu} + v'_\mu v'_\nu,$$
$$\sum_\lambda \epsilon_{\mu\nu}(v',\lambda)\epsilon_{\rho\sigma}(v',\lambda) = \frac{1}{2}(P_{\mu\rho}P_{\nu\sigma} + P_{\mu\sigma}P_{\nu\rho}) - \frac{1}{3}P_{\mu\nu}P_{\rho\sigma},$$

where the projection operator $P_{\mu\nu} = g_{\mu\nu} - v'_\mu v'_\nu$.

Now we can write down the contribution of each resonance to the hadronic tensor in the heavy quark limit (let us remind the spectroscopic notation: $n^{2S+1}L_J$, here $n$ is the radial quantum number, $S$ is the combined spin of the two quarks inside the meson, $L$ is the orbital momentum of the light quark and $J$ is the spin of the meson):
for pseudoscalar $D \to |1^1S_0\rangle$ and vector $D^* \to |1^3S_1\rangle$:

$$\langle D|V_\mu|B\rangle = \sqrt{2M_B}\eta_1(w)(v_\mu + v'_\mu),$$
$$\langle D|A_\mu|B\rangle = 0,$$
$$\langle D^*|A_\mu|B\rangle = \sqrt{2M_B}\eta_1(w)\{(w+1)\epsilon_\mu - \epsilon\cdot v v'_\mu\},$$
$$\langle D^*|V_\mu|B\rangle = i\sqrt{2M_B}\eta_1(w)\epsilon_{\mu\delta\alpha\beta}\epsilon^\delta v^\alpha v'^\beta,$$
$$w_{\mu\nu}(D) = \eta_1^2(w)\{v_\mu v_\nu + (v_\mu v'_\nu + v'_\mu v_\nu) + v'_\mu v'_\nu\},$$
$$= \eta_1^2(w)\frac{1}{M_X^2}\{(M_B + M_X)^2 v_\mu v_\nu - (M_B + M_X)(q_\nu v_\mu + q_\mu v_\nu) + q_\mu q_\nu\},$$

$$\begin{aligned}
w_{\mu\nu}(D^*) &= \eta_1^2(w)\{-2w(1+w)g_{\mu\nu} - v_\mu v_\nu - 2i(1+w)\epsilon_{\mu\nu a\beta}v^\alpha v'^\beta \\
&\quad +(1+2w)(v_\mu v'_\nu + v'_\mu v_\nu) - v'_\mu v'_\nu\}. \\
&= \eta_1^2(w)\frac{1}{M_X^2}\{-2(M_B - q_0)(M_B + M_X - q_0)g_{\mu\nu} \\
&\quad +(3M_B^2 + 2M_B M_X - M_X^2 - 4M_B q_0)v_\mu v_\nu + 2i(M_B + M_X - q_0)\epsilon_{\mu\delta\alpha\beta}v^\alpha q^\beta \\
&\quad -(M_B + M_X - 2q_0)(q_\nu v_\mu + q_\mu v_\nu) - q_\mu q_\nu\},
\end{aligned} \qquad (C.1)$$

$$\begin{aligned}
w_{\mu\nu}(D) + w_{\mu\nu}(D^*) &= 2(w+1)\eta_1^2(w)\{-wg_{\mu\nu} - i\epsilon_{\mu\nu\alpha\beta}v^\alpha v'^\beta + (v_\nu v'_\mu + v_\mu v'_\nu)\} \\
&= \frac{2(M_B + M_X - q_0)}{M_X^2}\eta_1^2(w)\{-(M_B - q_0)g_{\mu\nu} \\
&\quad + 2M_B v_\mu v_\nu + i\epsilon_{\mu\nu\alpha\beta}v^\alpha q^\beta - (q_\nu v_\mu + q_\mu v_\nu)\},
\end{aligned} \qquad (C.2)$$

$$w^1 = 2(w+1)\eta_1^2(w) = \frac{2(M_B + M_X - q_0)}{M_X}\eta_1^2(w), \qquad (C.3)$$

$$w_1^1 = 0, \qquad (C.4)$$

$$w_2^1 = \eta_1^2(w)\frac{(M_B + M_X)^2}{M_X^2}, \qquad (C.5)$$

$$w_3^1 = 0, \qquad (C.6)$$

$$w_1^2 = 2\eta_1^2(w)\frac{(M_B - q_0)(M_B + M_X - q_0)}{M_X^2}, \qquad (C.7)$$

$$w_2^2 = \eta_1^2(w)\frac{(3M_B^2 + 2M_B M_X - M_X^2 - 4M_B q_0)}{M_X^2}, \qquad (C.8)$$

$$w_3^2 = 2\eta_1^2(w)\frac{(M_B + M_X - q_0)}{M_X^2}, \qquad (C.9)$$

for tensor $D_2^* \to |1^3P_2\rangle$ and axial-vector $D_1 \to \sqrt{\frac{3}{2}}|1^1P_1\rangle + \sqrt{\frac{1}{2}}|1^3P_1\rangle$:

$$\begin{aligned}
\langle D_2^*|V_\mu|B\rangle &= i\sqrt{2M_B}\sqrt{3}\eta_2(w)\epsilon_{\mu\alpha\beta\delta}\epsilon^{\alpha\nu}v_\nu v^\beta v'^\delta, \\
\langle D_2^*|A_\mu|B\rangle &= \sqrt{2M_B}\sqrt{3}\eta_2(w)\{(w+1)\epsilon_{\mu\alpha}v^\alpha - v'_\mu\epsilon_{\alpha\beta}v^\alpha v^\beta\}, \\
\langle D_1|V_\mu|B\rangle &= \sqrt{2M_B}\frac{1}{\sqrt{2}}\eta_2(w)\{-(w^2-1)\epsilon_\mu + (-3v_\mu + (w-2)v'_\mu)\epsilon\cdot v\}, \\
\langle D_1|A_\mu|B\rangle &= -\sqrt{2M_B}\frac{i}{\sqrt{2}}\eta_2(w)(w+1)\epsilon_{\mu\alpha\beta\delta}\epsilon^\alpha v^\beta v'^\delta,
\end{aligned}$$

$$\begin{aligned}
w_{\mu\nu}(D_2^*) &= \eta_2^2(w)\{-w(w-1)(1+w)^2 g_{\mu\nu} \\
&\quad +(1+w)(w-2)v_\mu v_\nu - i(w-1)(1+w)^2\epsilon_{\mu\nu\alpha\beta}v^\alpha v'^\beta
\end{aligned}$$

$$
\begin{aligned}
&+(1+w)(-2+2w+w^2)(v_\nu v'_\mu + v'_\nu v_\mu) + (1+w)(w-2)v'_\mu v'_\nu\} \\
&= \eta_2^2(w)\frac{(M_B+M_X-q_0)}{M_X^4}\{-(M_B-M_X-q_0)(M_B+M_X-q_0)(M_B-q_0)g_{\mu\nu} \\
&+(3M_B^3+2M_B^2M_X \\
&-3M_BM_X^2-2M_X^3-5M_B^2q_0-4M_BM_Xq_0-M_X^2q_0+2M_Bq_0^2)v_\mu v_\nu \\
&+i(M_B+M_X-q_0)(M_B-M_X-q_0)\epsilon_{\mu\nu\alpha\beta}v^\alpha q^\beta \\
&-(2M_B^2-2M_X^2-3M_Bq_0-2M_Xq_0+q_0^2)(q_\nu v_\mu + q_\mu v_\nu) \\
&+(M_B-2M_X-q_0)q_\mu q_\nu\},
\end{aligned}
$$

$$
\begin{aligned}
w_{\mu\nu}(D_1) &= \eta_2^2(w)\{-3w(w-1)(1+w)^2 g_{\mu\nu} \\
&\quad -(1+w)(w-2)v_\mu v_\nu - 3i(w-1)(1+w)^2\epsilon_{\mu\nu\alpha\beta}v^\alpha v'^\beta \\
&\quad +(1+w)(-2-2w+3w^2)(v_\nu v'_\mu + v'_\nu v_\mu) - (w-2)(1+w)v'_\mu v'_\nu\} \\
&= \eta_2^2(w)\frac{(M_B+M_X-q_0)}{M_X^4}\{-3(M_B-M_X-q_0)(M_B+M_X-q_0)(M_B-q_0)g_{\mu\nu} \\
&\quad +(5M_B^3-2M_B^2M_X \\
&\quad -5M_BM_X^2+2M_X^3-11M_B^2q_0+4M_BM_Xq_0+M_X^2q_0+6M_Bq_0^2)v_\mu v_\nu \\
&\quad +3i(M_B+M_X-q_0)(M_B-M_X-q_0)\epsilon_{\mu\nu\alpha\beta}v^\alpha q^\beta \\
&\quad -(2M_B^2-2M_X^2-5M_Bq_0+2M_Xq_0+3q_0^2)(q_\nu v_\mu + q_\mu v_\nu) \\
&\quad -(M_B-2M_X-q_0)q_\mu q_\nu\}, \quad\quad (C.10)
\end{aligned}
$$

$$
\begin{aligned}
w_{\mu\nu}(D_2^*)+w_{\mu\nu}(D_1) &= 4(w-1)(w+1)^2\eta_2^2(w)\{-wg_{\mu\nu}-i\epsilon_{\mu\nu\alpha\beta}v^\alpha v'^\beta + (v_\nu v'_\mu + v_\mu v'_\nu)\} \\
&= \frac{4(M_B+M_X-q_0)^2(M_B-M_X-q_0)}{M_X^4}\eta_2^2(w)\{-(M_B-q_0)g_{\mu\nu} \\
&\quad +2M_Bv_\mu v_\nu + i\epsilon_{\mu\nu\alpha\beta}v^\alpha q^\beta - (q_\nu v_\mu + q_\mu v_\nu)\}, \quad\quad (C.11)
\end{aligned}
$$

$$
w^2 = 4(w-1)(w+1)^2\eta_2^2(w) = \frac{4(M_B+M_X-q_0)^2(M_B-M_X-q_0)}{M_X^3}\eta_2^2(w); \quad (C.12)
$$

$$
w_1^3 = \eta_2^2(w)\frac{(M_B+M_X-q_0)^2}{M_X^4}(M_B-M_X-q_0)(M_B-q_0), \quad (C.13)
$$

$$
\begin{aligned}
w_2^3 &= \eta_2^2(w)\frac{(M_B+M_X-q_0)}{M_X^4}(3M_B^3+2M_B^2M_X-3M_BM_X^2-2M_X^3- \\
&\quad 5M_B^2q_0-4M_BM_Xq_0-M_X^2q_0+2M_Bq_0^2), \quad\quad (C.14)
\end{aligned}
$$

$$w_3^3 = \eta_2^2(w)\frac{(M_B + M_X - q_0)^2}{M_X^4}(M_B - M_X - q_0), \tag{C.15}$$

$$w_1^4 = 3\eta_2^2(w)\frac{(M_B + M_X - q_0)^2}{M_X^4}(M_B - M_X - q_0)(M_B - q_0), \tag{C.16}$$

$$w_2^4 = \eta_2^2(w)\frac{(M_B + M_X - q_0)}{M_X^4}(5M_B^3 - 2M_B^2 M_X - 5M_B M_X^2 + 2M_X^3 - 11M_B^2 q_0 +$$
$$4M_B M_X q_0 + M_X^2 q_0 + 6M_B q_0^2), \tag{C.17}$$

$$w_3^4 = 3\eta_2^2(w)\frac{(M_B + M_X - q_0)^2}{M_X^4}(M_B - M_X - q_0). \tag{C.18}$$

for scalar $D^{**} \to |1^3P_0\rangle$ and axial-vector $D_1^{**} \to \sqrt{\frac{3}{2}}|1^1P_1\rangle - \sqrt{\frac{1}{2}}|1^3P_1\rangle$:

$$\langle D^{**}|V_\mu|B\rangle = 0,$$
$$\langle D^{**}|A_\mu|B\rangle = -2\sqrt{2M_B}\eta_3(w)(v_\mu - v'_\mu),$$
$$\langle D_1^{**}|V_\mu|B\rangle = 2\sqrt{2M_B}\eta_3(w)\{(w-1)\epsilon_\mu - \epsilon \cdot v v'_\mu\},$$
$$\langle D_1^{**}|A_\mu|B\rangle = 2i\sqrt{2M_B}\eta_3(w)\epsilon_{\mu\delta\alpha\beta}\epsilon_\delta v_\beta v'_\delta,$$
$$w_{\mu\nu}(D^{**}) = 4\eta_3^2(w)\{-(v_\nu v'_\mu + v'_\nu v_\mu) + v_\mu v_\nu + v'_\mu v'_\nu\}$$
$$= 4\eta_3^2(w)\frac{1}{M_X^2}\{(M_B - M_X)^2 v_\mu v_\nu - (M_B - M_X)(q_\nu v_\mu + q_\mu v_\nu) + q_\mu q_\nu\},$$
$$w_{\mu\nu}(D_1^{**}) = 4\eta_3^2(w)\{-2w(w-1)g_{\mu\nu} - v_\mu v_\nu - 2i(w-1)\epsilon_{\mu\nu\alpha\beta}v^\alpha v'^\beta$$
$$+ (2w-1)(v_\nu v'_\mu + v'_\nu v_\mu) - v'_\mu v'_\nu\}$$
$$= 4\eta_3^2(w)\frac{1}{M_X^2}\{-2(M_B - M_X - q_0)(M_B - q_0)g_{\mu\nu}$$
$$+ (-(M_B + M_X)^2 + 4M_B(M_B - q_0))v_\mu v_\nu$$
$$+ 2i(M_B - M_X - q_0)\epsilon_{\mu\nu\alpha\beta}v^\alpha q^\beta$$
$$- (M_B - M_X - 2q_0)(q_\nu v_\mu + q_\mu v_\nu) - q_\mu q_\nu\}, \tag{C.19}$$

$$w_{\mu\nu}(D^{**}) + w_{\mu\nu}(D_1^{**}) = 8(w-1)\eta_3^2(w)\{-wg_{\mu\nu} - i\epsilon_{\mu\nu\alpha\beta}v^\alpha v'^\beta + (v_\nu v'_\mu + v_\mu v'_\nu)\}$$
$$= \frac{8(M_B - M_X - q_0)}{M_X^2}\eta_3^2(w)\{-(M_B - q_0)g_{\mu\nu}$$
$$+ 2M_B v_\mu v_\nu + i\epsilon_{\mu\nu\alpha\beta}v^\alpha q^\beta - (q_\nu v_\mu + q_\mu v_\nu)\}, \tag{C.20}$$

$$w^3 = 8(w-1)\eta_3^2(w) = \frac{8(M_B - M_X - q_0)}{M_X}\eta_3^2(w);$$
$$w_1^5 = 0, \tag{C.21}$$

$$w_2^5 = 4\eta_3^2(w)\frac{(M_B - M_X)^2}{M_X^2}, \tag{C.22}$$

$$w_3^5 = 0, \tag{C.23}$$

$$w_1^6 = 8\eta_3^2(w)\frac{(M_B - M_X - q_0)(M_B - q_0)}{M_X^2}, \tag{C.24}$$

$$w_2^6 = 4\eta_3^2(w)\frac{-(M_B + M_X)^2 + 4M_B(M_B - q_0)}{M_X^2}, \tag{C.25}$$

$$w_3^6 = 8\eta_3^2(w)\frac{(M_B - M_X - q_0)}{M_X^2}. \tag{C.26}$$

For the fourth doublet which is (pseudoscalar,vector), like the first one, the formulas are exactly the same as corresponding expressions for the first one, with the evident substitution $\eta_1 \to \eta_4$.

In the above equations $\eta_1(w)$, $\eta_2(w)$ and $\eta_3(w)$ are the universal Isur-Wise form-factors for the three resonance doublets respectively; $M_X$ are the masses of the corresponding resonance doublets; $v, v'$ are velocities of the initial and final heavy mesons ($v' = (M_B - q_0)/M_X$), $w = v \cdot v' = E_X/M_X = (M_B - q_0)/M_X$, (note that $w \geq 1$); $\epsilon_\nu$, $\epsilon_{\alpha\nu}$ are the polarization vectors and tensors for the corresponding (pseudo)vector and tensor mesons in the final state.

Functions $w^1, w^2, w^3$ are known from the solutions (6.35) to the sum rules equations (6.6-6.8), which in turn determines the form factors $\eta_1, \eta_2, \eta_3$. The structure functions for each of the resonances could be read off from the above equations.